\documentclass[12pt]{article}
\usepackage[T1]{fontenc}
\usepackage[utf8]{inputenc}
\usepackage[sc]{mathpazo}
\usepackage{graphicx}
\usepackage{xcolor}
\usepackage{amsmath}
\usepackage{booktabs}
\usepackage{longtable}
\usepackage{threeparttable}
\usepackage{microtype}
\usepackage{enumitem}
\usepackage{float}
\usepackage{setspace}
\usepackage{caption}
\usepackage{subcaption}
\usepackage{fvextra}
\usepackage{ragged2e}
\usepackage[top=1in,bottom=1in,left=1in,right=1in]{geometry}
\usepackage{pdflscape}
\usepackage{rotating}
\usepackage{titlesec}
\usepackage{fancyvrb}
\usepackage[
  backend=biber,
  style=authoryear,
  natbib=false,
  giveninits=true,
  maxcitenames=2,
  maxbibnames=99,
  uniquename=false,
  uniquelist=false
]{biblatex}
\usepackage[colorlinks]{hyperref}
\hypersetup{
  linkcolor=blue,
  citecolor=blue,
  urlcolor=blue
}

\titleformat{\section}{\large\bfseries}{\thesection}{0.75em}{}
\titleformat{\subsection}{\normalsize\bfseries}{\thesubsection}{0.75em}{}
\titleformat{\subsubsection}{\normalsize\itshape}{\thesubsubsection}{0.75em}{}
\titleformat{\paragraph}[runin]{\normalfont\bfseries}{\theparagraph}{0.75em}{}[.]
\titlespacing*{\section}{0pt}{1.0ex plus 0.2ex minus 0.1ex}{0.55ex plus 0.1ex}
\titlespacing*{\subsection}{0pt}{0.8ex plus 0.2ex minus 0.1ex}{0.35ex plus 0.1ex}
\titlespacing*{\subsubsection}{0pt}{0.6ex plus 0.2ex minus 0.1ex}{0.25ex plus 0.1ex}
\titlespacing*{\paragraph}{0pt}{0.6ex plus 0.2ex minus 0.1ex}{0.75em}

\setlist{itemsep=0.15em,topsep=0.25em,parsep=0pt,partopsep=0pt}
\AtBeginBibliography{\small\setstretch{1.0}\setlength{\itemsep}{0pt}\setlength{\parskip}{0pt}}
\DeclareFieldFormat[article]{journaltitle}{#1}
\renewbibmacro{in:}{}
\DeclareNameAlias{sortname}{family-given}
\DeclareNameAlias{default}{family-given}
\DefineVerbatimEnvironment{PromptVerbatim}{Verbatim}{
  breaklines=true,
  breakanywhere=true,
  fontsize=\small
}

\makeatletter
\def\@maketitle{%
  \newpage
  \null
  \vspace*{-4.0em}%
  \begin{center}%
    {\LARGE\scshape\@title\par}%
    \vskip 1.0em%
    {\normalsize\lineskip .5em\begin{tabular}[t]{c}\@author\end{tabular}\par}%
  \end{center}%
  \par\vskip 0.75em}
\makeatother

\newcommand{\EnterpriseWorkerCharacteristicsFirmCount}{1,764}

\newcommand{\EnterpriseWorkerCharacteristicsMessageCount}{17,446,551}
\newcommand{\EnterpriseTaskClassificationFirmCount}{973}

\newcommand{\EnterpriseTaskClassificationMessageCount}{8,696,657}
\newcommand{\EnterpriseWorkerCharacteristicsHorizonWeeks}{26}
\newcommand{\EnterpriseWorkerCharacteristicsHorizonMonths}{6}
\newcommand{\EnterpriseTaskClassificationHorizonWeeks}{26}
\newcommand{\EnterpriseTaskClassificationStartDate}{October 30, 2025}

\newcommand{\EnterpriseEngineeringTechnicalPractitionerSharePercent}{11}
\newcommand{\EnterpriseExecutiveFounderPartnerSharePercent}{9}
\newcommand{\EnterpriseFinanceAccountingRoleSharePercent}{5}

\newcommand{\EnterpriseSalesAccountRoleSharePercent}{4}
\newcommand{\EnterpriseEarlyCareerTraineeSharePercent}{7}
\newcommand{\EnterpriseICProfessionalSharePercent}{15}
\newcommand{\EnterpriseSeniorICPrincipalSharePercent}{14}
\newcommand{\EnterpriseManagerDirectorSharePercent}{24}
\newcommand{\EnterpriseExecutiveSenioritySharePercent}{10}

\newcommand{\CapIQPublicCompanyFinancialCapitalDigitalFitChangeBaseYear}{2024}

\newcommand{\CapIQUsageMergeDomesticPublicCompanyTickerCount}{534}
\newcommand{\CapIQUsageMergeTickerYearObservationCount}{521}
\newcommand{\CapIQUsageMergePublicCompanyTickerCount}{417}
\newcommand{\CapIQUsageMergeUnassignedTickerCount}{117}
\newcommand{\CapIQUsageMergeNoFiscalYearWindowTickerCount}{50}
\newcommand{\CapIQUsageMergeAfterLastFiscalYearWindowTickerCount}{67}
\newcommand{\CapIQUsageMergePositiveMessageTickerYearObservationCount}{509}
\newcommand{\CapIQUsageMergePositiveMessagePublicCompanyTickerCount}{410}

\newcommand{\CapIQUsageMergeNonAdopterPublicCompanyTickerCount}{11,784}
\newcommand{\CapIQAdoptionRevenuePerEmployeeCoefficientMinPP}{0.4}
\newcommand{\CapIQAdoptionRevenuePerEmployeeCoefficientMaxPP}{0.9}
\newcommand{\CapIQAdoptionSuperstarLogRevenueCoefficientPP}{1.1}
\newcommand{\CapIQAdoptionSuperstarTopRevenueQuartileCoefficientPP}{6.9}
\newcommand{\CapIQAdoptionSuperstarTopRevenueFivePercentCoefficientPP}{9.8}
\newcommand{\CapIQAdoptionSuperstarIndustryYearTopRevenueQuartileCoefficientPP}{7.2}
\newcommand{\CapIQAdoptionSuperstarIndustryYearTopRevenueFivePercentCoefficientPP}{11.3}
\newcommand{\CapIQAdoptionComplementFiscalYearStart}{2024}
\newcommand{\CapIQAdoptionComplementFiscalYearEnd}{2025}
\newcommand{\CapIQAdoptionComplementStockYear}{2021}
\newcommand{\CapIQAdoptionComplementSGADepreciationPercent}{20}
\newcommand{\CapIQAdoptionComplementRDDepreciationPercent}{15}
\newcommand{\CapIQAdoptionComplementSGAAllCoefficient}{0.020}
\newcommand{\CapIQAdoptionComplementSGANoTechHighRDCoefficient}{0.010}
\newcommand{\CapIQAdoptionComplementRDAllCoefficient}{0.004}
\newcommand{\CapIQAdoptionComplementRDNoTechHighRDCoefficient}{0.003}
\newcommand{\CapIQAdoptionComplementCapitalizedSoftwareAllCoefficient}{0.008}
\newcommand{\CapIQAdoptionComplementCapitalizedSoftwareNoTechHighRDCoefficient}{0.006}

\newcommand{\CapIQPublicCompanyRevenueAdopterChangeBaseMedianMillions}{2,275.1}

\newcommand{\CapIQPublicCompanyRevenueNonAdopterChangeBaseMedianMillions}{209.6}

\newcommand{\CapIQPublicCompanyTotalAssetsAdopterChangeBaseMedianMillions}{4,394.2}

\newcommand{\CapIQPublicCompanyTotalAssetsNonAdopterChangeBaseMedianMillions}{667.6}

\newcommand{\CapIQPublicCompanyEmployeesAdopterChangeBaseMedianWorkers}{2,934}

\newcommand{\CapIQPublicCompanyEmployeesNonAdopterChangeBaseMedianWorkers}{424}

\newcommand{\CapIQPublicCompanyNetPPEAdopterChangeBaseMedianMillions}{271.2}

\newcommand{\CapIQPublicCompanyNetPPENonAdopterChangeBaseMedianMillions}{43.7}

\newcommand{\CapIQPublicCompanyMarketValueAdopterChangeBaseMedianMillions}{4,997.2}

\newcommand{\CapIQPublicCompanyMarketValueNonAdopterChangeBaseMedianMillions}{316.4}

\newcommand{\CapIQPublicCompanyRDExpenseAdopterChangeBaseMedianMillions}{113.1}

\newcommand{\CapIQPublicCompanyRDExpenseNonAdopterChangeBaseMedianMillions}{9.9}

\begin{document}
\thispagestyle{empty}

\vspace*{\fill}

\begin{center}
{\fontsize{17}{19}\selectfont\bfseries
How Organizations Use AI: Evidence from ChatGPT\par}
\vspace{0.2em}

{\large
Aaron Chatterji\textsuperscript{1}
\quad
David Holtz\textsuperscript{2,1}
\quad
Neel Rakholia\textsuperscript{1}
\par}

\vspace{0.1em}

{\large
Prasanna Tambe\textsuperscript{3,1}
\quad
Gawesha Weeratunga\textsuperscript{1}
\par}

\vspace{0.4em}

{\normalsize
\textsuperscript{1}OpenAI
\quad
\textsuperscript{2}Columbia Business School
\par}
\vspace{0.1em}

{\normalsize
\textsuperscript{3}Wharton School, U. Pennsylvania
\par}
\vspace{0.1em}
Last Updated: \today
\end{center}
\begingroup
\renewcommand{\thefootnote}{}
\footnotemark
\footnotetext{Working paper, results are subject to change. The authors are especially grateful to David Deming for feedback and guidance during the development of this work. We also thank Nicholas Otis, Harrison Satcher, Cassandra Duchan-Solis, Drew Johnston, Laura Bisesto, Joyce Shi, Emma Kollek, Jake Stamell, David Zimmerman, Zach Parent, Brice Challamel, Steve Imm, Rob Friedlander, Jennifer Robinson, Erik Boxhoorn, Vinnie Monaco, Meng Jia Yang, Peter Zhang, Hendrik Buyle, Chris Liao, Nathalie Gazzaneo, Adam Cohen, Kevin Wadman, Ignacio Lopez-Gaffney, Wesley Pasfield, and the OpenAI Analytics \& Insights team for helpful comments and assistance. We also thank Shane Greenstein (discussant) and participants in the NBER 2026 Summer Institute meeting on Digital Economics and Artificial Intelligence, as well as attendees of the 2026 Wharton AI and the Future of Work Conference for valuable feedback. David Holtz and Prasanna Tambe contributed to this work in their capacity as paid contractors for OpenAI. Contact: Aaron Chatterji, \href{mailto:ronnie@openai.com}{ronnie@openai.com}; David Holtz, \href{mailto:david.holtz@columbia.edu}{david.holtz@columbia.edu}; Neel Rakholia, \href{mailto:neelrakholia@openai.com}{neelrakholia@openai.com}; Prasanna Tambe, \href{mailto:tambe@wharton.upenn.edu}{tambe@wharton.upenn.edu}; Gawesha Weeratunga, \href{mailto:gawesha@openai.com}{gawesha@openai.com}.}
\endgroup

\vspace{-.30in}

\begin{center}
{\bfseries Abstract\par}
\vspace{0.4em}

\begin{minipage}{0.87\textwidth}
\small\setstretch{1}\setlength{\parindent}{0pt}
We study how organizations use frontier generative AI by linking ChatGPT Enterprise account records to usage, worker roles, task classifications, and public-company financial data through March 2026. These linked data enable a privacy-preserving analysis of adoption, worker roles, and message-level tasks at scale: for instance, the worker-level sample we analyze at the six-month adoption horizon includes over 1,500 organizations and over 17 million messages. We document four facts about enterprise AI adoption and use. First, ChatGPT Enterprise usage has grown rapidly due to a combination of new firm adoption and growing intensity among existing adopters. Second, U.S.-based public company adoption is concentrated among larger, more valuable, and more R\&D- and SG\&A-intensive firms. Third, active use within adopting firms spans job functions and seniority levels, with especially high usage intensity among early-career workers. Fourth, ChatGPT Enterprise usage encompasses a broad range of knowledge work tasks, including writing, technical work, communication, and information synthesis. In aggregate, these results suggest that firms differ widely in the speed, breadth and purpose of their enterprise AI adoption, and that they are still actively learning how to integrate AI into organizational workflows.
\par\vspace{3pt}\noindent\textbf{JEL Codes:} O33, O30, L23, M15
\end{minipage}
\end{center}

\vspace*{\fill}

\clearpage
\section{Introduction}

Generative AI systems can perform a growing range of economically valuable tasks \parencite{eloundou2024gpts,patwardhan2025gdpval}, and individuals use consumer-facing generative AI chatbots for many work-related and personal activities \parencite{chatterji2025how,handa2025economictasks}. However, less is known about firm AI adoption: which workers account for observed use, how intensively active users engage with generative AI, and for which tasks. Understanding these patterns is important for interpreting recent findings about the impact of AI on productivity and employment (e.g., \cite{brynjolfsson2025canaries}; \cite{brynjolfsson2025generative}). Most evidence on firm AI adoption comes from worker and firm surveys \parencite{mcelheran2024ai,bick2024rapid,yotzov2026firm,bonney2026microstructure}. Although surveys provide broad coverage and can capture non-use, adoption barriers, and organizational context, self-reported usage is typically less detailed and may suffer from imperfect recall or reporting biases. 

This paper studies workplace AI adoption using internal data from ChatGPT Enterprise, OpenAI's centrally administered workplace product.\footnote{Over time, OpenAI's centrally administered workplace offering has expanded beyond ChatGPT to include additional products, such as Codex. For simplicity, and because ChatGPT accounts for the overwhelming majority of observed usage during our study period, we refer to this offering throughout as ``ChatGPT Enterprise.''} We first examine firm adoption by decomposing enterprise growth into within- and between-firm components and linking ChatGPT Enterprise accounts to public company financial data. We then study within-firm heterogeneity by combining usage data with employee job title information and message-level task classifications, which tells us how usage intensity and task adoption varies across worker groups six months after organizational adoption.

We document four stylized facts about enterprise AI adoption and usage. First, enterprise AI usage is growing rapidly, reflecting both increased use among existing customers and the arrival of new adopters. Aggregate output tokens produced by ChatGPT Enterprise customers grew roughly sevenfold between June 2025 and March 2026, and by nearly fourfold within a consistent cohort of firms that adopted between January 2024 and June 2025. Thus, about half of the growth in token consumption over this period occurred within already-adopting firms. Second, among U.S.-based public companies, ChatGPT Enterprise adopters are larger, more valuable, and more R\&D- and SG\&A-intensive than non-adopters. This pattern suggests that early enterprise AI adoption is associated with greater prior investment in intangible and organizational capabilities.

Third, usage within adopting firms is broadly distributed across job title classes and seniority levels but with heterogeneous intensity. For example, marketing and communications workers send more messages than executives, and early-career workers send many more messages than more senior employees. Fourth, ChatGPT Enterprise use spans many different tasks across workers and organizations, rather than being concentrated in a single workflow. The most common use cases are writing, communication, and information synthesis, but usage is also common in tasks such as research, planning, data analysis, legal and regulatory work, finance, and many other applications. This breadth is consistent with generative AI functioning as a general purpose technology for knowledge work \parencite{bresnahan1995gpt,bresnahan2024gpt, eloundou2024gpts}.

Taken together, these findings portray enterprise AI adoption as a broad but uneven organizational phenomenon. Adoption is concentrated among firms with greater scale and intangible investment, while use within adopting firms is distributed across many worker groups and knowledge work tasks but varies substantially in intensity. This heterogeneity highlights that the long-run economic value of enterprise AI adoption will depend on whether dispersed individual use develops into complementary organizational capabilities \parencite{bresnahan1996coinvention,bresnahan2002information,brynjolfsson2021productivity}. 

The remainder of the paper proceeds as follows: we first review the related literature and describe the data and measurement. We then follow the structure introduced above: Sections~\ref{Sec:Growth} and~\ref{Sec:Adoption} examine adoption and usage across firms, while Sections~\ref{Sec:Distribution} and~\ref{Sec:Tasks} examine the distribution and task structure of use within adopting organizations. In Section~\ref{Sec:Conclusion}, we conclude.

\section{Related Literature}

Our work contributes to four related literatures. First, we add to the literature on AI adoption and diffusion within firms. A central insight from research on general purpose technologies is that initial adoption does not imply effective deployment: realizing value requires experimentation, complementary investment, and organizational change, so use often diffuses gradually within firms \parencite{bresnahan1995gpt,bresnahan1996coinvention,bresnahan2002information, bresnahan2024gpt,brynjolfsson2021productivity,mansfield1963intrafirm, fuentelsaz2003intrafirm}. Consistent with this view, recent studies of digital technology adoption and use find that organizations continue to discover applications and improve their use after obtaining access \parencite{mcelheran2024ai,yotzov2026firm,bick2024rapid,bick2026mind,brand2024firm,kim2026mapping,anthropic2026aeiv5}.\footnote{Related studies connect firm-level AI investment and exposure to firm growth, product innovation, and market value \parencite{babina2024artificial,eisfeldt2023generative,yu2026adoption}.} The most closely related paper to ours is \textcite{bonney2026microstructure}, which distinguishes firm adoption from the subsequent deployment of AI across business functions and worker tasks. We advance this literature by examining which firm attributes predict enterprise AI adoption and, among adopters at a common point in their adoption cycles, measuring how deployment is distributed across workers and tasks.

Second, our paper contributes to research that measures AI usage with telemetry data. One strand uses data on the content of human–AI interactions to characterize the tasks, occupations, and modes of interaction represented in observed use \parencite{handa2025economictasks,anthropic2026aeiv4,anthropic2026aeiv6,chatterji2025how,tomlinson2025working}. A second strand uses API and product-activity data to characterize demand across applications and organizational settings and to examine how AI is incorporated into production workflows \parencite{demirer2025market,fradkin2025demand,appelmccrorytamkin2025geoapi,daniotti2026using,chen2026artificial,demirer2026writing}. Two recent papers are particularly closely related to ours. \Textcite{counts2026ai} use telemetry from Microsoft 365 Copilot to characterize aggregate workplace use and document how its task composition varies across occupations and industries. \Textcite{johnston2026shift} use OpenAI telemetry data to study the shift from conversational to agentic AI, including how Codex adoption, usage intensity, and task composition vary across organizational settings, worker roles, and levels of seniority. 
%Our paper differs from these studies because we combine AI usage data with firm and worker characteristics, which provides a detailed picture of adoption within and across firms. 

Third, we contribute to research on heterogeneous effects of AI across workers and tasks. This literature distinguishes between the activities for which AI is technically capable and the settings in which those capabilities translate into realized use and benefits. Early work takes a task-based approach and estimate potential exposure by comparing model capabilities with occupational task descriptions \parencite{eloundou2024gpts}.\footnote{Subsequent work emphasizes that technical exposure need not translate directly into realized adoption, because workers vary in comparative advantage and in the costs of using and verifying AI output \parencite{lindenlaub2026beyond}.} More recently, \textcite{patwardhan2025gdpval} evaluate frontier models on expert-constructed tasks spanning 44 occupations and nine sectors, providing evidence about where models can produce professional-quality deliverables. Experimental studies, in turn, show that AI's \textit{realized} effects depend on the worker, the task, and how AI-generated output is evaluated and implemented \parencite{noy2023experimental,brynjolfsson2025generative,dellacqua2026jagged, cui2026highskilled,otis2026uneven}. We complement this work by documenting the deployment decisions that bridge the gap between capability and realized effects: which worker groups account for active use, how intensively users in each group engage with AI, and which tasks account for their activity.

Finally, our paper contributes to research on the relationship between AI, organizational structure, and the division of labor. Knowledge-based theories of the firm view organizational hierarchies as mechanisms for allocating problems among workers, managers, and specialized experts \parencite{garicano2000hierarchies,garicano2006organization}. Technologies that change the cost of acquiring or communicating knowledge can therefore change where expertise is located and how tasks are divided within the firm \parencite{bloom2014distinct}. Recent research applies this logic to AI by examining how it changes the value of expertise and which sequences of work can be delegated to AI \parencite{autor2025expertise,demirer2026chaining}, and within-firm evidence shows that intensive AI use can expand the range of tasks workers perform, accelerate learning, and shift work toward supervising and evaluating AI-generated output \parencite{huang2025aiwork}. The paper most closely related to ours in its organizational focus is \textcite{kim2026ainative}, which shows that AI-native startups are smaller, flatter, and more engineering-intensive than comparable firms. Whereas much of this work examines AI-native organizations and highly technical workers, we study how AI is deployed within the existing structures of established firms across a wide range of industries. By comparing use across job functions, seniority levels, and managerial positions, we provide evidence on where a general purpose AI technology enters the organizational hierarchy and how its role varies across organizational contexts.

\section{Data and Measurement}

\begin{figure}[htpb]
\centering
\caption{CONSTRUCTION OF CHATGPT ENTERPRISE ADOPTION AND USAGE SAMPLES}
\label{fig:dataset-construction}

\includegraphics[width=0.80\linewidth]{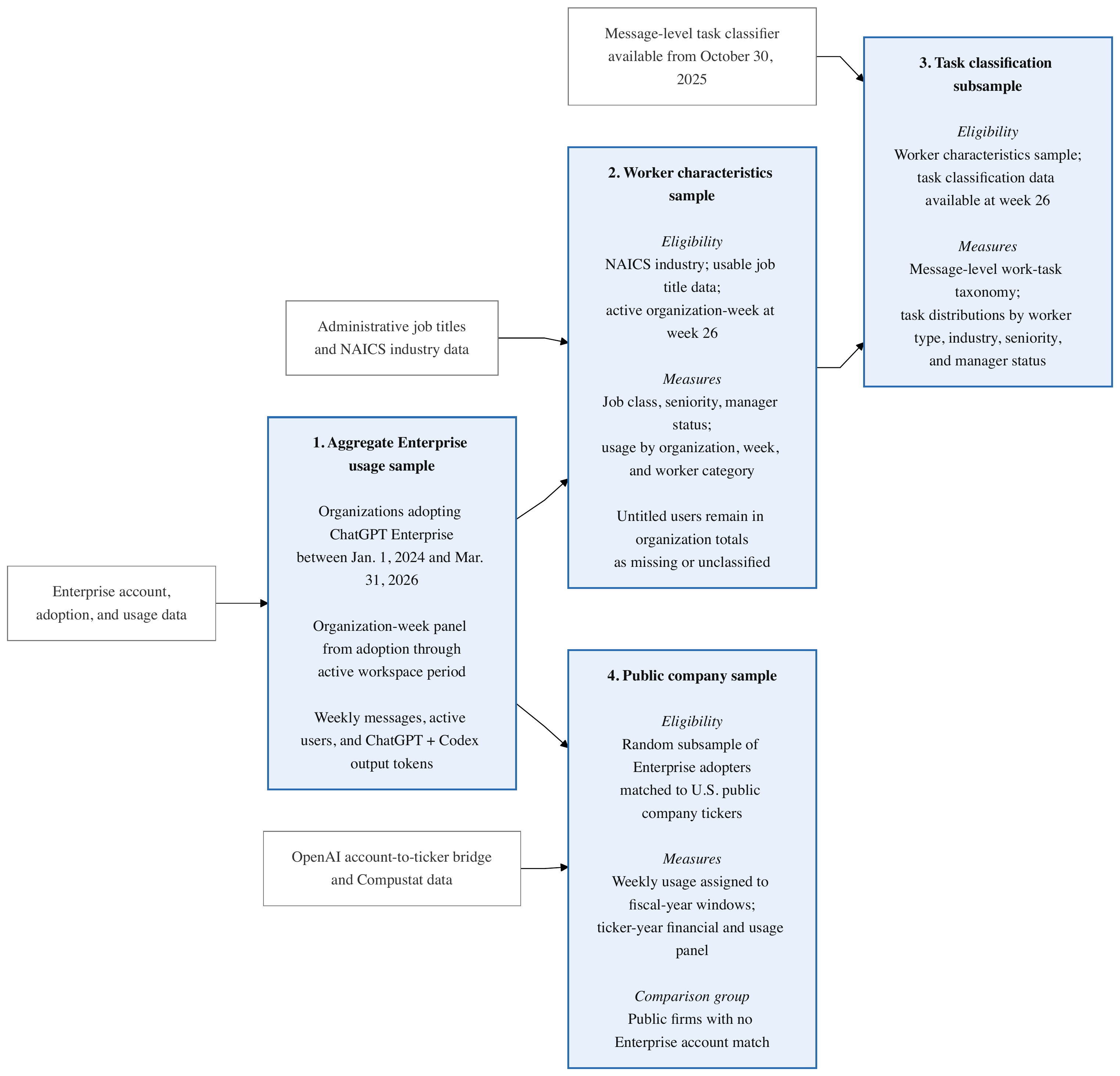}

\vspace{0.75em}
\begin{minipage}{0.85\linewidth}
\footnotesize
\emph{Note:} This figure summarizes the construction of the four samples used in the analysis. Blue boxes denote analysis samples, while white boxes denote additional data sources used to construct or augment them. Arrows indicate sample restrictions and data linkages. 
\end{minipage}
\end{figure}

Our analysis draws on four related but distinct samples: an aggregate enterprise usage sample, a smaller sample with employee job title and firm industry information, a further time-limited subset of the job title and industry sample used for task-classification analysis, and a public company sample linked to the Compustat database from S\&P Global Market Intelligence.\footnote{All Compustat data are copyright \copyright 2019, S\&P Global Market Intelligence. Reproduction of any information, data or material, including ratings (``Content'') in any form is prohibited except with the prior written permission of the relevant party. Such party, its affiliates and suppliers (“Content Providers”) do not guarantee the accuracy, adequacy, completeness, timeliness or availability of any Content and are not responsible for any errors or omissions (negligent or otherwise), regardless of the cause, or for the results obtained from the use of such Content. In no event shall Content Providers be liable for any damages, costs, expenses, legal fees, or losses (including lost income or lost profit and opportunity costs) in connection with any use of the Content.} We describe the construction of each sample in the following subsections and summarize their relationships in Figure~\ref{fig:dataset-construction}.

For our analysis of ChatGPT Enterprise usage data, we use de-identified data and report results only in aggregate. Message content is classified using automated systems, and job title metadata is mapped to broad job title class, seniority, and people manager categories. No researcher manually reviewed individual enterprise customer messages for this study. For our financial analysis of public companies, we securely link aggregate organizational usage data to public-company financial information from Compustat.

\subsection{ChatGPT Enterprise Usage Data} \label{Sec:usage_data}

Our primary data source is an organization-week panel of ChatGPT Enterprise adoption and usage, constructed from organizations whose ChatGPT Enterprise adoption dates range from January 1, 2024 to March 31, 2026. The data capture adoption of a paid, centrally administered ChatGPT Enterprise workspace, rather than use through personal accounts, the API, or other subscription plans. We observe each organization's enterprise account identifier, adoption date, and product usage over time.

Organizations enter the panel in the week they adopt ChatGPT Enterprise and they remain in the panel while their workspace is active. Organization-weeks with an active workspace but no observed product activity are retained with zero measured usage. For each organization-week, we measure messages sent, active users, and generated output tokens, including tokens generated through both ChatGPT and Codex. Weeks are indexed relative to each organization's adoption date. This aggregate ChatGPT Enterprise usage sample is used to measure adoption and product use over time.

\subsection{Job Titles, Firm Industries, and Task Classifications}
\label{sec:role}

For analyses of usage by worker characteristics, we also construct a sample of ChatGPT Enterprise organizations for which we observe both firm industry and high-quality employee job title information. Starting from the ChatGPT Enterprise usage sample described above, we retain organizations that can be assigned to a broad industry category using NAICS classifications. We further require that at least some user activity within the organization can be linked to a non-empty administrative job title.\footnote{Job title information is observed at a single point in time and may therefore not capture changes in workers' roles over the full study period.} Because the corresponding analyses measure usage \EnterpriseWorkerCharacteristicsHorizonMonths{} months (\EnterpriseWorkerCharacteristicsHorizonWeeks{} weeks) after adoption, we additionally require an observed, active organization-week at that horizon. The resulting worker characteristics sample contains \EnterpriseWorkerCharacteristicsFirmCount{} organizations and \EnterpriseWorkerCharacteristicsMessageCount{} messages.

Within this sample, we normalize the available job title strings and classify them into broad job title classes, seniority levels, and people manager categories.\footnote{Our main text analyses focus on job title class and seniority; results for people manager status appear in Appendix~\ref{sec:add_figures}.} These user-title and firm-industry measures are then linked to ChatGPT Enterprise usage and aggregated by organization, week, and worker category. Appendix~\ref{app:job-title-classification} describes the normalization, classification, and validation of our job title measures. Importantly, job title coverage within included organizations is incomplete. Active users without usable job title information remain in organization-level usage totals
and denominators but are classified as missing or unclassified in analyses of
heterogeneous use by worker type.

We also separately construct a task classification subsample of this dataset. We classify ChatGPT Enterprise messages into a taxonomy of work tasks using a message-level classifier that is available beginning on \EnterpriseTaskClassificationStartDate. Appendix~\ref{app:task-classification} provides information about the task taxonomy produced by this classifier. The task classification sample is restricted to organizations that satisfy the worker characteristics sample requirements above and have task classification data available at the week \EnterpriseTaskClassificationHorizonWeeks{} horizon. This produces a task-classification sample of \EnterpriseTaskClassificationFirmCount{} organizations and \EnterpriseTaskClassificationMessageCount{} classified messages. We use this task classification sample for both analysis of the overall task distribution and for analyses of the task distributions by job title class, seniority level, people manager status, and industry.

\subsection{Public Company Sample and Summary Statistics}

For analyses of U.S. public company characteristics and financial outcomes, we begin with U.S. public firms in Compustat and identify ChatGPT Enterprise adopters by mapping ChatGPT Enterprise accounts to public-company identifiers using a combined curated and LLM-assisted account-to-ticker crosswalk.\footnote{The public-company analyses are limited to U.S. firms because the Compustat data used here are restricted to U.S. public companies.} We then draw a random sample from the resulting set of ChatGPT Enterprise accounts with public-company identifiers.\footnote{We draw a random sample to limit disclosure of commercially sensitive information about OpenAI's enterprise business.} We define non-adopters as public firms with no ChatGPT Enterprise ticker-bridge match.\footnote{We classify as non-adopters firms that we cannot match to a ChatGPT Enterprise account; this classification does not imply that they use no AI products. These firms may use enterprise language-model products from other providers or access ChatGPT outside an enterprise workspace. If such use is correlated with firm characteristics in the same direction as ChatGPT Enterprise adoption, this measurement error could attenuate estimated differences between adopters and non-adopters. We do exclude public firms that match OpenAI customer records for other products (like ChatGPT Business) but not a ChatGPT Enterprise account, which avoids the use of known OpenAI customers as untreated controls.} The resulting financial panel includes firm and fiscal-year identifiers as well as the Compustat variables reported in Table~\ref{tab:capiq-measures}.\footnote{Importantly, this public company sample does not require job title, industry, or task classification coverage. It is therefore distinct from the samples described in Section~\ref{sec:role}, although both begin with the ChatGPT Enterprise usage panel.} 

We link weekly ChatGPT Enterprise activity to this panel by assigning each usage week to the Compustat fiscal year in which it falls and aggregating usage to the ticker-year level. For usage-intensity analyses, we measure weekly messages per employee, weekly output tokens per employee, and weekly active users (WAU) per employee. The first two measures capture usage volume relative to firm size, while WAU captures the breadth of participation in the ChatGPT Enterprise workspace.

This procedure yields a usage-linked panel of \CapIQUsageMergeTickerYearObservationCount{} ticker-year observations for \CapIQUsageMergePublicCompanyTickerCount{} public-company tickers.\footnote{We retain a usage-matched ticker only when at least one observed week of ChatGPT Enterprise usage falls within an observed Compustat fiscal-year window. Of \CapIQUsageMergeDomesticPublicCompanyTickerCount{} usage-matched domestic public-company tickers, \CapIQUsageMergeUnassignedTickerCount{} do not meet this condition: \CapIQUsageMergeNoFiscalYearWindowTickerCount{} have no observed Compustat fiscal-year window, and \CapIQUsageMergeAfterLastFiscalYearWindowTickerCount{} have ChatGPT Enterprise usage only after the last available Compustat fiscal-year window.} Of these usage-linked ticker-years, \CapIQUsageMergePositiveMessageTickerYearObservationCount{} observations, covering \CapIQUsageMergePositiveMessagePublicCompanyTickerCount{} tickers, have positive annual ChatGPT Enterprise message volume. The non-adopter comparison group contains \CapIQUsageMergeNonAdopterPublicCompanyTickerCount{} public-company tickers. These counts describe U.S.-based public-company adopters for which we observe matched ChatGPT Enterprise usage; sample sizes in regression tables may differ because the set of Compustat variables required as non-missing covariates varies across specifications.

\section{Four Facts about Enterprise AI Usage}

Using the datasets described above, we document four facts about the growth and composition of ChatGPT Enterprise use. First, aggregate use has grown rapidly, including within cohorts of organizations that adopted at different times. Second, early adoption is concentrated among larger, more productive firms with greater prior investment in intangible and organizational complements. Third, use is broadly distributed across job functions and seniority levels, but its intensity varies systematically across worker groups. Finally, ChatGPT Enterprise use spans a broad range of knowledge work tasks, while task mix varies across industries, job functions, and levels of seniority.

\subsection{ChatGPT Enterprise usage has grown rapidly} \label{Sec:Growth}

\begin{figure}[htbp]
\centering
\caption{ENTERPRISE OUTPUT TOKEN GROWTH}
\label{fig:token-demand}
\vspace{0.25em}
\includegraphics[width=\linewidth]{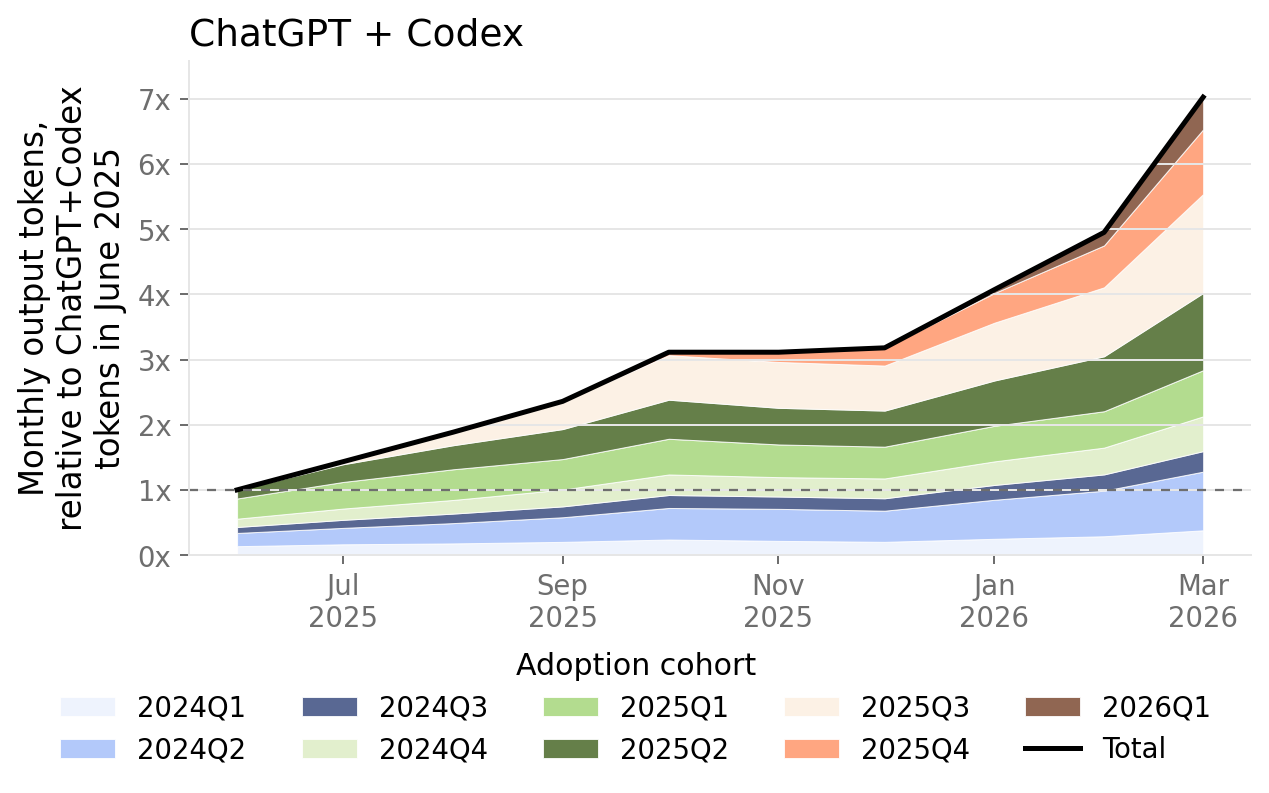}
\vspace{1em}
\begin{minipage}{0.85\linewidth}
\footnotesize
\emph{Note:} This figure plots monthly ChatGPT and Codex output tokens for firms in the ChatGPT Enterprise analysis sample, stacked by the quarter in which each firm first adopted ChatGPT Enterprise. Tokens are counted only after each firm’s ChatGPT Enterprise adoption date. Values are indexed to total combined ChatGPT and Codex output tokens in June 2025, with the dashed horizontal line marking 1.0 and the black line showing the aggregate total across cohorts.
\end{minipage}
\end{figure}

We first study the growth of ChatGPT Enterprise usage, both overall and within fixed adoption cohorts. Using the organization-week panel described in Section~\ref{Sec:usage_data}, Figure~\ref{fig:token-demand} plots total output tokens relative to June 2025, overall and separately by quarter-year adoption cohorts.\footnote{ChatGPT Enterprise increasingly includes access to Codex as part of the product bundle. During our sample period, however, enterprise token output is overwhelmingly generated by ChatGPT and related non-agentic AI tools. Appendix Figure~\ref{fig:token-demand-app} shows that the aggregate growth documented here is driven primarily by non-Codex output. For an analysis of Codex usage within organizations, we refer the reader to \cite{johnston2026shift}.}

Aggregate output tokens increased sevenfold between June 2025 and March 2026. This growth was not driven solely by the addition of new organizations: output tokens also increased substantially within existing adoption cohorts. Among firms that had adopted by June 2025, for example, output tokens increased roughly fourfold over the same period. Thus, enterprise demand continued to deepen after organizations entered the product, alongside continued growth in the number of adopters. We also find that usage accelerated in early 2026 across all adoption cohorts. Because organizations that adopted at different times experienced this acceleration simultaneously, the increase appears to reflect developments affecting ChatGPT Enterprise customers broadly rather than only the normal expansion of use following adoption. 

\subsection{Early enterprise AI adopters are larger, more valuable, and more heavily invested in intangibles and organizational complements} \label{Sec:Adoption}

We next examine how firms that adopt ChatGPT Enterprise differ from non-adopters and which firm characteristics are associated with usage intensity among adopters.

\begin{figure}[htpb]
\centering
\caption{SUMMARY STATISTICS: COMPARISON OF CHATGPT ENTERPRISE ADOPTERS WITH OTHER PUBLIC FIRMS}
\label{fig:public-company-financial-capital-digital-fit}

\includegraphics[width=0.85\linewidth]{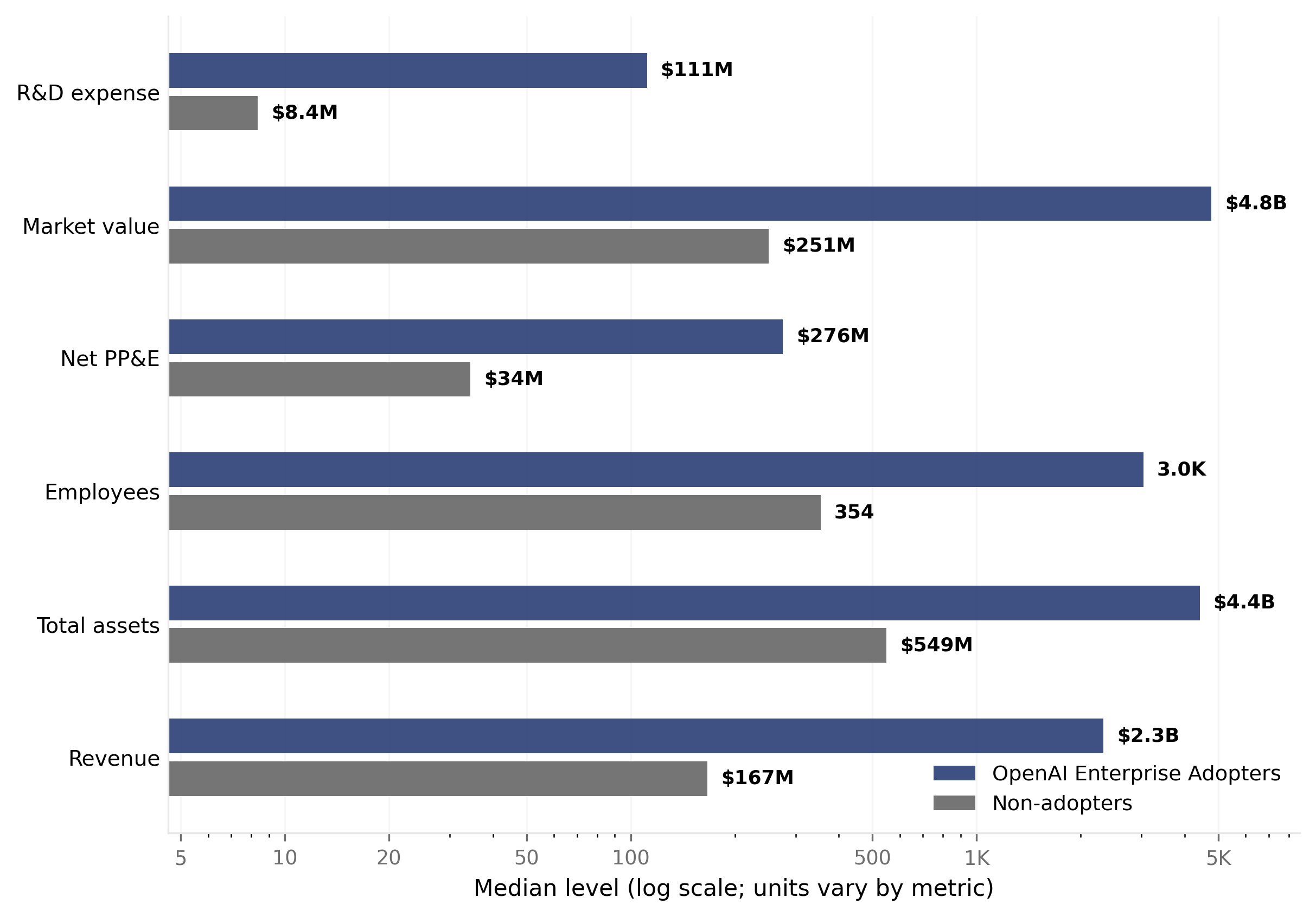}

\vspace{0.75em}
\begin{minipage}{0.85\linewidth}
\footnotesize
\emph{Note:} This figure compares median 2024 Compustat firm characteristics for public firms linked to ChatGPT Enterprise accounts with those of other public firms. Dollar-denominated variables are reported in millions of dollars, and employment is measured in workers. The underlying data combine Compustat annual business metrics with ChatGPT Enterprise account-to-ticker matches.
\end{minipage}
\end{figure}

Figure~\ref{fig:public-company-financial-capital-digital-fit} provides descriptive statistics from a comparison of \CapIQPublicCompanyFinancialCapitalDigitalFitChangeBaseYear{} firm characteristics for ChatGPT Enterprise adopters and non-adopters in the Compustat public-company sample. Across all measures, adopters are substantially larger. Median revenue is \$\CapIQPublicCompanyRevenueAdopterChangeBaseMedianMillions{}M for adopters versus \$\CapIQPublicCompanyRevenueNonAdopterChangeBaseMedianMillions{}M for non-adopters. Median total assets are \$\CapIQPublicCompanyTotalAssetsAdopterChangeBaseMedianMillions{}M versus \$\CapIQPublicCompanyTotalAssetsNonAdopterChangeBaseMedianMillions{}M, and median employment is \CapIQPublicCompanyEmployeesAdopterChangeBaseMedianWorkers{} workers versus \CapIQPublicCompanyEmployeesNonAdopterChangeBaseMedianWorkers{} workers. Adopters also have larger capital stocks, greater market valuations, and greater R\&D spending. Median net property, plant, and equipment (PP\&E) assets are \$\CapIQPublicCompanyNetPPEAdopterChangeBaseMedianMillions{}M for adopters compared with \$\CapIQPublicCompanyNetPPENonAdopterChangeBaseMedianMillions{}M for non-adopters. Median market value is \$\CapIQPublicCompanyMarketValueAdopterChangeBaseMedianMillions{}M versus \$\CapIQPublicCompanyMarketValueNonAdopterChangeBaseMedianMillions{}M, and the median research and development expenses are \$\CapIQPublicCompanyRDExpenseAdopterChangeBaseMedianMillions{}M among adopters, compared with \$\CapIQPublicCompanyRDExpenseNonAdopterChangeBaseMedianMillions{}M among non-adopters. These patterns indicate that early ChatGPT Enterprise adopters are not representative of the average public firm; they are larger, more capitalized, more valuable, and more R\&D-intensive.

\subsubsection{Financial characteristics}

These unadjusted differences motivate the regression analysis in Table~\ref{tab:adoption-outcome-financial-characteristics}, which relates ChatGPT Enterprise adoption to lagged financial characteristics measured at the public firm-year level. Importantly, our estimates describe conditional associations between these pre-adoption firm characteristics and the probability that a public firm is observed as a ChatGPT Enterprise adopter, and should not be interpreted causally.

Across specifications, firms with higher revenue per employee are more likely to adopt ChatGPT Enterprise. A one-log-point increase in lagged revenue per employee is associated with roughly \CapIQAdoptionRevenuePerEmployeeCoefficientMinPP{} to \CapIQAdoptionRevenuePerEmployeeCoefficientMaxPP{} percentage points higher adoption probability. This association remains positive and statistically significant after accounting for assets per employee, PP\&E per employee, employment, year fixed effects, and industry fixed effects, indicating that it is not solely a difference between larger firms or more capital-intensive industries. Firm scale is also strongly associated with adoption: lagged log employment is positive and statistically significant in all specifications, including those with additional capital-intensity controls and finer NAICS4 industry fixed effects. By contrast, PP\&E per employee is generally negatively associated with adoption conditional on scale and the other included financial characteristics. Thus, among otherwise comparable public firms, ChatGPT Enterprise adoption is less concentrated in firms with more physical-capital-intensive production. These patterns persist when excluding technology firms, indicating that they are not primarily driven by the information sector. In summary, ChatGPT Enterprise adoption is more likely among larger, higher revenue-per-worker public firms with relatively lower physical capital intensity.

\subsubsection{Usage intensity among adopters}

\begin{figure}[hbtp]
\centering
\caption{FINANCIAL MEASURES BY USAGE INTENSITY: ECDF}
\label{fig:usage-level-financial-observed-index-ecdf}

\vspace{1em}

\textbf{Panel A. Distribution of observed-value index}\par
\vspace{0.25em}
\includegraphics[width=0.8\linewidth]{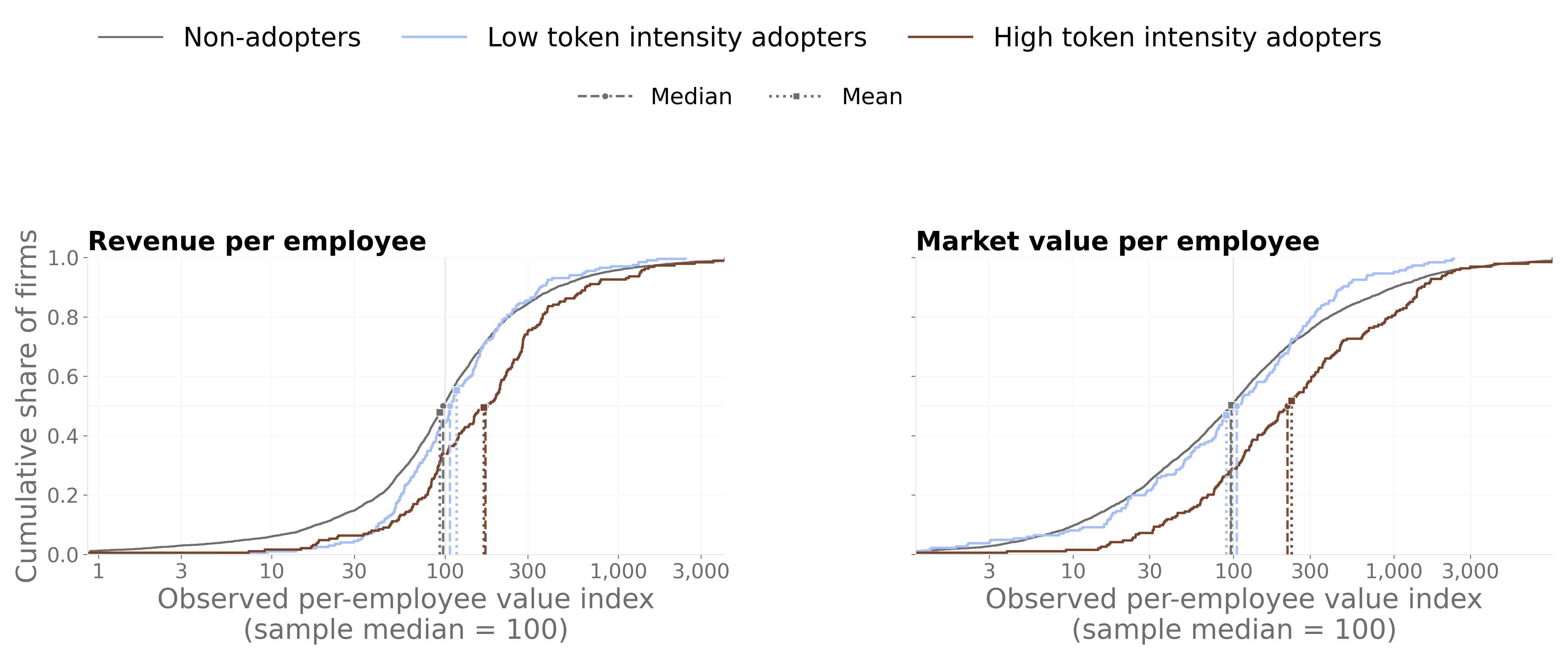}

\vspace{2em}

\textbf{Panel B. Distribution of industry-size-adjusted residuals}\par
\vspace{0.25em}
\includegraphics[width=.8\linewidth]{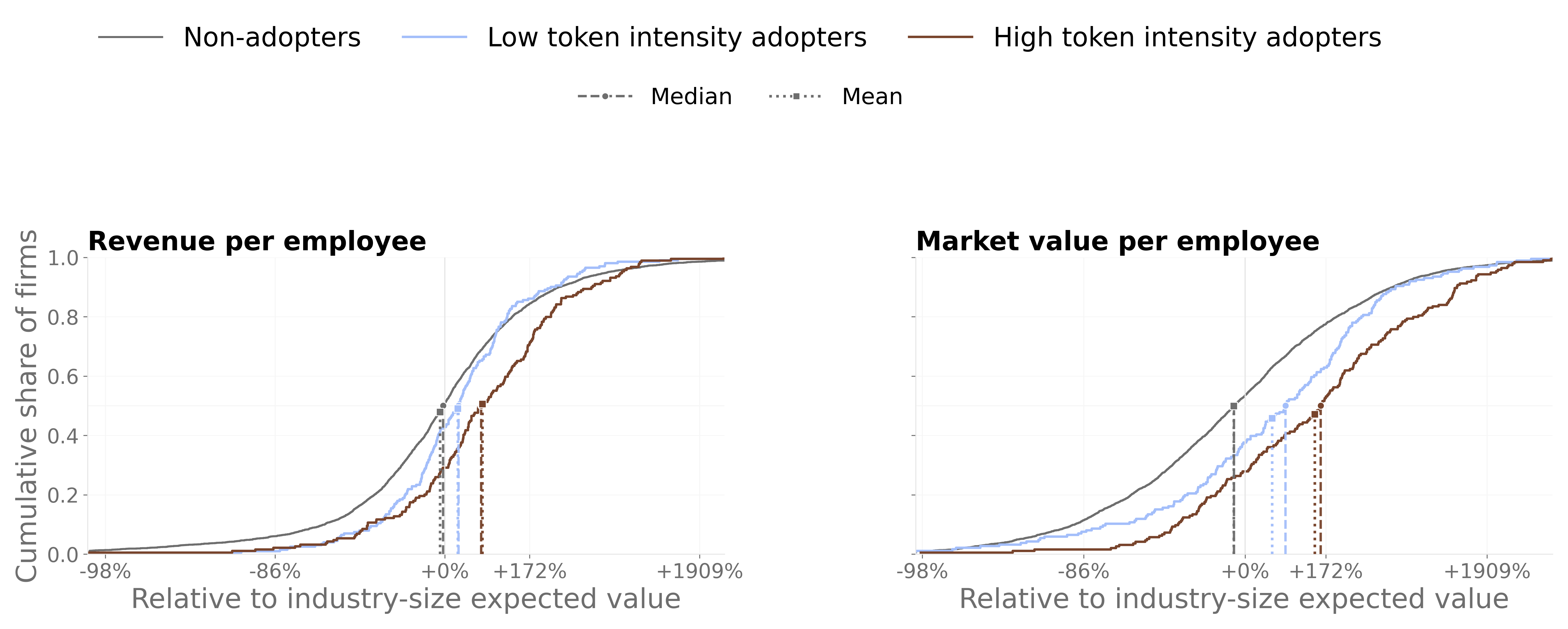}

\vspace*{\fill}

\clearpage

\vspace{0.5em}
\begin{minipage}{\linewidth}
\footnotesize
\emph{Note:} This figure compares 2025 financial outcomes across public firms by ChatGPT Enterprise usage intensity. Non-adopters are public firms not linked to a ChatGPT Enterprise account. Among adopters with positive observed usage, low- and high-intensity groups are defined using a median split of weekly output tokens per employee. Panel A reports per-employee financial outcomes indexed to the sample median, which is set to 100. Panel B reports outcomes after adjusting for industry and firm-size. Step functions show empirical CDFs; dashed and dotted vertical lines indicate group medians and means, respectively. The underlying data combine the ChatGPT Enterprise usage panel with Compustat annual business metrics.
\end{minipage}
\end{figure}

We next move from the extensive margin of adoption to the intensive margin, asking whether financial characteristics also vary with the level of ChatGPT Enterprise use among adopters. Figure~\ref{fig:usage-level-financial-observed-index-ecdf} provides descriptive evidence by comparing the distribution of revenue per employee and market value per employee by adoption intensity, measured by weekly output tokens per employee. Panel A indexes each outcome to the non-adopter median. For both revenue per employee and market value per employee, high-intensity adopters are shifted to the right of non-adopters and low-intensity adopters. This indicates that, among public firms, the firms using ChatGPT Enterprise most intensively tend to have higher revenue productivity and higher market valuation per worker.

Panel B provides a covariate-adjusted version of this comparison, accounting for industry and firm size. This adjustment asks whether high-intensity adopters look different not only because they are in larger or more productive sectors, but also relative to observably similar firms in the same broad industry and size class. The rightward shift for high-intensity adopters remains visible, especially for market value per employee. Low-intensity adopters sit closer to non-adopters, while high-intensity adopters are more likely to appear in the upper part of the adjusted outcome distribution. Thus, the relationship between usage intensity and financial performance is not explained solely by firm size: conditional on adopting, more intensive ChatGPT Enterprise use is concentrated among firms with stronger per-employee financial outcomes.

Table~\ref{tab:usage-outcome-financial-characteristics} provides regression-based evidence, relating lagged financial characteristics to four measures of usage intensity among U.S.-based public-company adopters with positive ChatGPT Enterprise activity: messages per active usage week per employee, weekly active users per employee, output tokens per employee, and messages per weekly active user.\footnote{For usage intensity, no single measure captures the whole intensive margin. Messages per active week per employee captures how much use occurs when a firm is actively using the product; WAU per employee captures the breadth of adoption across the workforce; output tokens per employee captures the volume of model-generated work; and messages per WAU captures intensity among active users. Looking across these measures helps distinguish whether a firm’s usage is broad-based, heavy among a small set of users, or concentrated in particular high-output workflows.} Revenue per employee has positive but imprecisely estimated associations with some usage margins. The point estimates are positive for messages per active usage week per employee and weekly active users per employee, but neither coefficient is statistically significant; revenue per employee is not meaningfully associated with output tokens per employee or messages per active user once other controls are included. The pattern is consistent with broader diffusion across employees and active weeks, but the estimates are too imprecise to support a strong conclusion. Firm size has the opposite association with per-employee usage intensity: lagged log employment is negative and statistically significant for messages per active week per employee, weekly active users per employee, and output tokens per employee. This pattern is consistent with a mechanical or organizational scaling effect: larger firms are more likely to adopt, but conditional on adoption, measured use per employee is lower.

\subsubsection{Firm scale and adoption}

Although larger firms have lower measured usage per employee conditional on adoption, they are more likely to adopt ChatGPT Enterprise in the first place. We next ask whether adoption is especially concentrated among the largest public firms. Table~\ref{tab:adoption-superstar-analysis} shows that a one-log-point increase in lagged revenue is associated with a \CapIQAdoptionSuperstarLogRevenueCoefficientPP{} percentage point higher probability of adoption. This relationship is most pronounced at the top of the revenue distribution: firms in the top revenue quartile are \CapIQAdoptionSuperstarTopRevenueQuartileCoefficientPP{} percentage points more likely to adopt, and firms in the top 5 percent are \CapIQAdoptionSuperstarTopRevenueFivePercentCoefficientPP{} percentage points more likely to adopt.

This pattern is not simply a consequence of large firms operating in large industries. When scale is measured relative to other firms in the same industry-year cell, following the approach in \cite{autor2020superstar}, firms in the top revenue quartile within their NAICS2-by-year cell are \CapIQAdoptionSuperstarIndustryYearTopRevenueQuartileCoefficientPP{} percentage points more likely to adopt, while firms in the top 5 percent are \CapIQAdoptionSuperstarIndustryYearTopRevenueFivePercentCoefficientPP{} percentage points more likely to adopt. Thus, even within industries, adoption is more common among the largest firms. Importantly, these estimates are limited to U.S.-based public companies, which are already large relative to the broader population of businesses. They therefore describe variation among relatively large firms and do not establish how ChatGPT Enterprise adoption varies among small private firms, startups, or mid-market firms; the relationship could be steeper, flatter, or nonlinear elsewhere in that part of the firm-size distribution.

\subsubsection{Pre-existing complements and enterprise adoption}

Firm scale is unlikely to be the only reason some public firms are more likely to adopt ChatGPT Enterprise. Larger firms may also have accumulated organizational, technical, and managerial capabilities that help them identify valuable use cases, redesign workflows, train workers, and integrate the tool into existing business processes. Table~\ref{tab:adoption-outcome-complement-characteristics} therefore examines whether ChatGPT Enterprise adoption is associated with pre-existing investments in organizational and intangible capital. 

We measure these complements using stocks of SG\&A, R\&D, and capitalized software per employee, transformed as log one plus the employee-normalized stock. The measures capture accumulated investment in organizational capabilities, innovation, and software infrastructure, respectively—forms of intangible capital that have been emphasized as complements to computing in the broader literature (e.g., \Cite{brynjolfsson2021productivity}). We use stocks rather than one-year spending flows to capture capacity built up before ChatGPT Enterprise adoption. The adoption regressions cover fiscal years \CapIQAdoptionComplementFiscalYearStart{}--\CapIQAdoptionComplementFiscalYearEnd{}, while all complement stocks are measured in fiscal year \CapIQAdoptionComplementStockYear{}. For SG\&A and R\&D, we construct stocks by cumulating historical Compustat spending flows using a perpetual-inventory approach: SG\&A expense is depreciated at \CapIQAdoptionComplementSGADepreciationPercent{} percent annually and R\&D expense at \CapIQAdoptionComplementRDDepreciationPercent{} percent annually. Capitalized software is measured directly using the Compustat capitalized-software stock. Each stock is converted to dollars and normalized by employment before entering the regressions.

The strongest and most robust association is with SG\&A stock per employee. In the full sample, its coefficient is \CapIQAdoptionComplementSGAAllCoefficient{} and statistically significant; in the sample excluding technology and high-R\&D sectors, it remains positive at \CapIQAdoptionComplementSGANoTechHighRDCoefficient{}, though it is not statistically significant. R\&D stock per employee is also positively associated with adoption in both samples, with coefficients of \CapIQAdoptionComplementRDAllCoefficient{} in the full sample and \CapIQAdoptionComplementRDNoTechHighRDCoefficient{} in the sample excluding technology and high-R\&D sectors. Capitalized software is positive and statistically significant in the full sample, with a coefficient of \CapIQAdoptionComplementCapitalizedSoftwareAllCoefficient{}, but is positive and imprecisely estimated in the non-tech/high-R\&D-excluded sample, with a coefficient of \CapIQAdoptionComplementCapitalizedSoftwareNoTechHighRDCoefficient{}.

Taken together, the results suggest that ChatGPT Enterprise adoption is associated not only with firm scale, but also with the organizational and intangible assets firms have accumulated beforehand. Alongside the earlier findings---that adoption rises with revenue productivity and especially firm scale, and that more productive adopters tend to use ChatGPT Enterprise more broadly per employee---this pattern is consistent with the view that generative AI, like other general purpose technologies, depends on complementary capabilities within firms. Larger and more organizationally intensive firms may be better positioned both to identify valuable use cases and to deploy and integrate the technology into existing workflows.

\subsection{Enterprise AI use is broadly distributed across worker groups but uneven in intensity} \label{Sec:Distribution}

We next examine the subset of firms for which we observe high-quality job title information, as described in Section~\ref{sec:role}. We focus on job title classes and seniority levels and study two margins: the share of weekly active users in each category and usage intensity conditional on active use.\footnote{In Figure~\ref{fig:adoption-variation-intensive_pm}, we also study how both of these margins vary with respect to whether or not a worker is a people manager.}

Two limitations of this approach are important for interpretation. First, job title coverage is not universal and varies across firms, so these estimates describe observed use within the covered job title sample rather than the full workforce of all adopting firms. Second, the composition of active users is not the same as a role-specific adoption rate, because we do not observe the denominator of all employees by role. Accordingly, the worker-composition results measure each category's share of observed weekly active users; they do not measure the share of employees in that category who adopt ChatGPT Enterprise or whether the category is overrepresented among users relative to its workforce share. Similarly, the usage-intensity estimates compare activity among active users and do not account for differences across categories in the probability of becoming active.

\subsubsection{The Extensive Margin of Use Across Job Title Class and Seniority}

\begin{figure}
\centering
\caption{COMPOSITION OF ACTIVE USERS BY WORKER TYPE}
\label{fig:adoption-variation-extensive_fig4}

\textbf{Panel A. Job title class}\par
\vspace{0.25em}
\includegraphics[width=.7\linewidth]{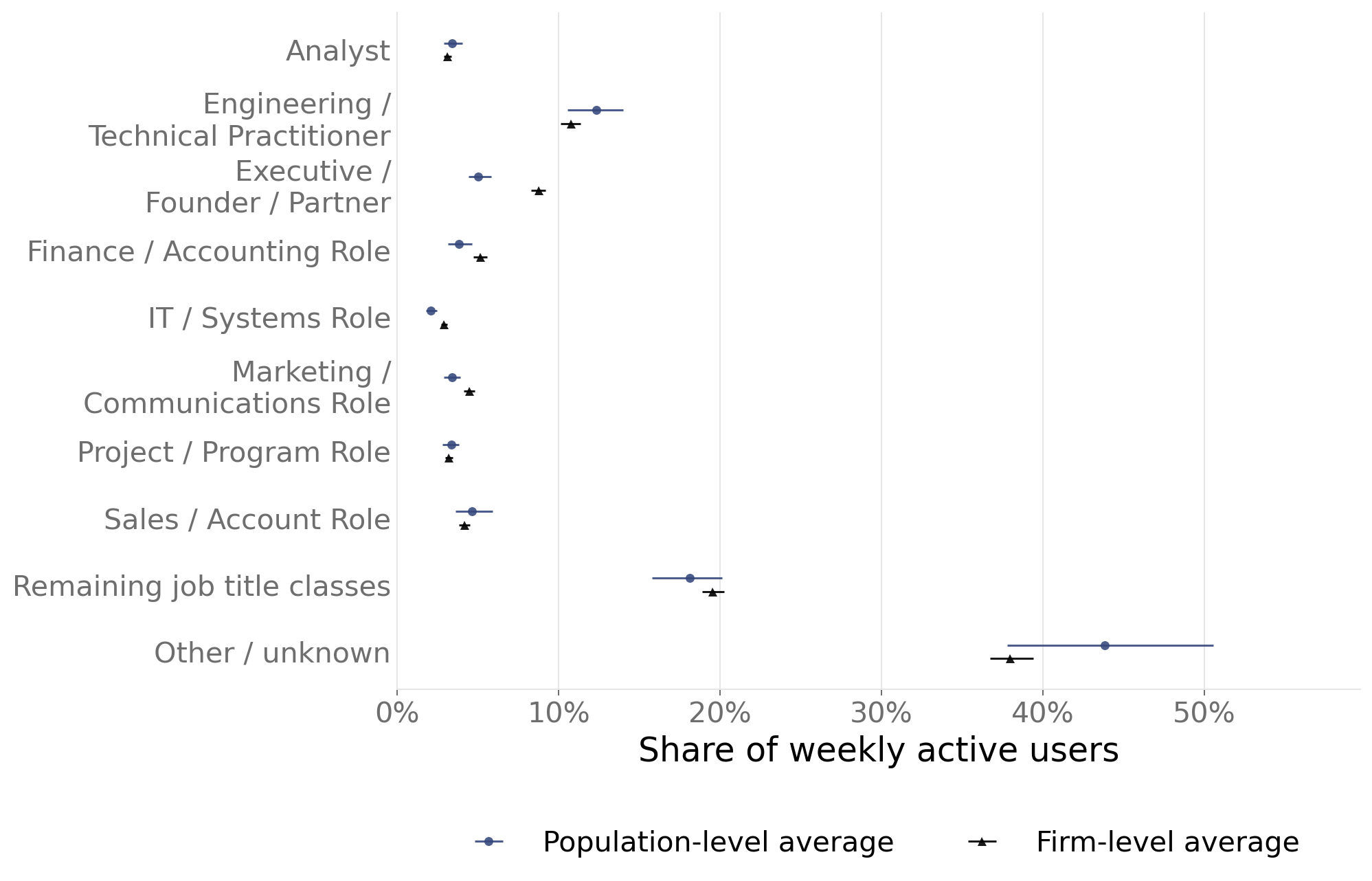}

\vspace{1em}

\textbf{Panel B. Seniority level}\par
\vspace{0.25em}
\includegraphics[width=.7\linewidth]{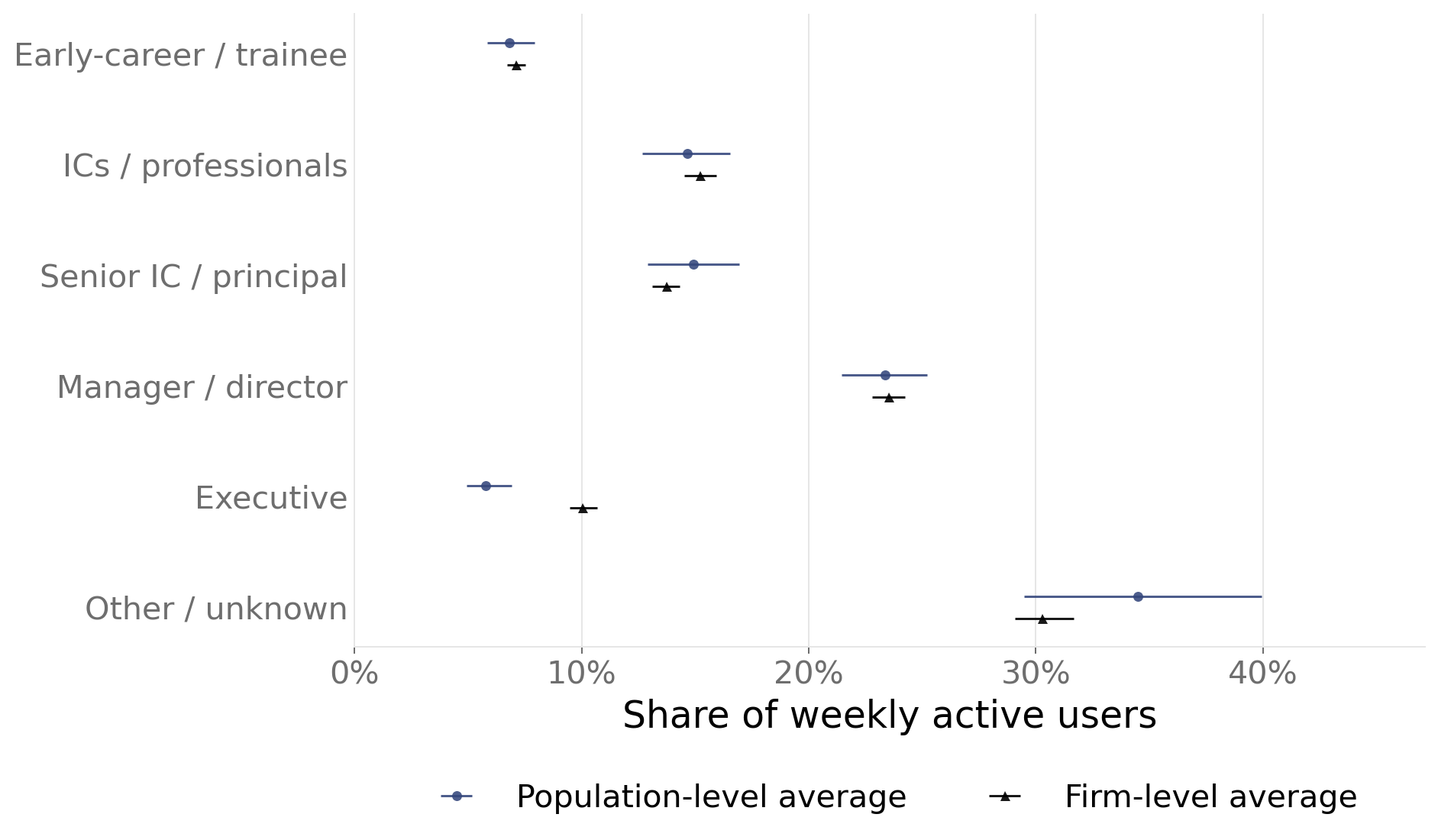}

\vspace{0.5em}
\begin{minipage}{\linewidth}
\footnotesize

\emph{Note:} This figure reports the distribution of weekly active ChatGPT Enterprise users across worker categories six months after firm adoption. Panel A groups users by inferred job title class. Panel B groups users by inferred seniority level. Blue points report the population-level average, computed as each category's share of all weekly active users across firms. Black points report the firm-level average, computed as each category's share of weekly active users within each firm and then averaged across firms, giving each firm equal weight. Horizontal bars show 95\% firm-bootstrap confidence intervals. ``Remaining job title classes'' pools inferred job title classes outside the individually displayed categories. ``Other / unknown'' includes users with missing, ambiguous, malformed, or otherwise low-information job titles that cannot be assigned to a substantive class. ``IC'' denotes an individual contributor.
\end{minipage}
\end{figure}

Figure~\ref{fig:adoption-variation-extensive_fig4} reports two ways of summarizing active-user composition six months after organizational adoption. The population-level estimate pools weekly active users across organizations, giving greater weight to firms with more active users. The firm-level estimate first calculates each category's share within an organization and then averages those shares across organizations, giving each firm equal weight. The former describes the composition of observed active users in the sample as a whole, while the latter describes the composition of the average firm.

Panel A reports the distribution across job title classes. Both estimates show that observed active use extends across a range of organizational functions and is not confined to technical workers. At the average firm, engineering and technical practitioners account for approximately \EnterpriseEngineeringTechnicalPractitionerSharePercent{}\% of weekly active users after six months, while executives, founders, and partners account for \EnterpriseExecutiveFounderPartnerSharePercent{}\%. Finance and accounting and marketing and communications each account for approximately \EnterpriseFinanceAccountingRoleSharePercent{}\%, and sales and account management accounts for approximately \EnterpriseSalesAccountRoleSharePercent{}\%. This broad functional distribution is consistent with ChatGPT being adopted across the organizational structure rather than within a single occupational domain. 

Panel B reports the corresponding distribution across inferred seniority levels. At the average firm, managers and directors account for approximately \EnterpriseManagerDirectorSharePercent{}\% of weekly active users after six months, followed by individual contributors (ICs) and professionals at \EnterpriseICProfessionalSharePercent{}\% and senior ICs and principals at \EnterpriseSeniorICPrincipalSharePercent{}\%. Executives account for approximately \EnterpriseExecutiveSenioritySharePercent{}\%, while early-career workers and trainees account for \EnterpriseEarlyCareerTraineeSharePercent{}\%. The seniority distribution similarly shows that ChatGPT Enterprise use spans multiple levels of the organizational hierarchy, rather than being concentrated among either junior employees or senior leadership.

\subsubsection{The Intensive Margin of Use Across Job Title Class and Seniority}

\begin{figure}[htbp]
\centering
\caption{DIFFERENCES IN AI USAGE INTENSITY ACROSS WORKER ROLES}
\label{fig:adoption-variation-intensive_fig5}

\textbf{Panel A. Job title class}\par
\vspace{0.25em}
\includegraphics[width=.7\linewidth]{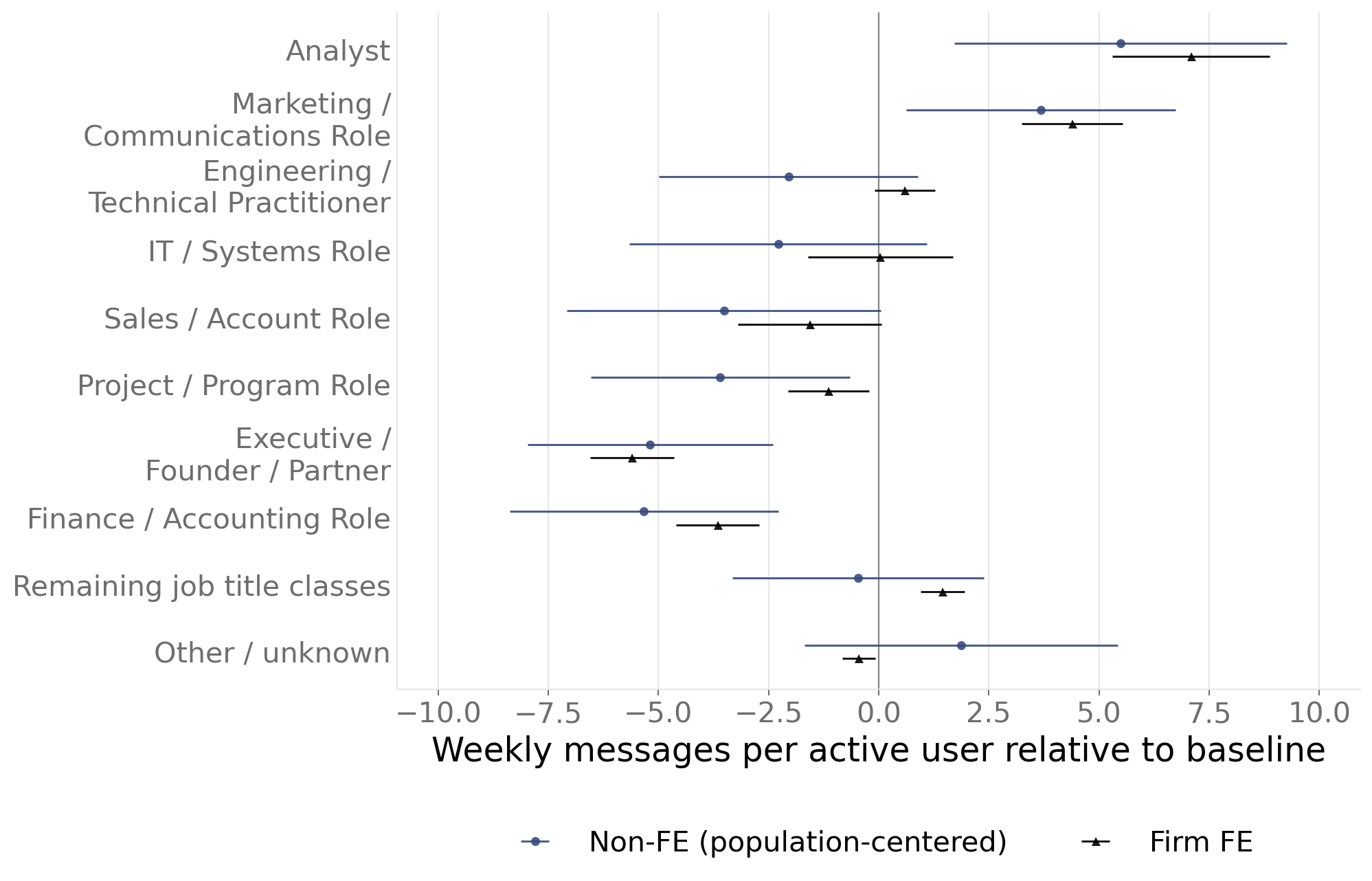}

\vspace{1em}

\textbf{Panel B. Seniority level}\par
\vspace{0.25em}
\includegraphics[width=.7\linewidth]{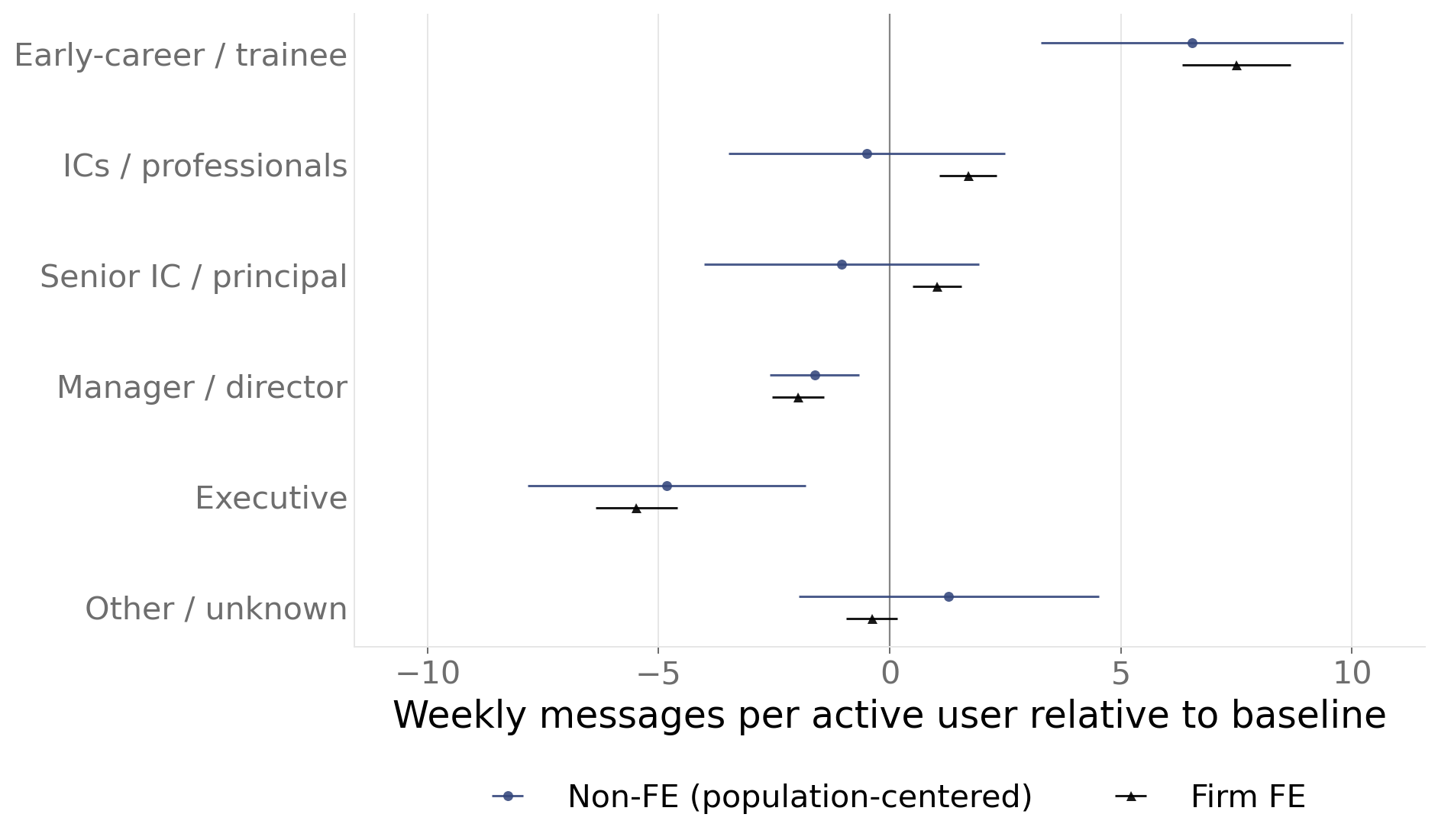}

\vspace{0.5em}
\begin{minipage}{\linewidth}
\footnotesize
\emph{Note:} This figure reports differences in weekly ChatGPT Enterprise usage intensity across worker categories six months after firm adoption. Panel A groups users by inferred job title class. Panel B groups users by inferred seniority level. Usage intensity is measured as weekly messages per active user within each worker category. Blue points report population-level estimates, comparing each worker category to the average active user across firms. Black points report firm fixed effect estimates, comparing each worker category to other active users within the same firm. Positive values indicate more messages per active user than the relevant baseline; negative values indicate fewer. Horizontal bars show 95\% confidence intervals with firm-clustered standard errors. ``Remaining job title classes'' pools inferred job title classes outside the individually displayed categories. ``Other / unknown'' includes users with missing, ambiguous, malformed, or otherwise low-information job titles that cannot be assigned to a substantive class. ``IC'' denotes an individual contributor.
\end{minipage}
\end{figure}

Figure \ref{fig:adoption-variation-intensive_fig5} reports differences in weekly ChatGPT Enterprise usage intensity across worker categories, measured as the difference in weekly messages per active user from the relevant baseline. Panel A reports differences by job title class, while Panel B reports differences by inferred seniority level. In each panel, the specifications without firm fixed effects compare each group to the average active user in the sample, while the firm fixed effects specifications absorb differences in average usage intensity across firms and compare workers to other active users within the same organization.

Panel A of Figure \ref{fig:adoption-variation-intensive_fig5} reports variation in weekly messages per active user by job title class among active users. The results show that job title classes that account for larger shares of observed weekly active users are not necessarily those with the highest usage intensity conditional on active use. In particular, analysts and marketing and communications workers send more weekly messages than the average active user in their firms, even though they do not account for the largest shares of observed weekly active users (Figure~\ref{fig:adoption-variation-extensive_fig4}, Panel A). By contrast, executives, founders, and partners send fewer messages than other active users within the same firm. These patterns indicate that active-user composition and usage intensity capture distinct dimensions of enterprise adoption.

Panel B of Figure \ref{fig:adoption-variation-intensive_fig5} reports usage intensity by inferred seniority level. The clearest pattern is a strong negative seniority gradient in message volume. Among adopters, early-career workers and trainees send roughly eight to nine more weekly messages than the average active user within the same firm, while managers, directors, and executives send fewer messages. This pattern is especially relevant in light of recent evidence on generative AI and early-career labor-market outcomes (e.g. \cite{brynjolfsson2025canaries}), because it identifies early-career workers as particularly intensive users conditional on active use. At the same time, message volume should be interpreted as a measure of usage intensity rather than as a complete measure of economic importance.

Overall, high-intensity use appears both in specific functional groups, such as analysts and marketing and communications workers, and at particular points in the career hierarchy, especially among early-career workers and trainees.\footnote{In Figures~\ref{fig:industry-job-title-variation_app}-\ref{fig:industry-people-manager-variation-app}, we show how both the composition of ChatGPT Enterprise users and the intensity of usage covary across worker type and industry.} This heterogeneity motivates the task-level analysis below, which examines whether these differences in usage intensity correspond to differences in the kinds of work for which employees use ChatGPT.

\subsection{Enterprise AI use spans a broad set of knowledge work tasks within organizations} \label{Sec:Tasks}

We now turn from who uses ChatGPT Enterprise to what kinds of work they use it for. We classify Enterprise messages into a task taxonomy using an automated classifier described in Appendix~\ref{app:task-classification} and summarize task composition using two complementary measures. The first is the share of weekly active users who perform a task at least once during the week. This task-prevalence measure captures the breadth of exposure to each task category. Because users may perform multiple task types in the same week, the category shares are not mutually exclusive and therefore do not sum to one. The second is the share of weekly messages assigned to each task category. This message-share measure captures the intensity of use by task and shows which activities account for the largest share of observed interaction with ChatGPT Enterprise. 

\subsubsection{Overall Enterprise Task Composition}

\begin{figure}[htbp]
\centering
\caption{DISTRIBUTION OF AI USE ACROSS TASKS}
\label{fig:task-distribution}

\textbf{Panel A. Task Prevalence Among Active Users}\par
\vspace{0.25em}
\includegraphics[width=.65\linewidth]{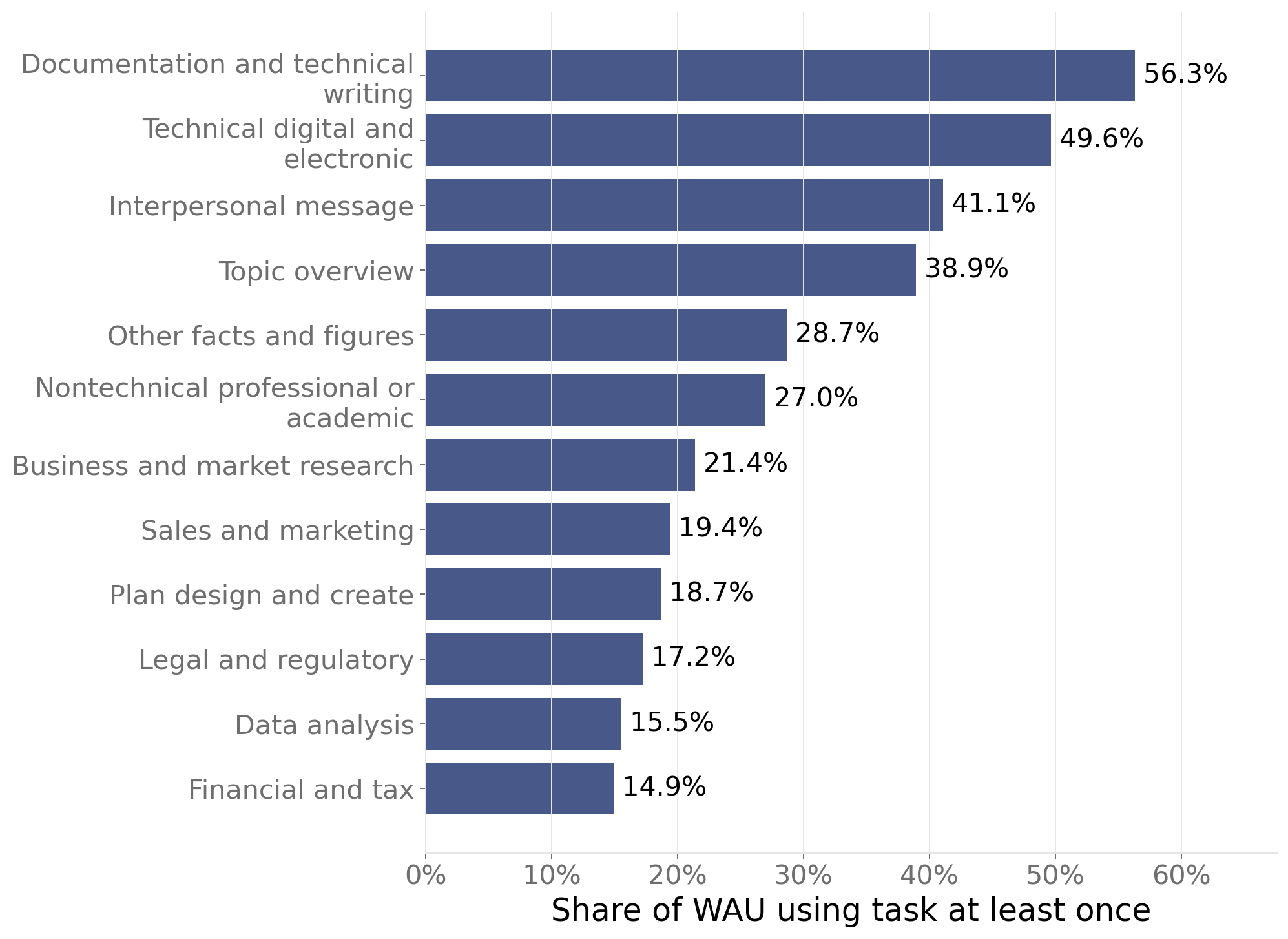}

\vspace{1em}

\textbf{Panel B. Distribution of Messages Across Tasks}\par
\vspace{0.25em}
\includegraphics[width=.65\linewidth]{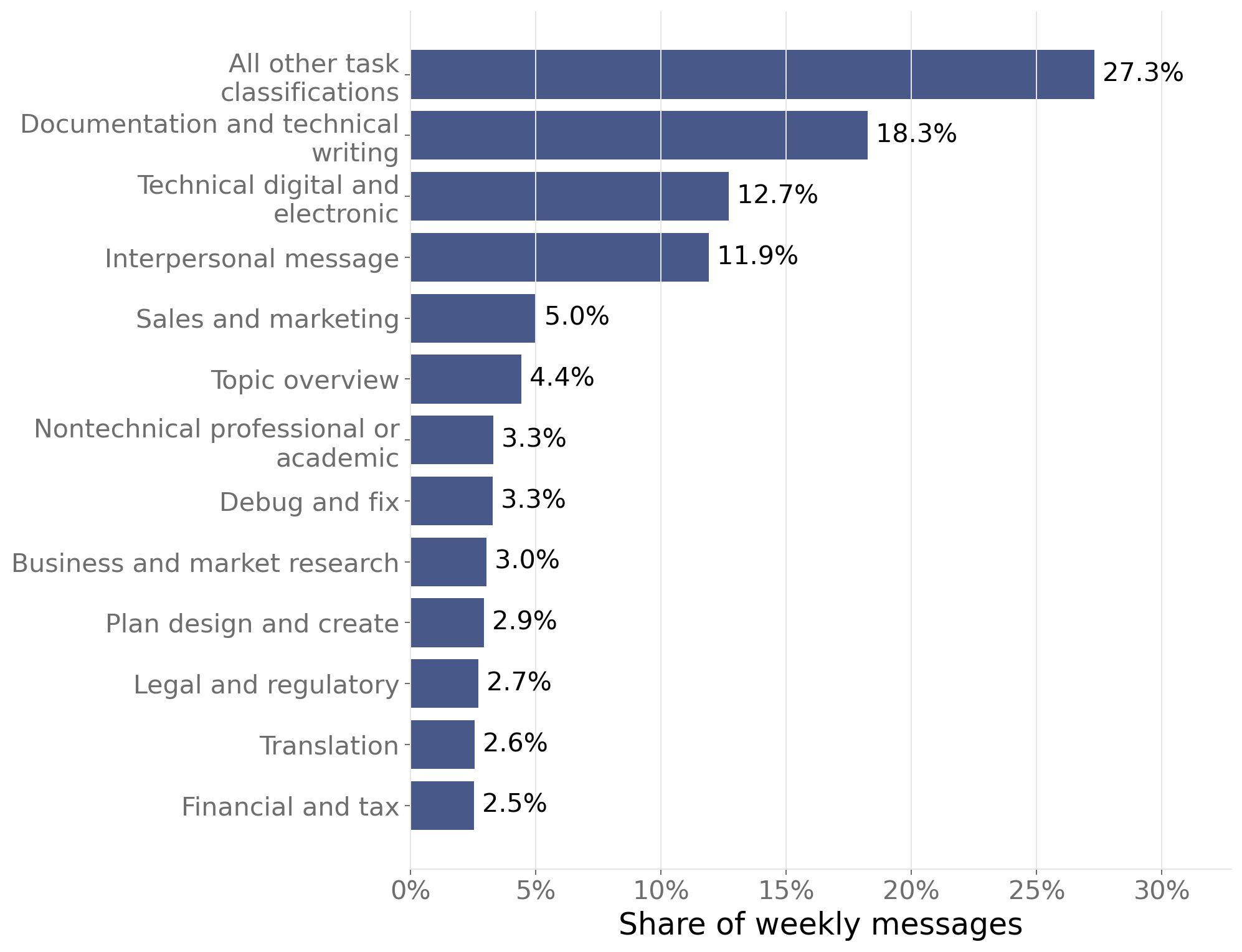}

\vspace{0.5em}
\begin{minipage}{\linewidth}
\footnotesize
\emph{Note:} The figure describes the distribution of ChatGPT Enterprise use across tasks among ChatGPT Enterprise firms with task-classification data, measured six months after adoption. Each message is assigned to one of 60 task categories using the classifier described in the text. Panel A reports the share of weekly active users who submitted at least one message in each of the 12 task categories with the greatest user reach. Because users may use ChatGPT for multiple tasks during the same week, these shares are not mutually exclusive and do not sum to 100 percent. Panel B reports the share of classified weekly messages assigned to the 12 largest task categories by message volume. ``All other task classifications'' pools the remaining categories. Categories are selected separately in each panel based on the corresponding measure.
\end{minipage}
\end{figure}

Figure~\ref{fig:task-distribution} reports the overall task composition of ChatGPT Enterprise use. Panel A shows that more than half of active users perform documentation or technical-writing tasks, nearly half perform technical digital work, and large shares use ChatGPT Enterprise for messages, topic overviews, facts and figures, professional work, research, sales and marketing, planning, legal work, data analysis, and financial or tax tasks. The central pattern is not the dominance of a single application, but the breadth of task exposure among active users.

Panel B shows that message volume is more concentrated than task incidence. Documentation and technical writing, technical digital work, and message drafting account for large shares of total messages, while several categories that reach many users account for relatively small shares of message volume. Topic overviews, business and market research, legal and regulatory work, and financial and tax tasks are widespread but comparatively less message-intensive. This distinction matters for interpreting enterprise AI use. A task can be economically relevant because it reaches many workers, because it accounts for large amounts of usage, or both. User reach and usage depth are therefore separate margins of task-level diffusion. Panel B also shows a substantial residual category of other task classifications. This is useful evidence in itself, in that it indicates that ChatGPT Enterprise use has a long tail: workers apply the tool to many activities that do not fit cleanly into the largest task categories. 

\subsubsection{Task Differences Across Industries}

\begin{figure}[htbp]
\centering
\caption{DIFFERENCES IN AI TASK USE ACROSS INDUSTRIES}
\label{fig:industry-task-distribution}

\textbf{Panel A. Task Prevalence Among Active Users}\par
\vspace{0.25em}
\includegraphics[width=.75\linewidth]{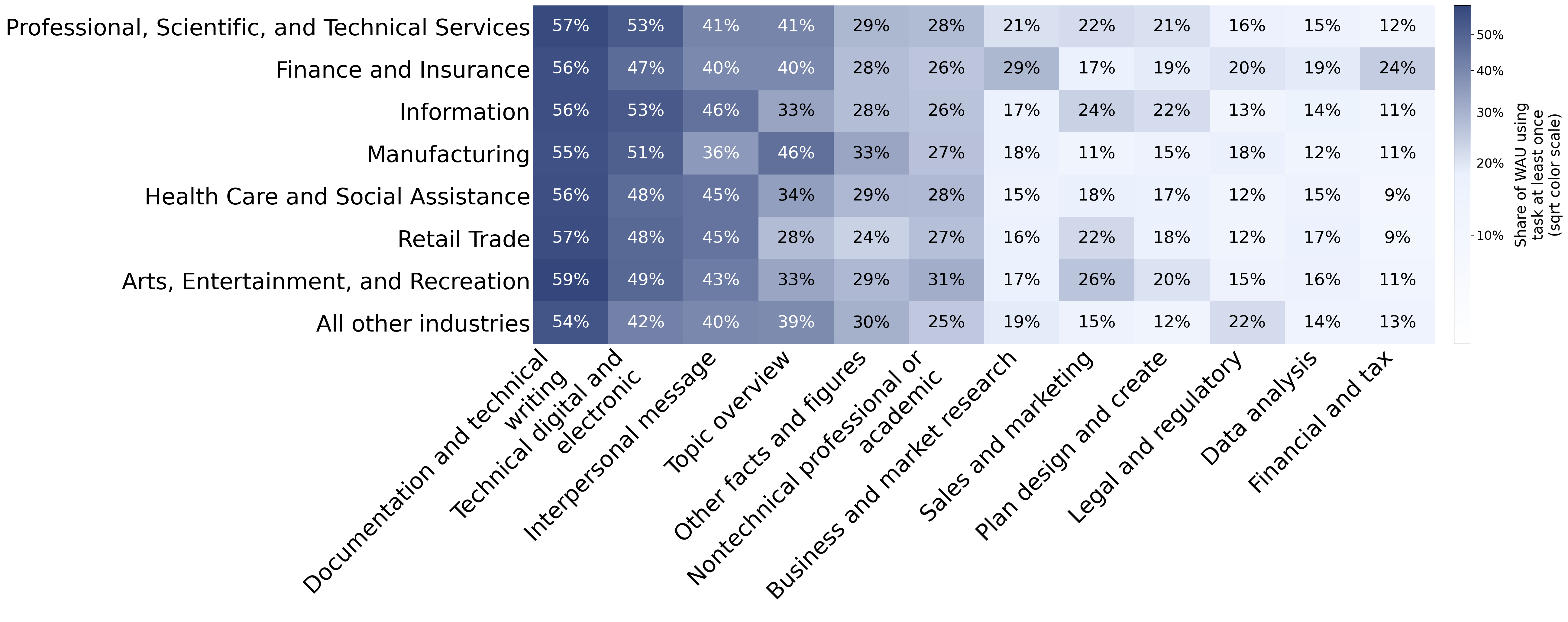}

\vspace{1em}

\textbf{Panel B. Distribution of Messages Across Tasks}\par
\vspace{0.25em}
\includegraphics[width=.75\linewidth]{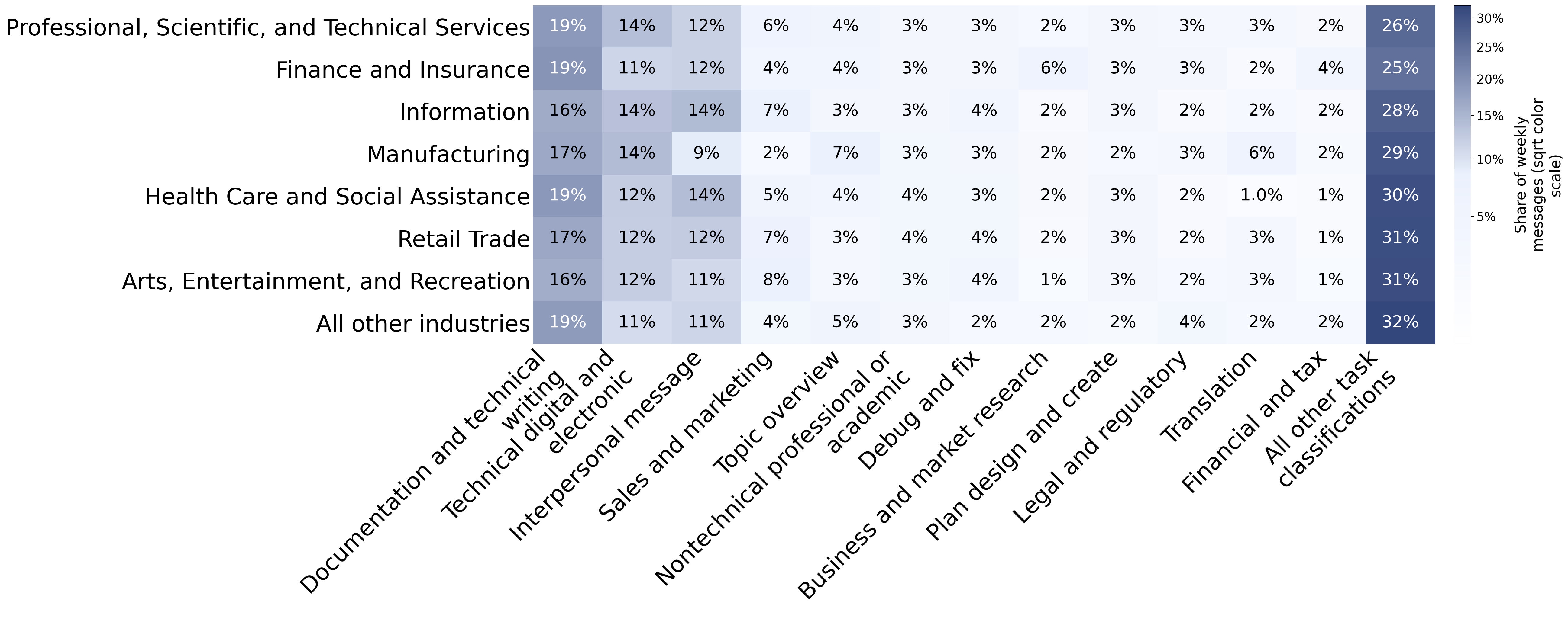}

\vspace{0.5em}
\begin{minipage}{\linewidth}
\footnotesize
\emph{Note:} The figure compares the distribution of ChatGPT Enterprise use across tasks and two-digit NAICS sectors among ChatGPT firms with task-classification data, measured six months after adoption. Each message is assigned to one of 60 task categories using the classifier described in the text. Panel A reports the share of weekly active users in each industry who submitted at least one message in each of the 12 task categories with the greatest user reach. Because users may use ChatGPT for multiple tasks during the same week, the shares in each row are not mutually exclusive and do not sum to 100 percent. Panel B reports the share of classified weekly messages in each industry assigned to the 12 largest task categories by message volume. ``All other task classifications'' pools the remaining task categories, while ``All other industries'' pools NAICS two-digit sectors not displayed separately. Task categories are selected separately in each panel based on the corresponding measure.
\end{minipage}
\end{figure}

Figure~\ref{fig:industry-task-distribution} examines task composition by two-digit NAICS sector (referred to hereafter as ``industry''), separately for task prevalence and message shares. Panel A reports the share of weekly active users in each industry who use ChatGPT Enterprise for a given task at least once, while Panel B reports the share of weekly messages in each industry assigned to each task.

Panel A shows both commonality and industry variation. The broad task structure is similar across industries: documentation and technical writing, technical digital work, and communication are among the most common tasks in every major industry group. At the same time, task prevalence varies in ways that are consistent with differences in the underlying task content of work across industries. For example, financial and tax-related tasks are substantially more prevalent in finance and insurance than in other industries, business and market research is also especially common in finance and insurance, and sales and marketing tasks are more prevalent in arts, entertainment, and recreation, information, and retail trade than in manufacturing. Panel B shows that these industry differences are more muted when tasks are weighted by message volume. Across industries, messages are concentrated in a similar set of categories, especially documentation and technical writing, technical digital work, and communication. In other words, industry differences appear more strongly on the extensive task margin, i.e., which tasks active users try at least once, than on the intensive task margin, i.e., which tasks account for the largest shares of total messages.

\subsubsection{Task Differences by Job Title Class and Seniority}

\begin{figure}[htbp]
\centering
\caption{DIFFERENCES IN AI TASK USE ACROSS JOB TITLE CLASSES}
\label{fig:industry-job-title-task-variation}

\textbf{Panel A. Task Prevalence Among Active Users}\par
\vspace{0.25em}
\includegraphics[width=.75\linewidth]{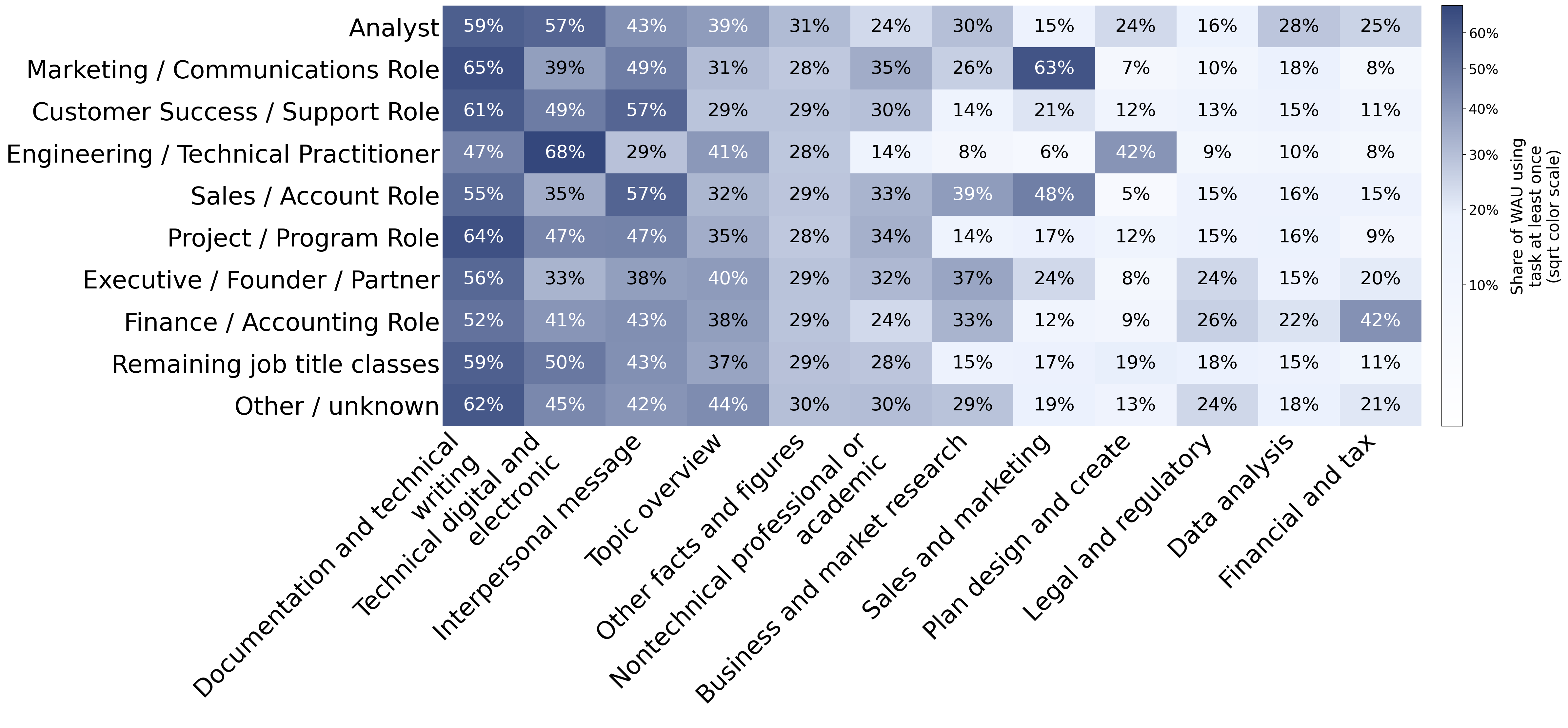}

\vspace{1em}

\textbf{Panel B. Distribution of Messages Across Tasks}\par
\vspace{0.25em}
\includegraphics[width=.75\linewidth]{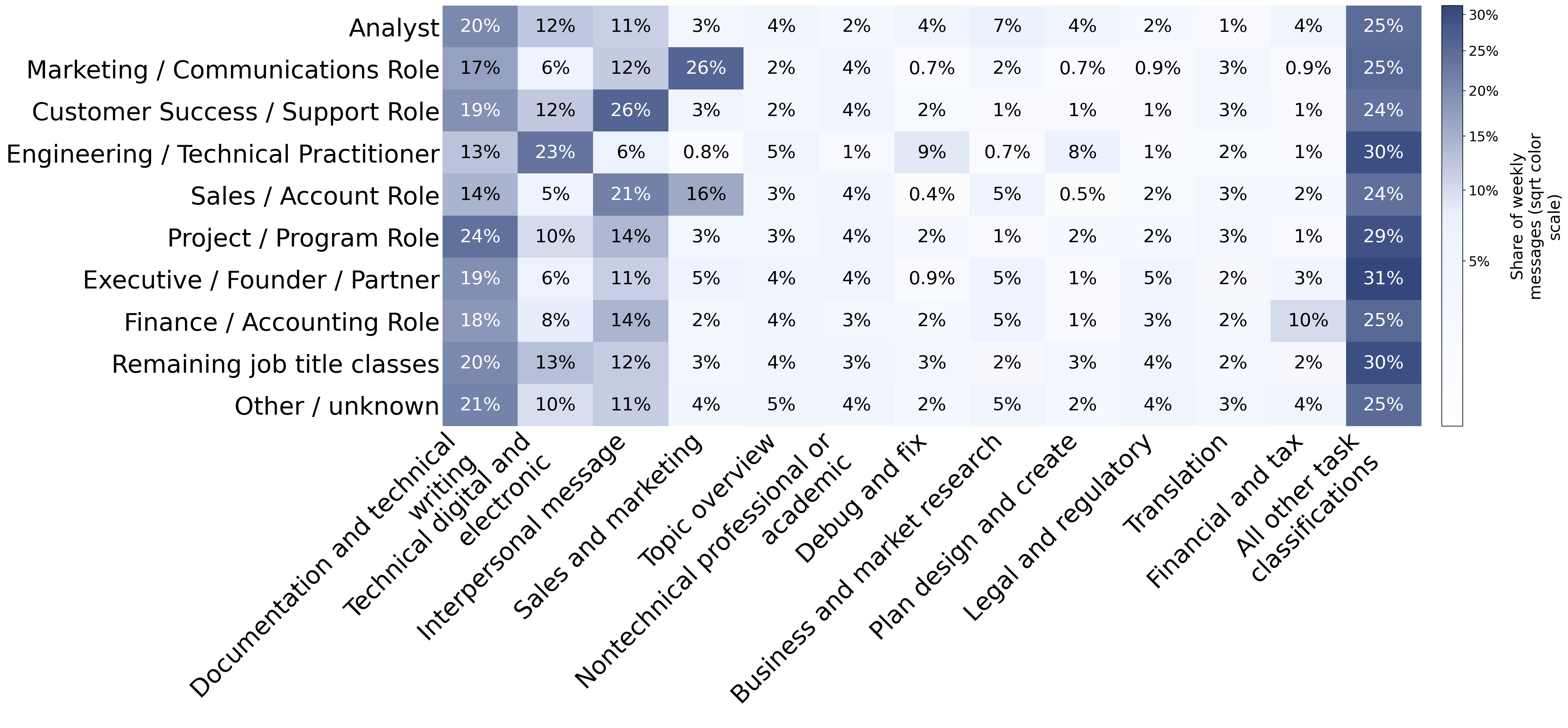}

\vspace{0.5em}
\begin{minipage}{\linewidth}
\footnotesize
\emph{Note:} The figure compares the distribution of ChatGPT Enterprise use across tasks and inferred job title classes among firms with task-classification and job title data, measured six months after adoption. Each message is assigned to one of 60 task categories using the classifier described in the text. Panel A reports the share of weekly active users in each job title class who submitted at least one message in each of the 12 task categories with the greatest user reach. Because users may use ChatGPT for multiple tasks during the same week, the shares in each row are not mutually exclusive and do not sum to 100 percent. Panel B reports the distribution of classified weekly messages across the 12 largest task categories by message volume within each job title class. ``All other task classifications'' pools the remaining task categories. ``Remaining job title classes'' pools substantive inferred job title classes outside the individually displayed categories, while ``Other / unknown'' includes users whose job titles are missing, ambiguous, malformed, or otherwise cannot be assigned to a substantive class. Task categories are selected separately in each panel using their prevalence in the full task-classification sample. 
\end{minipage}
\end{figure}

\begin{figure}[htbp]
\centering
\caption{DIFFERENCES IN AI TASK USE ACROSS SENIORITY LEVELS}
\label{fig:industry-seniority-task-variation}

\textbf{Panel A. Task Prevalence Among Active Users}\par
\vspace{0.25em}
\includegraphics[width=.75\linewidth]{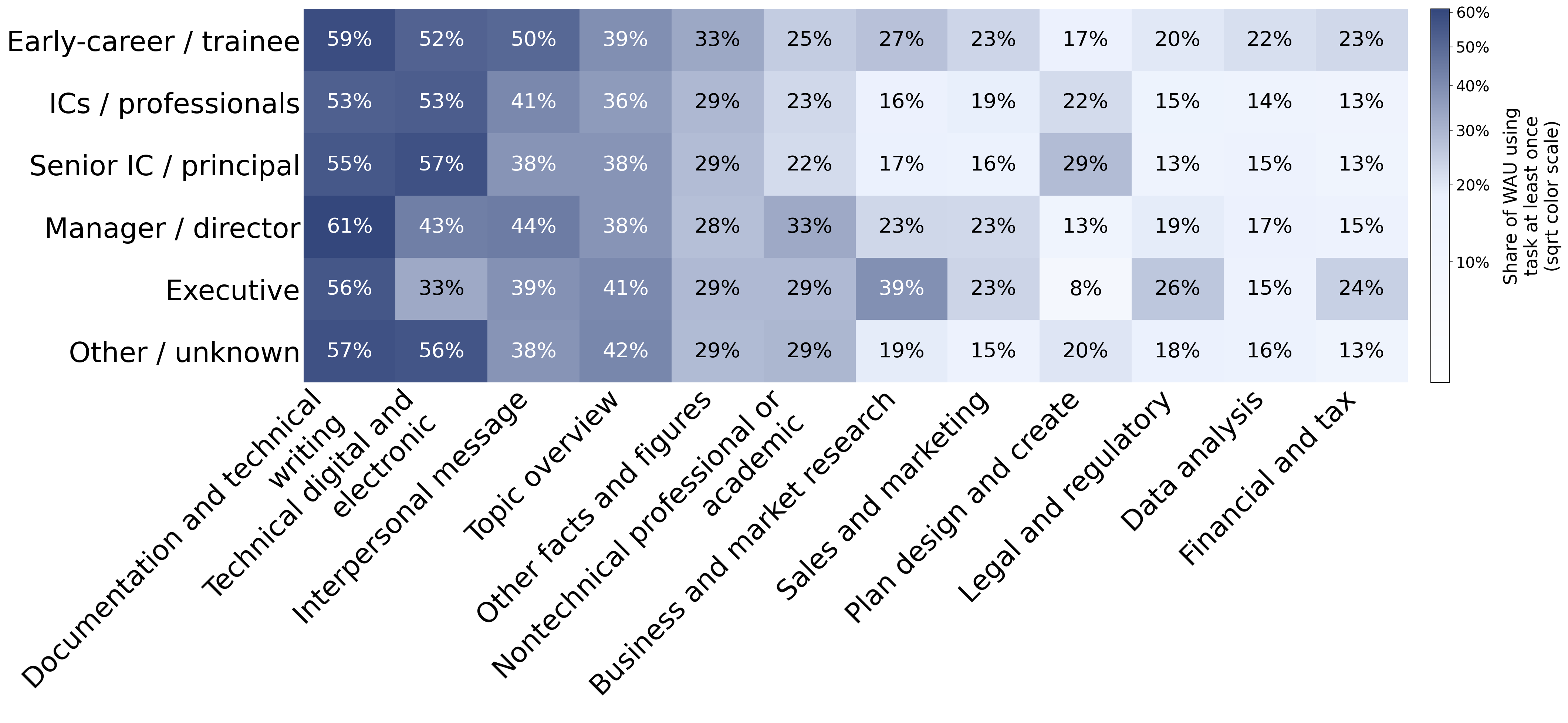}

\vspace{1em}

\textbf{Panel B. Distribution of Messages Across Tasks}\par
\vspace{0.25em}
\includegraphics[width=.75\linewidth]{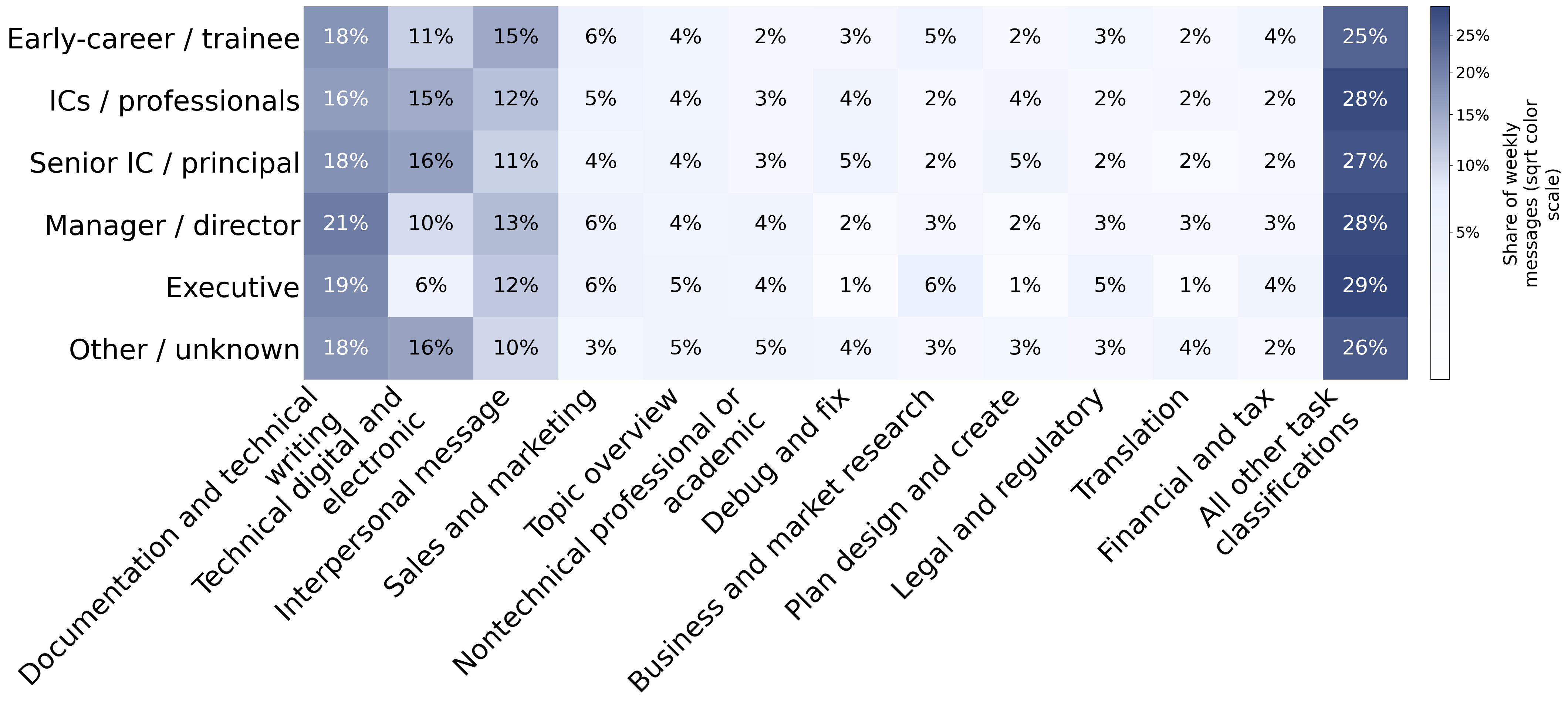}

\vspace{0.5em}
\begin{minipage}{\linewidth}
\footnotesize
\emph{Note:} The figure compares the distribution of ChatGPT Enterprise use across tasks and inferred seniority levels among firms with task-classification and job title data, measured six months after adoption. Each message is assigned to one of 60 task categories using the classifier described in the text. Panel A reports the share of weekly active users at each seniority level who submitted at least one message in each of the 12 task categories with the greatest user reach. Because users may use ChatGPT for multiple tasks during the same week, the shares in each row are not mutually exclusive and do not sum to 100 percent. Panel B reports the distribution of classified weekly messages across the 12 largest task categories by message volume within each seniority group. ``All other task classifications'' pools the remaining task categories. ``Contractor / Temporary'' identifies users whose job titles indicate employment status rather than a conventional seniority level. ``Other / unknown'' includes users whose job titles are missing, ambiguous, malformed, or otherwise do not provide enough information to infer seniority. ``IC'' denotes an individual contributor. Task categories are selected separately in each panel using their prevalence in the full task classification sample. 
\end{minipage}
\end{figure}

Figures~\ref{fig:industry-job-title-task-variation} and \ref{fig:industry-seniority-task-variation} examine task composition by job title class and inferred seniority, respectively.\footnote{We report task composition across people manager status in Figure~\ref{fig:industry-people-manager-task-variation}.} As above, Panel A in each figure reports the share of active users in each group who use ChatGPT Enterprise for a task at least once during the week, while Panel B reports the share of messages assigned to each task. 

Figure~\ref{fig:industry-job-title-task-variation} shows both broad commonality and role-specific specialization. Documentation and technical writing, communication, and technical or digital tasks appear across many job title classes. At the same time, task prevalence varies in ways that align with job responsibilities: engineering and technical practitioners are especially likely to use ChatGPT Enterprise for technical digital work and debugging, finance and accounting workers for financial and tax tasks, and sales, account, marketing, and communications roles for sales and marketing tasks. The message-share panel shows a related pattern, but also makes clear that these role-specific tasks do not overpower the small set of core tasks that are performed by all roles.

Figure~\ref{fig:industry-seniority-task-variation} shows a similar structure across the organizational hierarchy. Workers at different seniority levels use overlapping task categories, but with different relative emphasis. Early-career workers, individual contributors, managers, and executives all use ChatGPT Enterprise for common categories such as documentation and technical writing, technical digital work, communication, and information-oriented tasks. At the same time, seniority is associated with differences in the breadth and mix of task use. For instance, early-career workers and individual contributors have high prevalence in several common production-oriented categories, whereas executives are relatively more represented in categories such as topic overviews, facts and figures, legal and regulatory work, and financial or tax-related tasks.

\section{Conclusion} \label{Sec:Conclusion}

A growing literature argues that AI, and especially large language models, have the characteristics of a general purpose technology \parencite{goldfarb2023could,eloundou2024gpts}. For such technologies to affect production, firms must do more than obtain access: they must discover valuable use cases, encourage use across workers, and integrate the technology into existing workflows. This paper studies that process using administrative data from ChatGPT Enterprise. One central message is that access to the same underlying system does not imply uniform use: firms differ in whether and when they adopt, workers differ in how intensively they use the tool, and task use varies across industries, job title classes, and seniority levels.

This interpretation has several implications. First, the earliest U.S.-based public company adopters are not average firms; they are larger, more intangible-intensive, and more highly valued. The relationship between adoption and firm capabilities also points toward the importance of complements. Firms with greater scale and accumulated intangible investments may be better positioned to identify valuable applications, support workers in using the technology, and integrate it into business processes. Second, diffusion may initially reinforce existing firm heterogeneity. If larger and more intangible-intensive firms adopt earlier and are better positioned to integrate the technology into work, generative AI could widen differences in productivity or value creation across firms even when the underlying models are broadly available. Third, adopting firms differ substantially in the breadth of participation, the intensity of use, and the task mix to which the technology is applied. These margins matter because the economic role of generative AI depends not only on whether a firm has access, but also on where the technology enters the organization of work.

Importantly, these results should be interpreted in light of the scope of the data. The analyses measure usage only within OpenAI's ChatGPT Enterprise product, not usage of other AI systems, API-based tools, internally built applications, or personal accounts. The worker-level results are based on observed administrative job titles, which are incomplete and do not provide denominators for the full workforce in each role. The task results are based on classified message content and do not measure downstream work products, productivity effects, or changes in organizational routines. Finally, the public company analyses are limited to the selected subset of U.S.-based enterprise organizations that can be linked to financial data. Future work should connect enterprise AI telemetry to measures of output, organizational change, and longer-run firm performance, and should examine how adoption, usage intensity, worker composition, and task mix evolve as generative AI continues to become more widespread.

Even with these limitations, the patterns documented in this paper point to a central feature of enterprise AI diffusion: adoption is only the beginning of deployment. The rapid adoption of generative AI by firms should therefore not be equated with immediate productivity transformation. General purpose technologies rarely generate immediate, economy-wide gains; their impact unfolds through a slower process of co-invention in which firms discover use cases, invest in complements, and reorganize production so that a new capability becomes reliable in everyday work \parencite{griliches1957hybrid,mansfield1961imitation,hall2003adoption,jovanovic2005general,bresnahan1995gpt}. We are still in the early stages of that process. Firms are not merely deciding whether to use generative AI; they are learning where it belongs in their organizational workflow. The economic effects of generative AI will depend on how that learning and decision-making process unfolds across firms, workers, and tasks.

\clearpage
\section{Figures}

\vspace*{\fill}

\vspace*{\fill}

\clearpage

\vspace*{\fill}

\vspace*{\fill}

\clearpage

\vspace*{\fill}

\vspace*{\fill}

\vspace*{\fill}

\vspace*{\fill}

\vspace*{\fill}

\vspace*{\fill}

\vspace*{\fill}

\vspace*{\fill}

\vspace*{\fill}

\vspace*{\fill}

\vspace*{\fill}

\vspace*{\fill}

\clearpage

\section{Tables}
  
\begin{table}[H]
\centering
\caption{FINANCIAL CHARACTERISTICS AND ENTERPRISE ADOPTION}
\label{tab:adoption-outcome-financial-characteristics}
{\footnotesize\setstretch{1}
\renewcommand{\arraystretch}{1.15}
\setlength{\tabcolsep}{2pt}
\begin{tabular*}{\textwidth}{@{\extracolsep{\fill}}lcccc@{}}
\toprule
 & (1) & (2) & (3) & (4) \\
\midrule
DV & \shortstack{Adopter} & \shortstack{Adopter} & \shortstack{Adopter} & \shortstack{Adopter} \\
Controls & \shortstack{Base.} & \shortstack{Addl.} & \shortstack{Addl.} & \shortstack{Addl.} \\
Sample & \shortstack{All} & \shortstack{All} & \shortstack{All} & \shortstack{No\\tech} \\
\midrule
L. log rev./emp. & 0.009*** & 0.006*** & 0.004* & 0.005** \\
 & (0.002) & (0.002) & (0.002) & (0.002) \\
L. log assets/emp. &  & 0.010*** & 0.013*** & 0.009*** \\
 &  & (0.003) & (0.003) & (0.003) \\
L. log PP\&E/emp. &  & -0.007*** & -0.001 & -0.007*** \\
 &  & (0.002) & (0.002) & (0.002) \\
L. log emp. & 0.013*** & 0.015*** & 0.019*** & 0.013*** \\
 & (0.001) & (0.001) & (0.001) & (0.001) \\
\midrule
Obs. & 8,229 & 8,229 & 8,229 & 7,379 \\
$R^2$ & 0.053 & 0.055 & 0.109 & 0.046 \\
FYs & 2024-2025 & 2024-2025 & 2024-2025 & 2024-2025 \\
Year FE & Yes & Yes & Yes & Yes \\
Ind. FE & \shortstack{NAICS2} & \shortstack{NAICS2} & \shortstack{NAICS4} & \shortstack{NAICS2} \\
\bottomrule
\end{tabular*}
\par
\vspace{4pt}
\begin{minipage}{\textwidth}
\raggedright\scriptsize\textit{Note:} Each column reports a linear probability model estimated on public firm-years with positive lagged employment and positive lagged firm-characteristic values. The dependent variable equals one when a public firm's first OpenAI Enterprise adoption date falls in the current fiscal year; controls are public firm-years not linked through the OpenAI ticker bridge. The table reports the industry fixed-effect level used in each column. Baseline columns control for lagged log employment. Additional-control columns additionally control for lagged log assets per employee, lagged log positive PP\&E per employee, and indicators for no positive and missing lagged PP\&E; the indicator coefficients are included in the model but omitted from the table. The no-tech column excludes firms with two-digit NAICS code 51. Standard errors clustered by Compustat gvkey are in parentheses. *, **, and *** indicate significance at the 10\%, 5\%, and 1\% levels.\par
\end{minipage}
}
\end{table}

\clearpage
\begin{table}[p]
\centering
\caption{FINANCIAL CHARACTERISTICS AND USAGE INTENSITY}
\label{tab:usage-outcome-financial-characteristics}
{\footnotesize\setstretch{1}
\renewcommand{\arraystretch}{1.15}
\setlength{\tabcolsep}{2pt}
\begin{tabular*}{\textwidth}{@{\extracolsep{\fill}}lcccc@{}}
\toprule
 & (1) & (2) & (3) & (4) \\
\midrule
DV & \shortstack{Msgs/\\act. wk/emp} & \shortstack{WAU/emp} & \shortstack{Tokens/emp} & \shortstack{Msgs/WAU} \\
\midrule
L. log rev./emp. & 0.062 & 0.006 & 0.036 & -0.014 \\
 & (0.044) & (0.007) & (0.094) & (0.021) \\
L. log assets/emp. & 0.061 & 0.012* & 0.107 & 0.008 \\
 & (0.045) & (0.008) & (0.099) & (0.019) \\
L. log PP\&E/emp. & -0.079** & -0.010* & -0.087 & -0.018 \\
 & (0.033) & (0.006) & (0.070) & (0.014) \\
L. log emp. & -0.266*** & -0.032*** & -0.667*** & -0.002 \\
 & (0.019) & (0.003) & (0.048) & (0.008) \\
\midrule
Obs. & 478 & 478 & 396 & 482 \\
$R^2$ & 0.480 & 0.362 & 0.471 & 0.100 \\
FYs & 2024-2025 & 2024-2025 & 2024-2025 & 2024-2025 \\
Year FE & Yes & Yes & Yes & Yes \\
Ind. FE & NAICS2 & NAICS2 & NAICS2 & NAICS2 \\
\bottomrule
\end{tabular*}
\par
\vspace{4pt}
\begin{minipage}{\textwidth}
\raggedright\scriptsize\textit{Note:} Each column reports an OLS regression of transformed current fiscal-year Enterprise usage intensity on lagged public-company financial characteristics, estimated on public tickers with positive fiscal-year Enterprise usage, positive lagged employment, and positive lagged firm-characteristic values. Dependent variables are transformed as log one plus usage intensity. Msgs/act. wk/emp is messages per active usage week per employee; WAU/emp is mean weekly active users per employee; Tokens/emp is mean weekly output tokens per employee; Msgs/WAU is messages per weekly active user. Output-token usage is measured over weeks with complete output-token coverage. All columns control for lagged log employment, lagged log assets per employee, lagged log positive PP\&E per employee, and indicators for no positive and missing lagged PP\&E; the indicator coefficients are included in the model but omitted from the table. Standard errors clustered by Compustat gvkey are in parentheses. *, **, and *** indicate significance at the 10\%, 5\%, and 1\% levels.\par
\end{minipage}
}
\end{table}

\clearpage
\begin{table}[p]
\centering
\caption{FIRM SCALE AND ENTERPRISE ADOPTION}
\label{tab:adoption-superstar-analysis}
{\footnotesize\setstretch{1}
\renewcommand{\arraystretch}{1.15}
\begin{tabular*}{\textwidth}{@{\extracolsep{\fill}}lccccc@{}}
\toprule
 & (1) & (2) & (3) & (4) & (5) \\
\midrule
DV & Adopter & Adopter & Adopter & Adopter & Adopter \\
Scale & Revenue & Revenue & Revenue & Ind.-yr rev. & Ind.-yr rev. \\
\midrule
L. log revenue & 0.011*** &  &  &  &  \\
 & (0.001) &  &  &  &  \\
Top 25\% by L. rev. &  & 0.069*** &  &  &  \\
 &  & (0.007) &  &  &  \\
Top 5\% by L. rev. &  &  & 0.098*** &  &  \\
 &  &  & (0.016) &  &  \\
Top 25\% by L. rev., ind.-yr &  &  &  & 0.072*** &  \\
 &  &  &  & (0.006) &  \\
Top 5\% by L. rev., ind.-yr &  &  &  &  & 0.113*** \\
 &  &  &  &  & (0.017) \\
\midrule
Obs. & 9,391 & 9,391 & 9,391 & 9,391 & 9,391 \\
$R^2$ & 0.051 & 0.046 & 0.035 & 0.048 & 0.040 \\
FYs & 2024-2025 & 2024-2025 & 2024-2025 & 2024-2025 & 2024-2025 \\
Year FE & Yes & Yes & Yes & Yes & Yes \\
Ind. FE & NAICS2 & NAICS2 & NAICS2 & NAICS2 & NAICS2 \\
\bottomrule
\end{tabular*}
\par
\vspace{4pt}
{\raggedright\scriptsize\textit{Note:} Each column reports a linear probability model estimated on public firm-years with positive lagged scale values. The dependent variable equals one when a public firm's first OpenAI Enterprise adoption date falls in the current fiscal year; controls are public firm-years not linked through the OpenAI ticker bridge. Column 1 uses lagged log revenue. Columns 2--5 use indicators for being in the top tail of lagged revenue, computed within fiscal year using lagged total revenue; industry-year indicators are computed within fiscal-year and two-digit NAICS cells. All columns include fiscal-year and two-digit NAICS industry fixed effects. Standard errors clustered by Compustat gvkey are in parentheses. *, **, and *** indicate significance at the 10\%, 5\%, and 1\% levels.\par}
}
\end{table}

\clearpage
\vspace*{\fill}
\begin{table}[H]
\centering
\caption{\large\MakeUppercase{Intangible assets and enterprise adoption}}
\label{tab:adoption-outcome-complement-characteristics}
\vspace{6pt}
\footnotesize
\begin{tabular}{lcccccc}
\toprule
  & (1) & (2) & (3) & (4) & (5) & (6) \\
\midrule
Dependent variable & Adopter & Adopter & Adopter & Adopter & Adopter & Adopter \\
Complement measure & \shortstack{SG\&A\\stock} & \shortstack{R\&D\\stock} & \shortstack{Capitalized\\software} & \shortstack{SG\&A\\stock} & \shortstack{R\&D\\stock} & \shortstack{Capitalized\\software} \\
Sample & All & All & All & \shortstack{No tech/\\high R\&D} & \shortstack{No tech/\\high R\&D} & \shortstack{No tech/\\high R\&D} \\
\midrule
Log(1 + comp./emp.) & 0.020*** & 0.004*** & 0.008** & 0.010 & 0.003*** & 0.006 \\
 & (0.005) & (0.001) & (0.003) & (0.007) & (0.001) & (0.006) \\
L. log revenue/emp. & 0.006** & 0.008*** & 0.012 & 0.010 & 0.019*** & 0.021 \\
 & (0.003) & (0.002) & (0.009) & (0.007) & (0.006) & (0.019) \\
L. log assets/emp. & 0.004 & 0.010*** & 0.020 & 0.002 & -0.001 & 0.005 \\
 & (0.004) & (0.003) & (0.012) & (0.007) & (0.006) & (0.020) \\
L. log PP\&E/emp. & -0.005** & -0.005** & -0.017** & -0.005 & -0.005* & -0.010 \\
 & (0.002) & (0.002) & (0.008) & (0.003) & (0.003) & (0.012) \\
L. log employment & 0.020*** & 0.015*** & 0.017*** & 0.017*** & 0.015*** & 0.023*** \\
 & (0.002) & (0.001) & (0.004) & (0.002) & (0.002) & (0.007) \\
\midrule
Observations & 5,943 & 7,117 & 1,076 & 3,247 & 4,029 & 477 \\
$R^2$ & 0.064 & 0.058 & 0.075 & 0.055 & 0.057 & 0.085 \\
Year FE & Yes & Yes & Yes & Yes & Yes & Yes \\
Ind. FE & NAICS2 & NAICS2 & NAICS2 & NAICS2 & NAICS2 & NAICS2 \\
\bottomrule
\end{tabular}
\vspace{4pt}\par\raggedright\scriptsize\setstretch{1}\textit{Note:} Each column reports a linear probability model estimated on public firm-years with positive employment and finite nonnegative complement stock per employee. The dependent variable equals one when a public firm's first OpenAI Enterprise adoption date falls in the current fiscal year; controls are public firm-years not linked through the OpenAI ticker bridge. Entries report coefficients from regressions of adoption on log one plus complement stock per employee, lagged log revenue per employee, lagged log employment, lagged log assets per employee, lagged log positive PP\&E per employee, indicators for no positive and missing lagged PP\&E, fiscal-year fixed effects, and two-digit NAICS industry fixed effects. The no-tech/high-R\&D columns exclude NAICS2 sectors 32, 33, 51; high-R\&D sectors are selected using fiscal-year 2021 sector median R\&D stock per employee. Fiscal years: 2024-2025. Complement variables are measured in fiscal year 2021. SG\&A and R\&D stocks are accumulated from Compustat flow variables: SG\&A uses \texttt{xsga} with 20\% annual depreciation, and R\&D uses \texttt{xrd} with 15\% annual depreciation. Capitalized software uses the Compustat \texttt{capsft} level joined from the capitalized-software extract. Complement values are converted to dollars, divided by employees, and transformed as log one plus the employee-normalized value before entering the model. Complement-stock opening stock: zero-growth steady-state opening stock. Missing SG\&A and negative flows are left missing; missing R\&D and observed zero flows are treated as zero investment. Standard errors clustered by Compustat gvkey are in parentheses. *, **, and *** indicate significance at the 10\%, 5\%, and 1\% levels.\par
\end{table}

\vspace*{\fill}

\clearpage
{\small\setstretch{1.05}
\printbibliography
}

\clearpage

\appendix

\renewcommand{\thefigure}{A\arabic{figure}}
\renewcommand{\thetable}{A\arabic{table}}
\setcounter{figure}{0}
\setcounter{table}{0}

\section{Additional Tables}

\begin{table}[H]
\centering
\footnotesize
\caption{Compustat measures used in the analysis}
\label{tab:capiq-measures}
\begin{tabular}{p{0.22\linewidth} p{0.20\linewidth} p{0.32\linewidth} p{0.18\linewidth}}
\toprule
Measure & Source field(s) & Construction & Used for \\
\midrule
Revenue & \texttt{revt} & Annual Compustat revenue, reported in millions of dollars & Firm scale and revenue-per-employee outcomes \\
Total assets & \texttt{at} & Annual Compustat total assets, reported in millions of dollars & Firm scale and balance-sheet comparison \\
Employees & \texttt{emp} & Compustat employment, converted from thousands to workers & Denominator for per-employee measures \\
Net PP\&E & \texttt{ppent} & Annual Compustat net property, plant, and equipment, reported in millions of dollars & Physical capital comparison \\
Market value & \texttt{mkvalt} & Annual Compustat market value, reported in millions of dollars & Market-value and market-value-per-employee outcomes \\
R\&D expense & \texttt{xrd} & Annual Compustat research and development expense, reported in millions of dollars & Intangible investment comparison \\
Revenue per employee & \texttt{revt}, \texttt{emp} & Revenue divided by employees & Financial distribution and regression outcome \\
Market value per employee & \texttt{mkvalt}, \texttt{emp} & Market value divided by employees & Financial distribution and regression outcome \\
Enterprise usage intensity &
\shortstack[l]{\texttt{annual\_total}\\\texttt{messages},\\\texttt{active\_weeks},\\\texttt{emp}} &
Messages per active usage week per employee &
Usage-level financial comparisons \\
WAU per employee &
\shortstack[l]{\texttt{mean\_weekly}\\\texttt{active\_users},\\\texttt{emp}} &
Mean weekly active users divided by employees &
Industry usage-intensity coefficient figure \\
\bottomrule
\end{tabular}

\vspace{0.5em}
\begin{minipage}{0.94\linewidth}
\footnotesize
\emph{Note:} Financial variables are drawn from the Compustat annual business metrics panel. Dollar-denominated Compustat variables are reported in millions of dollars. Employment is converted from thousands of workers to workers before constructing per-employee measures. Enterprise usage measures are aggregated to ticker-fiscal-year observations before merging to the financial panel.
\end{minipage}
\end{table}
\newpage
\begin{longtable}{llr}
\caption{Top 5 actual job titles by job title class}\label{tab:enterprise-scim-job-title-validation-job-title-class}\\
\toprule
Job title class & Actual job title & User share \\
\midrule
\endfirsthead
\toprule
Job title class & Actual job title & User share \\
\midrule
\endhead
Administrative / executive support & Executive assistant & 15.4\% \\
Administrative / executive support & Administrative assistant & 4.5\% \\
Administrative / executive support & Student assistant & 3.4\% \\
Administrative / executive support & Senior executive assistant & 2.6\% \\
Administrative / executive support & Office manager & 2.0\% \\
\addlinespace
Analyst & Analyst & 7.9\% \\
Analyst & Business analyst & 3.9\% \\
Analyst & \textless{}redacted\textgreater{} & 2.8\% \\
Analyst & Senior analyst & 2.2\% \\
Analyst & Senior business analyst & 1.9\% \\
\addlinespace
Consultant / professional services & Consultant & 19.2\% \\
Consultant / professional services & Senior consultant & 6.4\% \\
Consultant / professional services & Management consultant & 6.0\% \\
Consultant / professional services & Management consulting manager & 4.1\% \\
Consultant / professional services & Associate consultant & 3.1\% \\
\addlinespace
Contractor / temporary & Contractor & 36.0\% \\
Contractor / temporary & Contingent worker & 9.4\% \\
Contractor / temporary & Contingent worker hourly & 6.3\% \\
Contractor / temporary & Temporary worker & 5.5\% \\
Contractor / temporary & Contractor - \textless{}redacted\textgreater{} & 2.5\% \\
\addlinespace
Customer success / support role & Customer success manager & 2.6\% \\
Customer success / support role & Senior customer success manager & 1.5\% \\
Customer success / support role & Customer service specialist & 1.2\% \\
Customer success / support role & Senior customer service representative & 1.1\% \\
Customer success / support role & Customer service representative & 0.9\% \\
\addlinespace
Data / analytics practitioner & Data engineer & 4.9\% \\
Data / analytics practitioner & Senior data scientist & 4.4\% \\
Data / analytics practitioner & Data scientist & 4.4\% \\
Data / analytics practitioner & Senior data engineer & 4.1\% \\
Data / analytics practitioner & Machine learning engineer & 2.3\% \\
\addlinespace
Design / creative role & Senior product designer & 4.3\% \\
Design / creative role & Product designer & 3.7\% \\
Design / creative role & Graphic designer & 2.1\% \\
Design / creative role & Designer & 1.9\% \\
Design / creative role & Senior designer & 1.7\% \\
\addlinespace
Education / academic role & Student & 67.1\% \\
Education / academic role & Student worker & 3.2\% \\
Education / academic role & Professor & 3.2\% \\
Education / academic role & Undergraduate & 2.5\% \\
Education / academic role & Associate professor & 2.4\% \\
\addlinespace
Engineering / technical practitioner & Software engineer & 8.6\% \\
Engineering / technical practitioner & Senior software engineer & 7.1\% \\
Engineering / technical practitioner & Staff software engineer & 1.8\% \\
Engineering / technical practitioner & Software engineer II & 1.5\% \\
Engineering / technical practitioner & \textless{}redacted\textgreater{} & 1.1\% \\
\addlinespace
Executive / founder / partner & Partner & 14.6\% \\
Executive / founder / partner & Managing director & 7.1\% \\
Executive / founder / partner & Executive director & 1.8\% \\
Executive / founder / partner & Associate partner & 1.7\% \\
Executive / founder / partner & General manager & 1.2\% \\
\addlinespace
Finance / accounting role & Senior accountant & 1.7\% \\
Finance / accounting role & Portfolio manager & 1.2\% \\
Finance / accounting role & Finance manager & 1.1\% \\
Finance / accounting role & Accountant & 1.0\% \\
Finance / accounting role & Trader & 0.9\% \\
\addlinespace
Generic rank / seniority only & Associate & 12.3\% \\
Generic rank / seniority only & Senior associate & 10.0\% \\
Generic rank / seniority only & Manager & 8.1\% \\
Generic rank / seniority only & Director & 5.7\% \\
Generic rank / seniority only & Senior manager & 5.2\% \\
\addlinespace
Healthcare / clinical role & CRA & 5.6\% \\
Healthcare / clinical role & \textless{}redacted\textgreater{} & 3.0\% \\
Healthcare / clinical role & Physician & 1.4\% \\
Healthcare / clinical role & \textless{}redacted\textgreater{} & 1.2\% \\
Healthcare / clinical role & \textless{}redacted\textgreater{} & 1.2\% \\
\addlinespace
IT / systems role & IT specialist & 1.3\% \\
IT / systems role & Systems administrator & 0.9\% \\
IT / systems role & IT manager & 0.9\% \\
IT / systems role & \textless{}redacted\textgreater{} & 0.8\% \\
IT / systems role & IT support specialist & 0.7\% \\
\addlinespace
Legal role & Associate attorney & 2.8\% \\
Legal role & Legal counsel & 2.7\% \\
Legal role & Paralegal & 2.6\% \\
Legal role & Senior legal counsel & 2.4\% \\
Legal role & Counsel & 2.1\% \\
\addlinespace
Marketing / communications role & Marketing manager & 1.7\% \\
Marketing / communications role & Product marketing manager & 0.8\% \\
Marketing / communications role & Senior marketing manager & 0.6\% \\
Marketing / communications role & Senior product marketing manager & 0.6\% \\
Marketing / communications role & Technical writer & 0.6\% \\
\addlinespace
People / recruiting role & Recruiter & 2.2\% \\
People / recruiting role & HR business partner & 1.7\% \\
People / recruiting role & Senior recruiter & 1.7\% \\
People / recruiting role & Talent acquisition partner & 1.2\% \\
People / recruiting role & HR manager & 1.0\% \\
\addlinespace
Product role & Senior product manager & 10.5\% \\
Product role & Product manager & 10.3\% \\
Product role & Product owner & 3.3\% \\
Product role & Principal product manager & 2.1\% \\
Product role & Director, product management & 1.9\% \\
\addlinespace
Project / program role & Project manager & 6.3\% \\
Project / program role & Senior project manager & 3.7\% \\
Project / program role & Project leader & 2.6\% \\
Project / program role & Program manager & 1.7\% \\
Project / program role & Senior program manager & 1.0\% \\
\addlinespace
Research / scientist role & Senior CRA & 4.3\% \\
Research / scientist role & Principal CRA & 2.3\% \\
Research / scientist role & Research assistant & 2.1\% \\
Research / scientist role & Senior scientist & 2.1\% \\
Research / scientist role & Research associate & 1.9\% \\
\addlinespace
Sales / account role & Account executive & 3.3\% \\
Sales / account role & Account manager & 3.0\% \\
Sales / account role & Senior account executive & 1.8\% \\
Sales / account role & Senior account manager & 1.6\% \\
Sales / account role & Sales development representative & 1.3\% \\
\addlinespace
Security role & Security engineer & 3.4\% \\
Security role & Senior security engineer & 2.5\% \\
Security role & Chief information security officer & 1.3\% \\
Security role & \textless{}redacted\textgreater{} & 1.1\% \\
Security role & Security analyst & 1.0\% \\
\bottomrule
\end{longtable}
\begin{flushleft}
\footnotesize \textit{Note:} Percentages are shares of title-user observations within each classifier value in the full SCIM job-title classification universe. Unknown and malformed classifier values are omitted. Normalized title variants are combined before applying the publication rule. Job-title detail observed in fewer than 2 workspaces is redacted. Ties are ordered alphabetically by actual job title.
\end{flushleft}

\newpage
\begin{longtable}{llr}
\caption{Top 5 actual job titles by seniority level}\label{tab:enterprise-scim-job-title-validation-seniority-level}\\
\toprule
Seniority level & Actual job title & User share \\
\midrule
\endfirsthead
\toprule
Seniority level & Actual job title & User share \\
\midrule
\endhead
C-suite / founder / owner / partner & Partner & 29.1\% \\
C-suite / founder / owner / partner & Managing director & 14.1\% \\
C-suite / founder / owner / partner & Executive director & 3.5\% \\
C-suite / founder / owner / partner & Managing director and partner & 2.4\% \\
C-suite / founder / owner / partner & Chief financial officer & 1.8\% \\
\addlinespace
Contractor / temporary & Contractor & 34.3\% \\
Contractor / temporary & Contingent worker & 9.0\% \\
Contractor / temporary & Contingent worker hourly & 6.0\% \\
Contractor / temporary & Temporary worker & 5.2\% \\
Contractor / temporary & IT BST outside consultant & 2.9\% \\
\addlinespace
Director / head & Director & 7.6\% \\
Director / head & Senior manager & 6.6\% \\
Director / head & Associate director & 1.0\% \\
Director / head & Associate partner & 0.8\% \\
Director / head & Senior director & 0.8\% \\
\addlinespace
Entry / associate & Associate & 15.2\% \\
Entry / associate & Analyst & 3.7\% \\
Entry / associate & Business analyst & 1.8\% \\
Entry / associate & \textless{}redacted\textgreater{} & 1.3\% \\
Entry / associate & Associate consultant & 1.2\% \\
\addlinespace
Individual contributor / professional & Software engineer & 6.5\% \\
Individual contributor / professional & Consultant & 3.8\% \\
Individual contributor / professional & Executive assistant & 1.4\% \\
Individual contributor / professional & Project manager & 1.3\% \\
Individual contributor / professional & Management consultant & 1.2\% \\
\addlinespace
Manager / team lead & Manager & 7.0\% \\
Manager / team lead & Management consulting manager & 1.2\% \\
Manager / team lead & Engineering manager & 1.0\% \\
Manager / team lead & [non-Latin text] & 0.8\% \\
Manager / team lead & Project leader & 0.8\% \\
\addlinespace
Senior IC / principal & Senior associate & 6.3\% \\
Senior IC / principal & Senior software engineer & 5.5\% \\
Senior IC / principal & Principal & 1.4\% \\
Senior IC / principal & Staff software engineer & 1.4\% \\
Senior IC / principal & Senior consultant & 1.3\% \\
\addlinespace
Student / trainee / intern & Student & 61.4\% \\
Student / trainee / intern & Intern & 6.1\% \\
Student / trainee / intern & Student worker & 3.0\% \\
Student / trainee / intern & Undergraduate & 2.3\% \\
Student / trainee / intern & Assistant professor & 2.0\% \\
\addlinespace
VP / senior executive & Vice president & 14.2\% \\
VP / senior executive & Senior vice president & 2.9\% \\
VP / senior executive & Corporate vice president & 2.8\% \\
VP / senior executive & Assistant vice president & 1.2\% \\
VP / senior executive & VP & 0.6\% \\
\bottomrule
\end{longtable}
\begin{flushleft}
\footnotesize \textit{Note:} Percentages are shares of title-user observations within each classifier value in the full SCIM job-title classification universe. Unknown and malformed classifier values are omitted. Normalized title variants are combined before applying the publication rule. Job-title detail observed in fewer than 2 workspaces is redacted. Ties are ordered alphabetically by actual job title.
\end{flushleft}

\newpage
\begin{longtable}{llr}
\caption{Top 5 actual job titles by manager status}\label{tab:enterprise-scim-job-title-validation-people-manager-signal}\\
\toprule
Manager status & Actual job title & User share \\
\midrule
\endfirsthead
\toprule
Manager status & Actual job title & User share \\
\midrule
\endhead
Likely individual contributor & Student & 6.7\% \\
Likely individual contributor & Associate & 3.5\% \\
Likely individual contributor & Software engineer & 3.0\% \\
Likely individual contributor & Senior associate & 2.9\% \\
Likely individual contributor & Senior software engineer & 2.5\% \\
\addlinespace
Likely people manager & Manager & 3.6\% \\
Likely people manager & Director & 2.6\% \\
Likely people manager & Partner & 2.5\% \\
Likely people manager & Senior manager & 2.3\% \\
Likely people manager & Vice president & 1.4\% \\
\addlinespace
Manager title but ambiguous & Project manager & 2.2\% \\
Manager title but ambiguous & Senior product manager & 1.9\% \\
Manager title but ambiguous & Product manager & 1.8\% \\
Manager title but ambiguous & \textless{}redacted\textgreater{} & 1.4\% \\
Manager title but ambiguous & Account executive & 1.4\% \\
\bottomrule
\end{longtable}
\begin{flushleft}
\footnotesize \textit{Note:} Percentages are shares of title-user observations within each classifier value in the full SCIM job-title classification universe. Unknown and malformed classifier values are omitted. Normalized title variants are combined before applying the publication rule. Job-title detail observed in fewer than 2 workspaces is redacted. Ties are ordered alphabetically by actual job title.
\end{flushleft}

\newpage

\begin{longtable}{@{}ll@{}}
\caption{Task classifier values}
\label{tab:task-classifier-values} \\
\toprule
First-level category & Second-level value \\
\midrule
\endfirsthead

\toprule
First-level category & Second-level value \\
\midrule
\endhead

\bottomrule
\endfoot

\texttt{advice} & \texttt{career} \\
 & \texttt{academic} \\
 & \texttt{legal and regulatory} \\
 & \texttt{financial and tax} \\
 & \texttt{beauty} \\
 & \texttt{cooking and recipes} \\
 & \texttt{spirituality} \\
\addlinespace

\texttt{analysis and calculations} & \texttt{data analysis} \\
 & \texttt{scientific math and engineering computation} \\
\addlinespace

\texttt{coding} & \texttt{plan design and create} \\
 & \texttt{debug and fix} \\
 & \texttt{modify and extend} \\
 & \texttt{interpret and explain} \\
\addlinespace

\texttt{commerce} & \texttt{product specs} \\
 & \texttt{product reviews and recommendations} \\
 & \texttt{facilitate purchase} \\
\addlinespace

\texttt{creative media} & \texttt{generate image} \\
 & \texttt{edit image} \\
 & \texttt{video audio or other} \\
\addlinespace

\texttt{education and learning} & \texttt{review and develop academic essay} \\
 & \texttt{help with homework assignment} \\
\addlinespace

\texttt{health and wellness} & \texttt{triage and symptoms} \\
 & \texttt{tests and results} \\
 & \texttt{care and treatments} \\
 & \texttt{nutrition and diet} \\
 & \texttt{fitness and exercise} \\
 & \texttt{mental and emotional} \\
\addlinespace

\texttt{how to and procedural guidance} & \texttt{technical digital and electronic} \\
 & \texttt{technical mechanical and vehicle} \\
 & \texttt{technical other} \\
 & \texttt{nontechnical personal} \\
 & \texttt{nontechnical professional or academic} \\
\addlinespace

\texttt{information} & \texttt{topic overview} \\
 & \texttt{entertainment and sports} \\
 & \texttt{current events and news} \\
 & \texttt{local travel and weather} \\
 & \texttt{business and market research} \\
 & \texttt{people} \\
 & \texttt{find file or document} \\
 & \texttt{other facts and figures} \\
\addlinespace

\texttt{support} & \texttt{goal setting and coaching} \\
 & \texttt{relationships} \\
 & \texttt{self disclosure and companionship} \\
\addlinespace

\texttt{writing and communication} & \texttt{fiction} \\
 & \texttt{games and role play} \\
 & \texttt{private personal message} \\
 & \texttt{public personal message} \\
 & \texttt{documentation and technical writing} \\
 & \texttt{instructional content} \\
 & \texttt{sales and marketing} \\
 & \texttt{career and job search} \\
 & \texttt{persuasion and argumentation} \\
 & \texttt{translation} \\

\end{longtable}

\clearpage
\section{Additional Figures} \label{sec:add_figures}

\begin{figure}[H]
\centering
\caption{ENTERPRISE OUTPUT TOKEN GROWTH BY PRODUCT}
\label{fig:token-demand-app}
\vspace{0.25em}
\includegraphics[width=\linewidth]{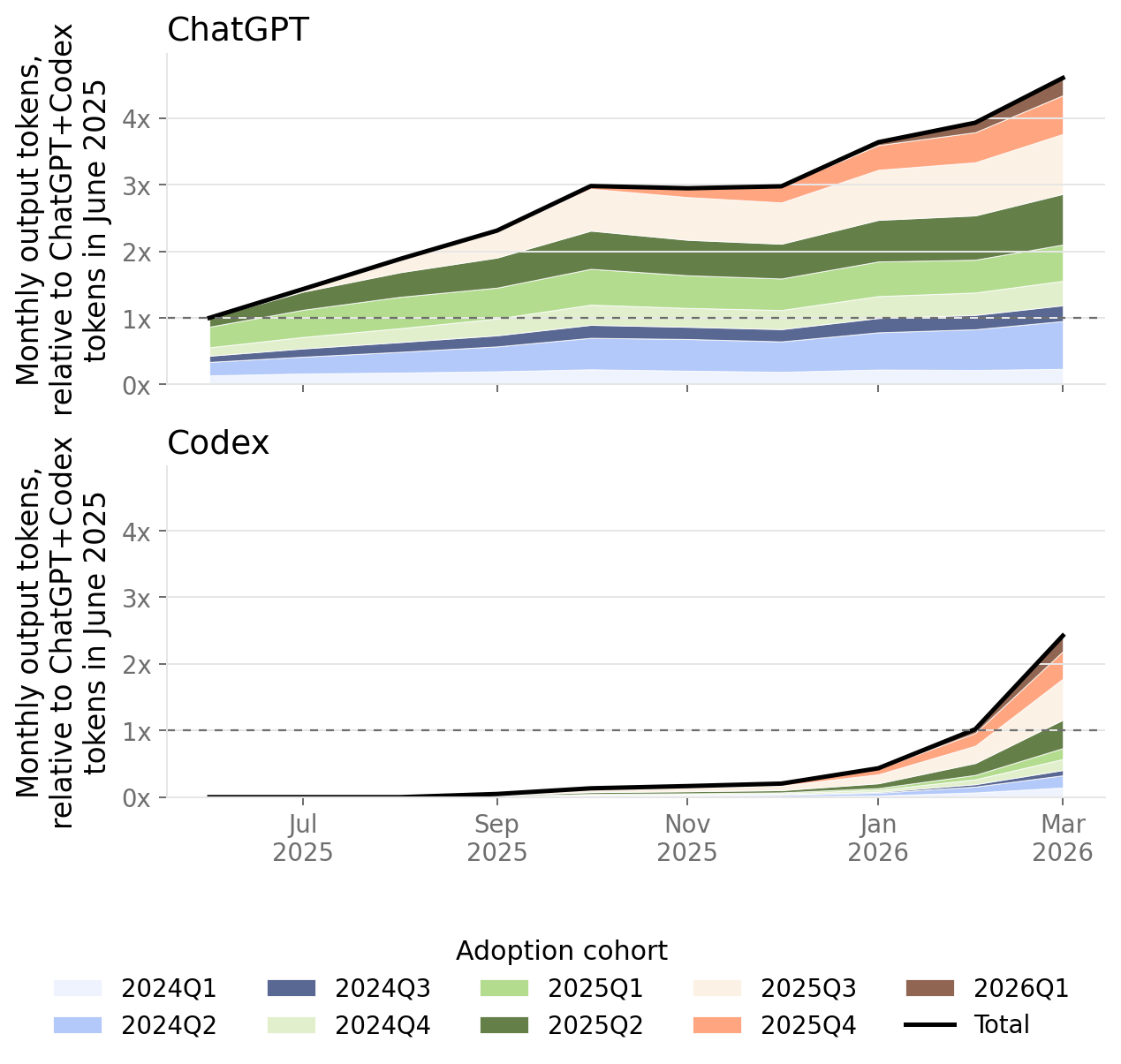}
\vspace{1em}
\begin{minipage}{\linewidth}
\footnotesize
\emph{Note:} This figure plots monthly ChatGPT and Codex output tokens for firms in the ChatGPT Enterprise analysis sample, shown separately by product and stacked by the quarter in which each firm first adopted ChatGPT Enterprise. Tokens are counted only after each firm’s ChatGPT Enterprise adoption date. Values in both panels are indexed to total combined ChatGPT and Codex output tokens in June 2025, with the dashed horizontal line marking 1.0 and the black line in each panel showing the aggregate total across cohorts for that product.
\end{minipage}
\end{figure}

\vspace*{\fill}

\begin{figure}[htbp]
\centering
\caption{COMPOSITION AND INTENSITY OF AI USE BY PEOPLE MANAGER STATUS}
\label{fig:adoption-variation-intensive_pm}

\textbf{Panel A. Composition of Active Users}\par
\vspace{0.25em}
\includegraphics[width=.7\linewidth]{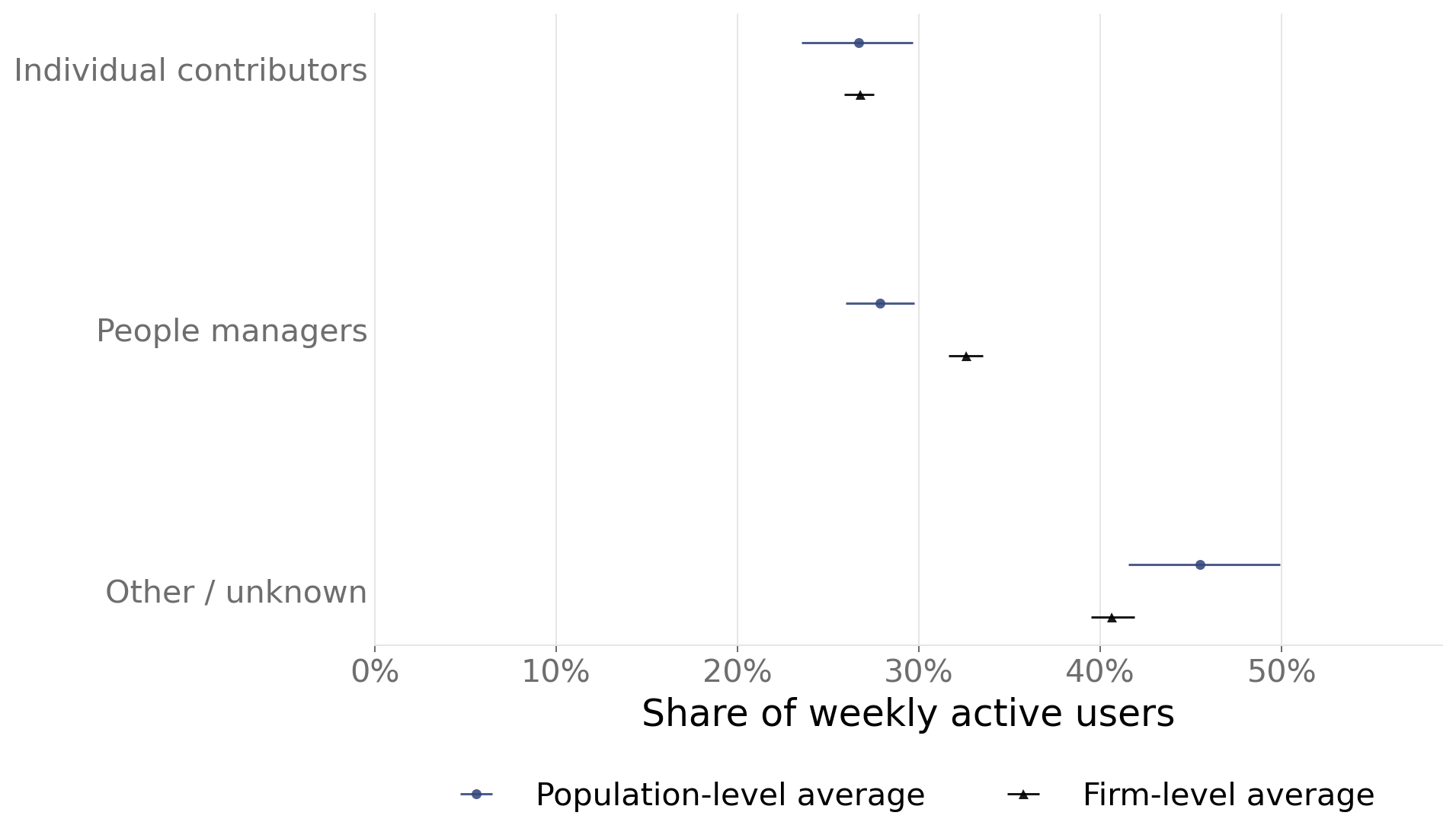}

\vspace{1em}

\textbf{Panel B. Usage Intensity}\par
\vspace{0.25em}
\includegraphics[width=.7\linewidth]{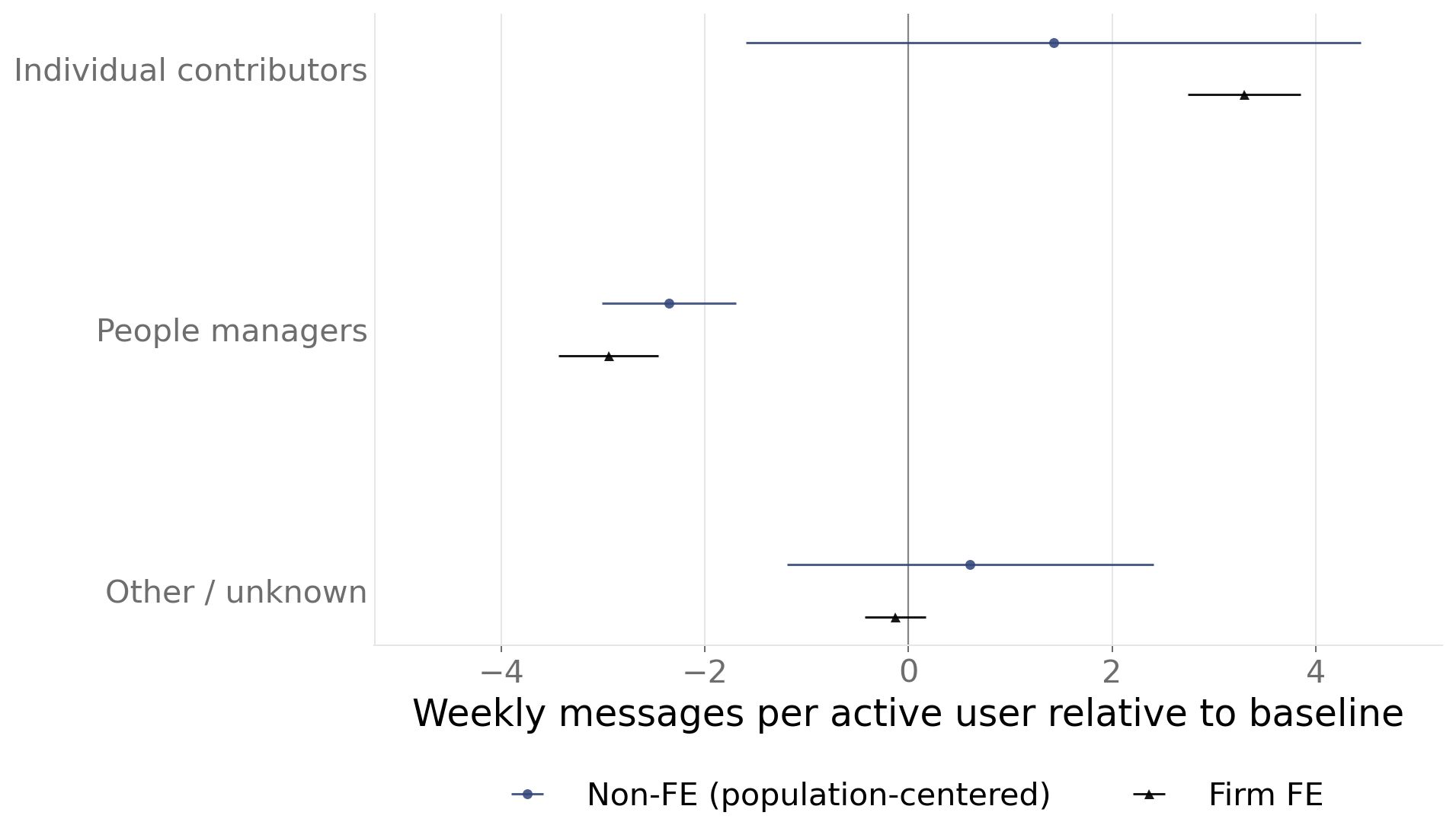}

\vspace{0.5em}
\begin{minipage}{\linewidth}
\footnotesize
\emph{Note:} This figure compares ChatGPT Enterprise use among individual contributors, people managers, and users whose manager status cannot be inferred, measured six months after firm adoption. Manager status is inferred from normalized job titles. Panel A reports each group's share of weekly active users. Blue points report the population-level average, computed as each group's share of all weekly active users across firms. Black points report the firm-level average, computed as each group's share of weekly active users within each firm and then averaged across firms, giving each firm equal weight. Horizontal bars show 95\% firm-bootstrap confidence intervals. Panel B reports differences in weekly messages per active user. Blue points report population-level estimates, comparing each group to the average active user across firms. Black points report firm fixed effect estimates, comparing each group to other active users within the same firm. Positive values indicate more intensive use than the relevant baseline; negative values indicate less intensive use. Horizontal bars in Panel B show 95\% confidence intervals with firm-clustered standard errors. ``Other / unknown'' includes users whose job titles are missing, ambiguous, malformed, or otherwise do not provide enough information to infer manager status.
\end{minipage}
\end{figure}

\vspace*{\fill}

\vspace*{\fill}

\begin{figure}[htbp]
\centering
\caption{DIFFERENCES IN AI USE ACROSS INDUSTRIES AND JOB TITLE CLASSES}
\label{fig:industry-job-title-variation_app}

\textbf{Panel A. Composition of Active Users}\par
\vspace{0.25em}
\includegraphics[width=.75\linewidth]{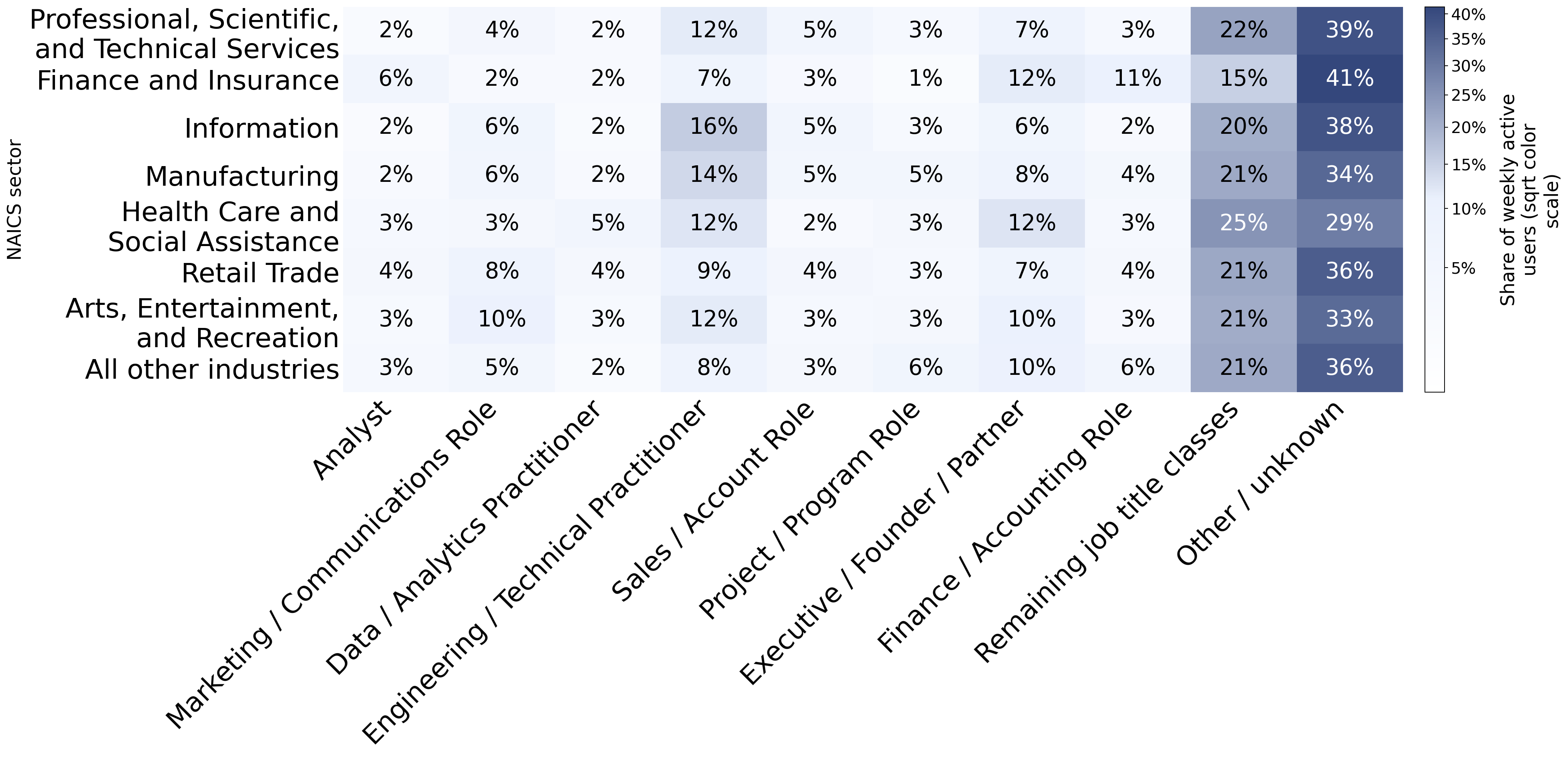}

\vspace{1em}

\textbf{Panel B. Usage Intensity}\par
\vspace{0.25em}
\includegraphics[width=.75\linewidth]{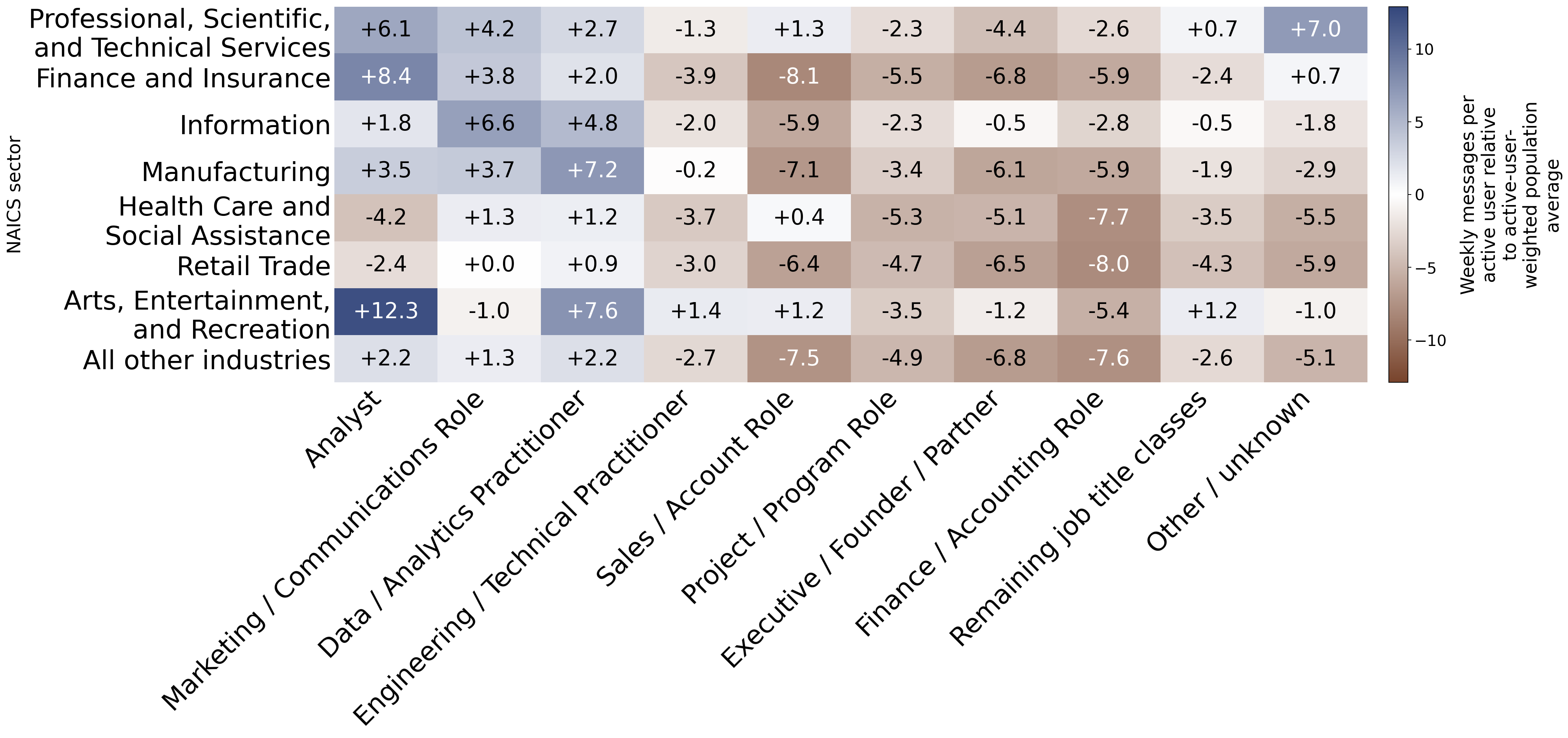}

\vspace{0.5em}
\begin{minipage}{\linewidth}
\footnotesize
\emph{Note:} The figure compares ChatGPT Enterprise use across two-digit NAICS sectors and inferred job title classes six months after firm adoption. Panel A reports the average share of weekly active users in each job title class within each industry; darker cells indicate that a larger share of active users in that industry come from that worker type. Panel B reports usage intensity for the same industry-by-role cells, measured as weekly messages per active user relative to the overall average active user in the sample. Positive values indicate more intensive use than average; negative values indicate less intensive use. ``Remaining job title classes'' pools inferred job title classes outside the individually displayed top categories. ``Other / unknown'' includes users with missing, ambiguous, malformed, or otherwise low-information job titles that cannot be assigned to a substantive class. A version of Panel B that accounts for firm fixed effects can be found in Panel A of Figure~\ref{fig:industry-job-title-variation-fe-app}.
\end{minipage}
\end{figure}

\vspace*{\fill}

\begin{figure}[htbp]
\centering
\caption{DIFFERENCES IN AI USE ACROSS INDUSTRIES AND SENIORITY LEVELS}
\label{fig:industry-seniority-variation}

\textbf{Panel A. Composition of Active Users}\par
\vspace{0.25em}
\includegraphics[width=.75\linewidth]{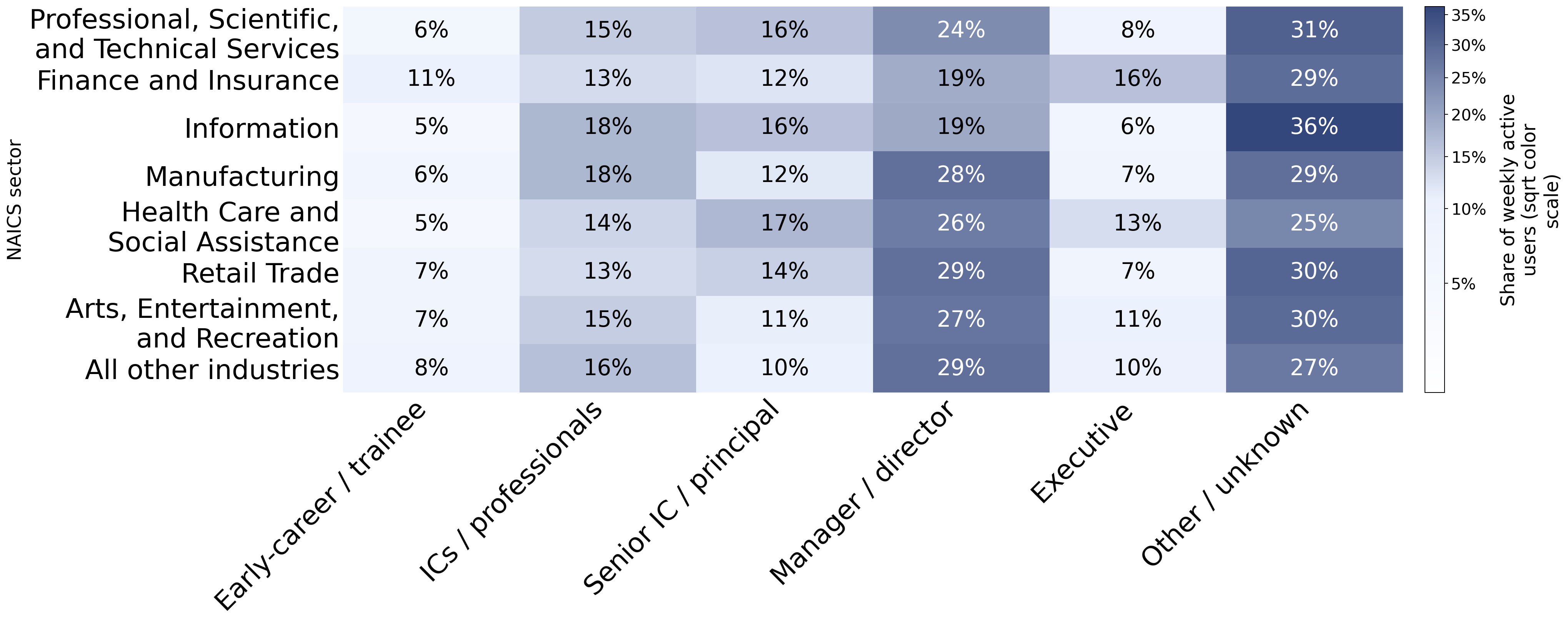}

\vspace{1em}

\textbf{Panel B. Usage Intensity}\par
\vspace{0.25em}
\includegraphics[width=.75\linewidth]{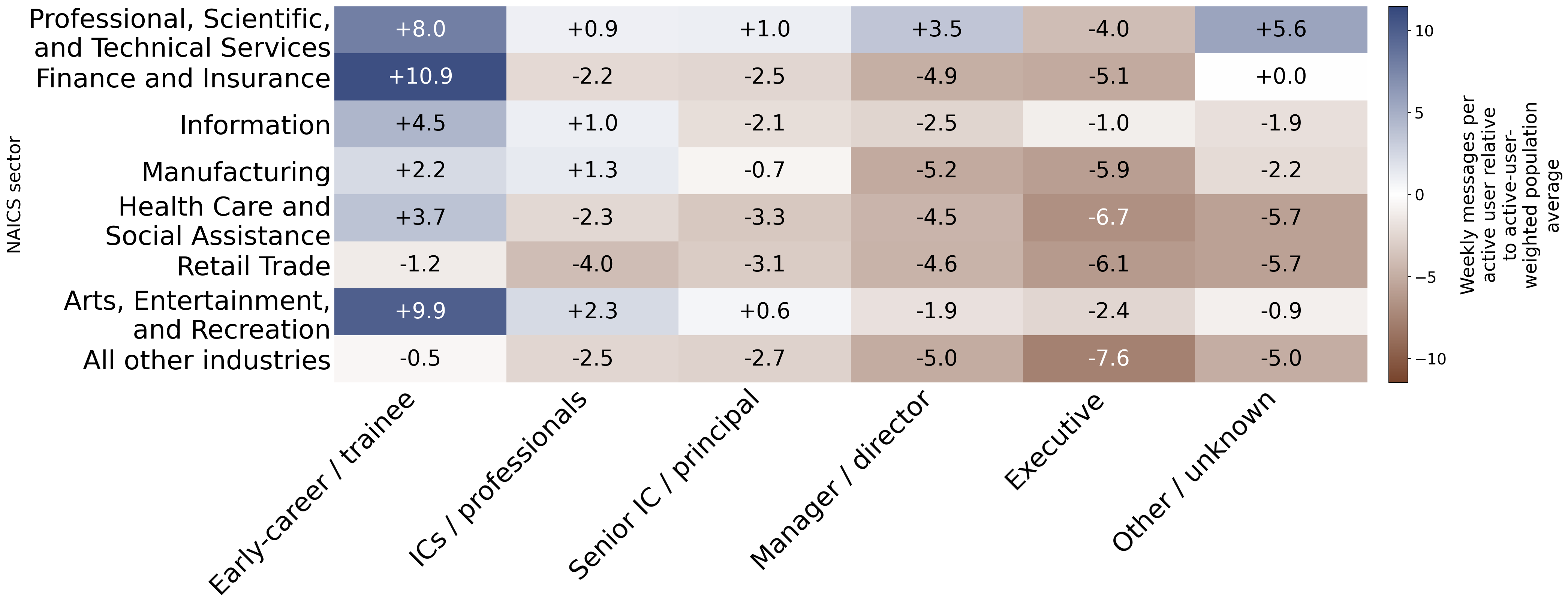}

\vspace{0.5em}
\begin{minipage}{\linewidth}
\footnotesize
\emph{Note:} The figure compares ChatGPT Enterprise use across two-digit NAICS sectors and inferred seniority levels six months after firm adoption. Panel A reports the average share of weekly active users at each seniority level within each industry; darker cells indicate that a larger share of active users in that industry belong to that seniority group. Panel B reports usage intensity for the same industry-by-seniority cells, measured as weekly messages per active user relative to the overall average active user in the sample. Positive values indicate more intensive use than average; negative values indicate less intensive use. ``Other / unknown'' includes users whose job titles are missing, ambiguous, malformed, indicate contractor or temporary status without conveying seniority, or otherwise do not provide enough information to infer a seniority level. ``IC'' denotes an individual contributor. A version of Panel B that accounts for firm fixed effects can be found in Panel B of Figure~\ref{fig:industry-job-title-variation-fe-app}.
\end{minipage}
\end{figure}

\vspace*{\fill}

\vspace*{\fill}

\begin{figure}[htbp]
\centering
\caption{DIFFERENCES IN AI USE ACROSS INDUSTRIES AND PEOPLE MANAGER STATUS}
\label{fig:industry-people-manager-variation-app}

\textbf{Panel A. Composition of Active Users}\par
\vspace{0.25em}
\includegraphics[width=.75\linewidth]{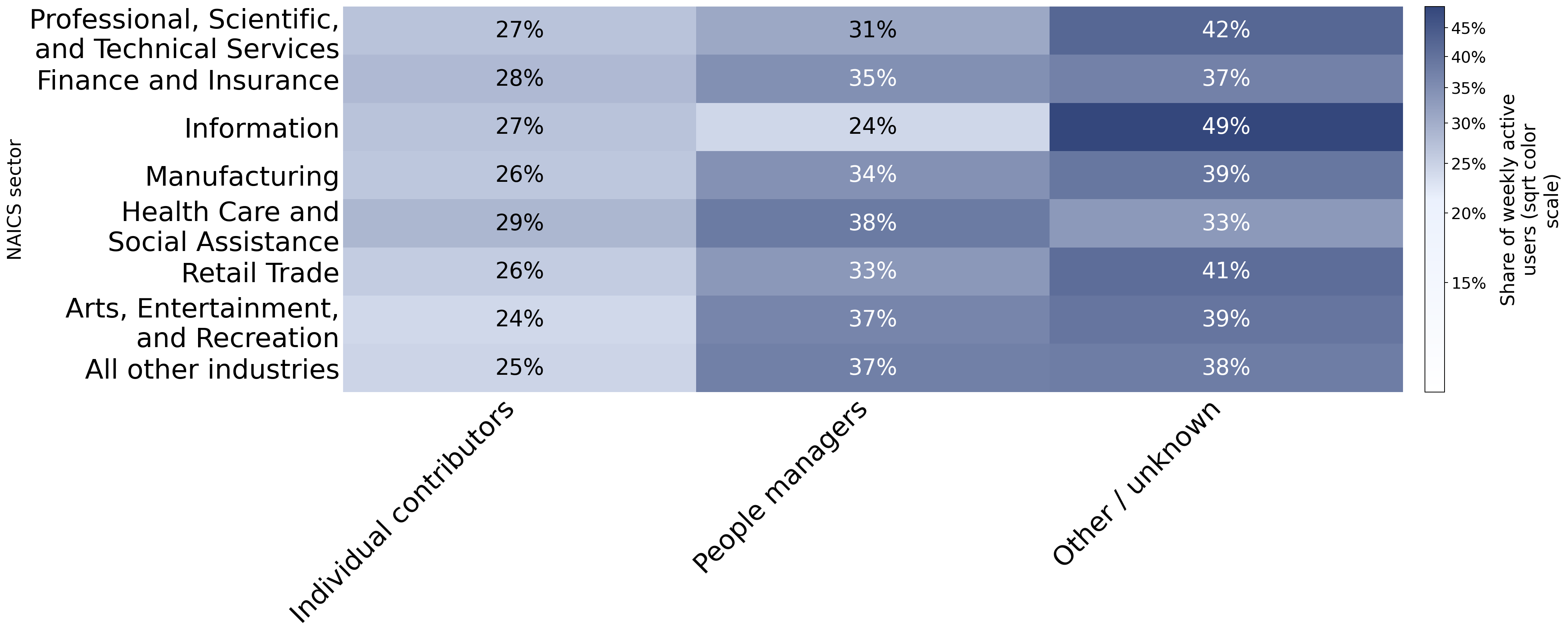}

\vspace{1em}

\textbf{Panel B. Intensity of Usage}\par
\vspace{0.25em}
\includegraphics[width=.75\linewidth]{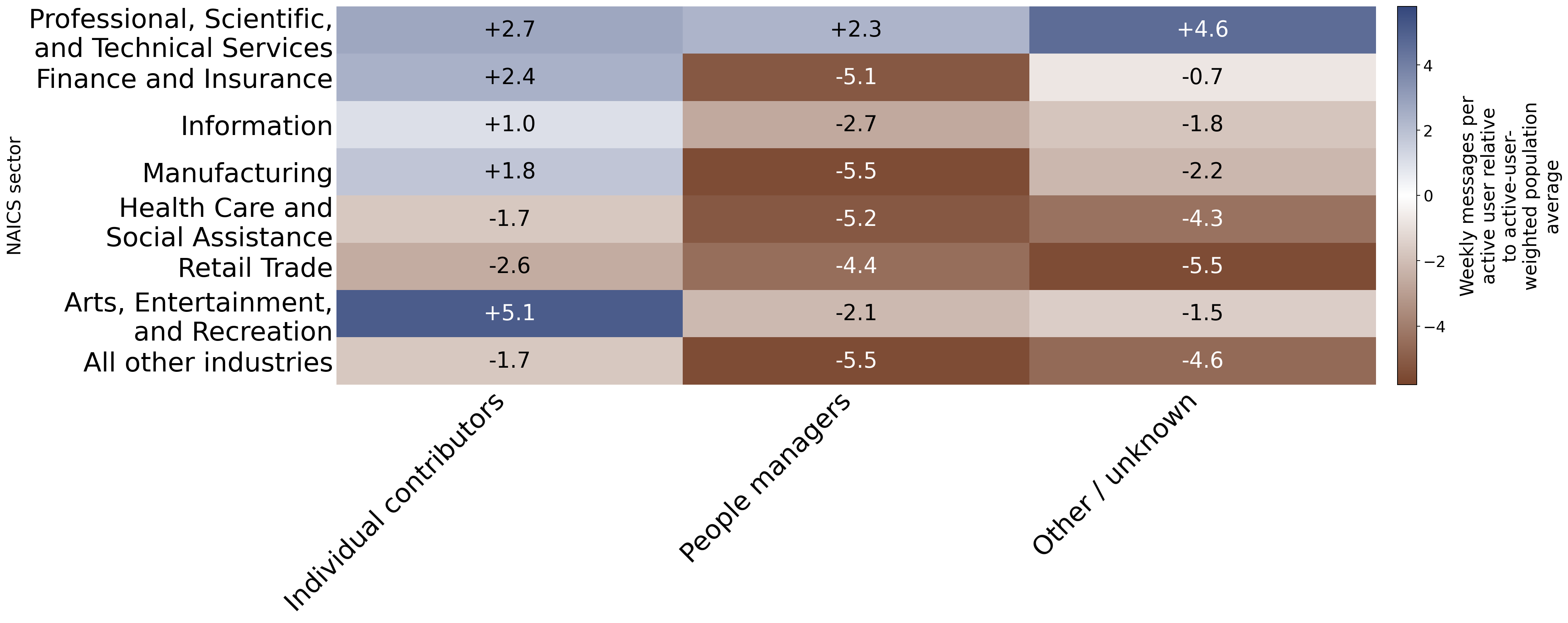}

\vspace{0.5em}
\begin{minipage}{\linewidth}
\footnotesize
\emph{Note:} This figure compares ChatGPT Enterprise use across two-digit NAICS sectors and inferred people manager status groups among firms with job title data, measured six months after adoption. Manager status is inferred from normalized job titles. Panel A reports the average share of weekly active users in each people manager status group within each industry; darker cells indicate that a larger share of active users belong to that group. Panel B reports usage intensity for the same industry-by-people-manager-status cells, measured as weekly messages per active user relative to the overall average active user in the sample. Positive values indicate more intensive use than average, while negative values indicate less intensive use. ``Other / unknown'' includes users whose job titles are missing, ambiguous, malformed, or otherwise do not provide enough information to infer manager status. ``All other industries'' pools two-digit NAICS sectors not displayed separately. Cell labels report percentages in Panel A and differences from the overall average in Panel B. A version of Panel B that accounts for firm fixed effects can be found in Figure~\ref{fig:people-manager-fe-app}.
\end{minipage}
\end{figure}

\vspace*{\fill}

\begin{figure}[htbp]
\centering
\caption{WITHIN-FIRM DIFFERENCES IN AI USAGE INTENSITY BY INDUSTRY, JOB TITLE CLASS AND SENIORITY LEVEL}
\label{fig:industry-job-title-variation-fe-app}

\textbf{Panel A. Job title class}\par
\vspace{0.25em}
\includegraphics[width=.75\linewidth]{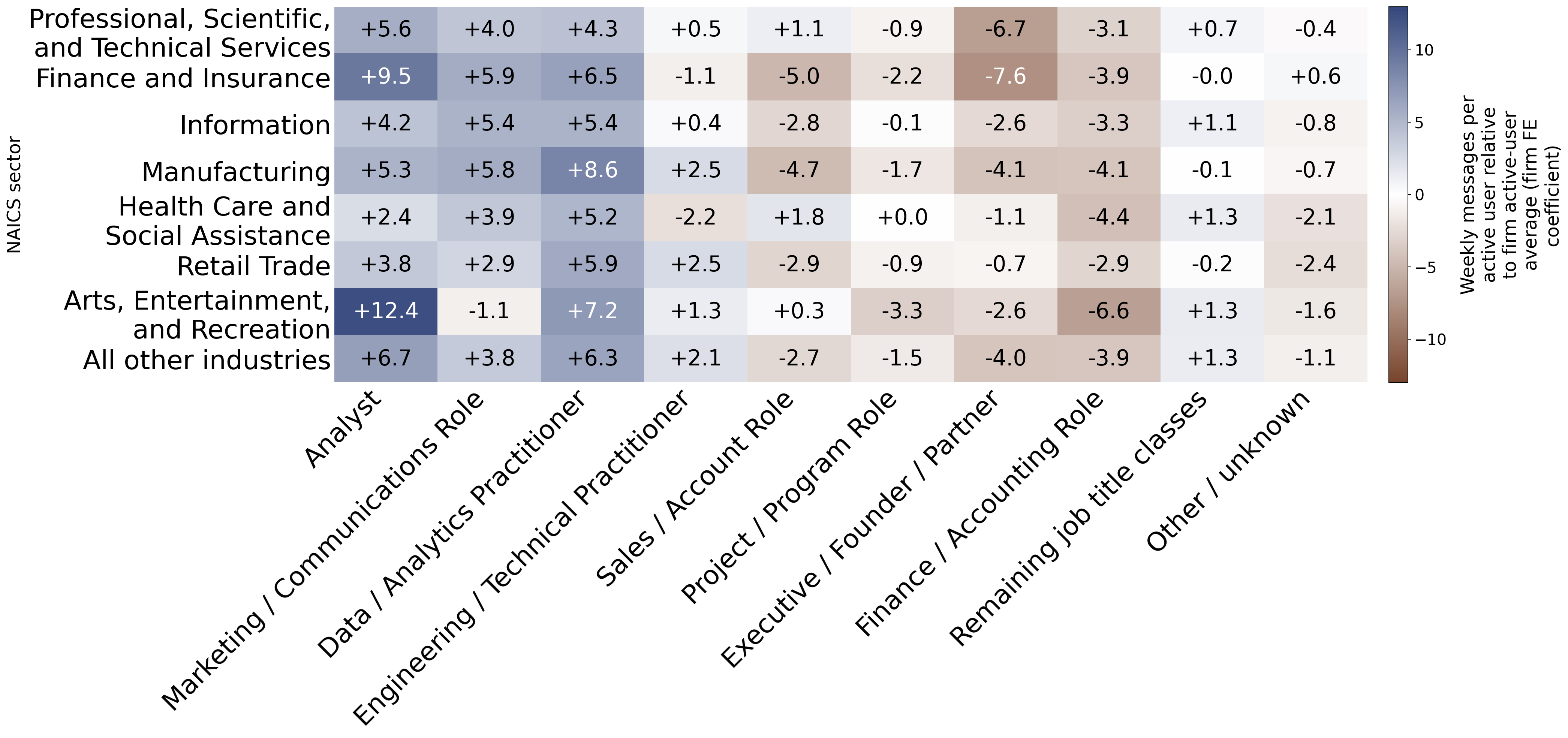}

\vspace{1em}

\textbf{Panel B. Seniority level}\par
\vspace{0.25em}
\includegraphics[width=.75\linewidth]{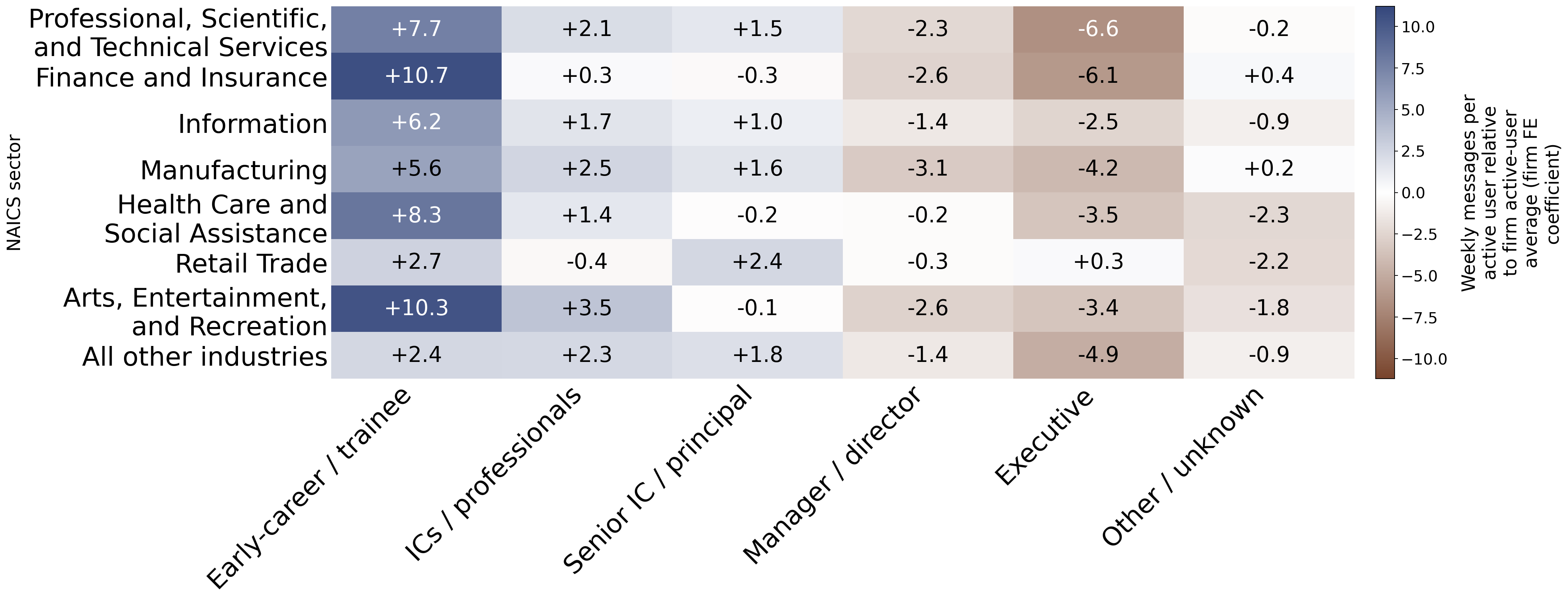}

\vspace{0.5em}
\begin{minipage}{\linewidth}
\footnotesize
\emph{Note:} This figure compares weekly ChatGPT Enterprise usage intensity across worker categories within firms and separately by two-digit NAICS sector, measured six months after adoption. Usage intensity is measured as weekly messages per active user. Panel A groups users by inferred job title class, while Panel B groups users by inferred seniority level. Each cell reports how usage intensity for that worker category differs from the average among active users at the same firm. Positive values indicate more intensive use than the within-firm average, while negative values indicate less intensive use. The estimates remove differences in overall usage intensity across firms and therefore describe within-firm differences across worker types, rather than differences in average usage across industries. ``Remaining job title classes'' pools substantive inferred classes outside the individually displayed categories. ``Other / unknown'' includes users whose job titles do not provide enough information to assign the relevant worker classification. ``IC'' denotes an individual contributor. ``All other industries'' pools two-digit NAICS sectors not displayed separately.
\end{minipage}
\end{figure}

\vspace*{\fill}

\vspace*{\fill}

\begin{figure}[htbp]
\centering
\caption{WITHIN-FIRM DIFFERENCES IN AI USAGE INTENSITY BY INDUSTRY AND PEOPLE MANAGER STATUS}
\label{fig:people-manager-fe-app}
\vspace{0.25em}
\includegraphics[width=\linewidth]{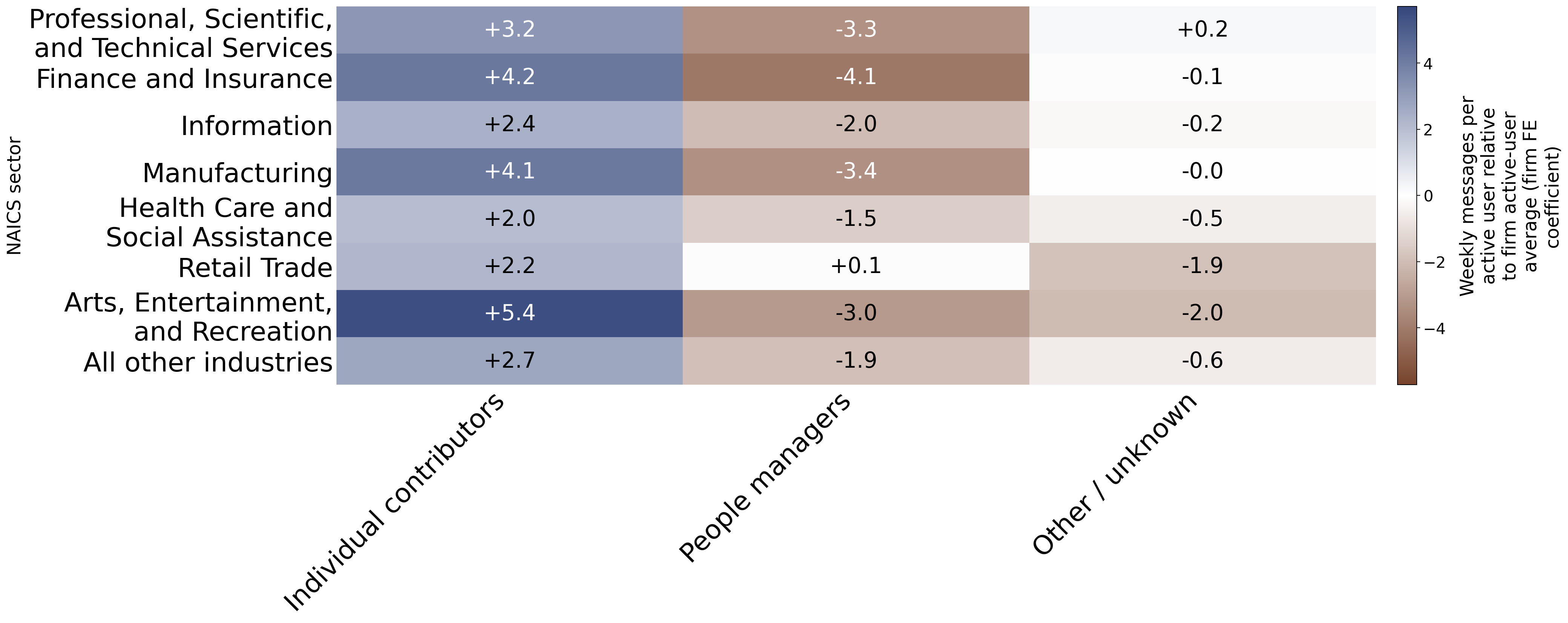}
\vspace{1em}
\begin{minipage}{\linewidth}
\footnotesize
\emph{Note:} This figure compares weekly ChatGPT Enterprise usage intensity across people manager status groups within firms and separately by NAICS two-digit sector, measured six months after adoption. People manager status is inferred from normalized job titles. Each cell reports how weekly messages per active user for that group differ from the average among active users at the same firm. Positive values indicate more intensive use than the within-firm average, while negative values indicate less intensive use. The estimates remove differences in overall usage intensity across firms and therefore describe within-firm differences by manager status, rather than differences in average usage across industries. ``Other / unknown'' includes users whose job titles are missing, ambiguous, malformed, or otherwise do not provide enough information to infer manager status. ``All other industries'' pools two-digit NAICS sectors not displayed separately.
\end{minipage}
\end{figure}

\vspace*{\fill}

\vspace*{\fill}

\begin{figure}[htbp]
\centering
\caption{DIFFERENCES IN AI TASK USE BY PEOPLE MANAGER STATUS}
\label{fig:industry-people-manager-task-variation}

\textbf{Panel A. Task Prevalence Among Active Users}\par
\vspace{0.25em}
\includegraphics[width=.75\linewidth]{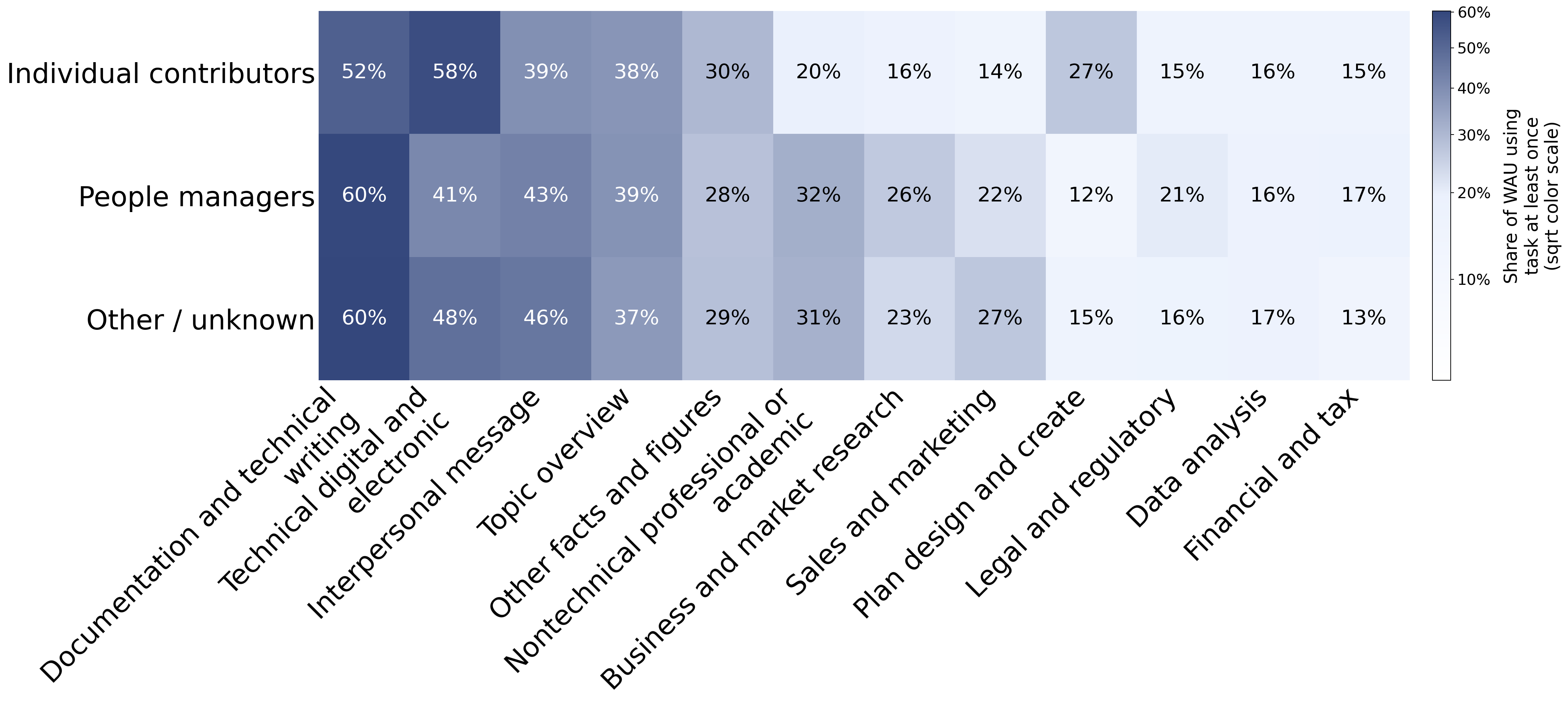}

\vspace{1em}

\textbf{Panel B. Distribution of Messages Across Tasks}\par
\vspace{0.25em}
\includegraphics[width=.75\linewidth]{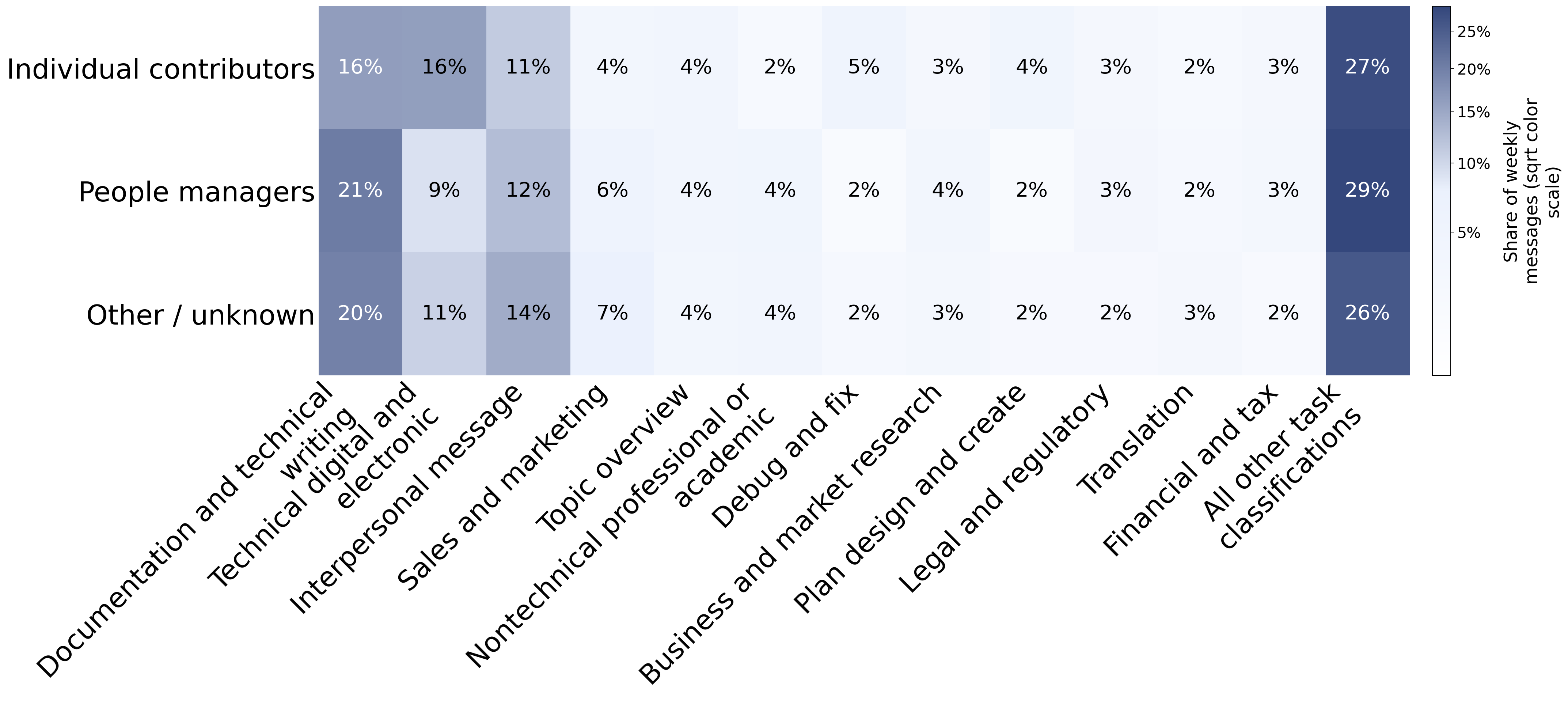}

\vspace{0.5em}
\begin{minipage}{\linewidth}
\footnotesize
\emph{Note:} This figure compares the distribution of ChatGPT Enterprise use across tasks and inferred people manager status groups among firms with task classification and job title data, measured six months after adoption. Each message is assigned to one of 60 task categories using the classifier described in the text, and manager status is inferred from normalized job titles. Panel A reports the share of weekly active users in each manager-status group who submitted at least one message in each of the 12 task categories with the greatest user reach. Because users may use ChatGPT for multiple tasks during the same week, the shares in each row are not mutually exclusive and do not sum to 100 percent. Panel B reports the distribution of classified weekly messages across the 12 largest task categories by message volume within each manager-status group. ``All other task classifications'' pools the remaining task categories. ``Other / unknown'' includes users whose job titles do not provide enough information to infer manager status. Task categories are selected separately in each panel using their prevalence in the full task-classification sample. 
\end{minipage}
\end{figure}

\newpage

\section{Job Title Classification}
\label{app:job-title-classification}

In this appendix, we detail the classifier that we use to classify organizational users' job titles. Note that this same classifier was used in \textcite{johnston2026shift}, and a nearly verbatim copy of this appendix appears in that paper as well. 

We use the gpt-5-mini model with minimal reasoning effort to classify organizational users' job titles. The classifier uses only the supplied job title, and returns structured outputs for inferred department, seniority, people manager status, and cleaned job title class. In the prompt below, we include only the definitions used for each field.

\subsection{Job Title Classification Prompt}

\begin{PromptVerbatim}
Classify this enterprise ChatGPT user job title. Use only the title below.

Field definitions:
- inferred_department is the likely department, function, or substantive work area. It is not seniority and it is not the literal title family.
- role_level is seniority or rank. It should be classified even when inferred_department is Other / Unknown.
- people_manager_signal is the likelihood that the title implies managing people. It is separate from role_level because many titles with Manager do not imply people management.
- job_title_class is the cleaned-up occupation or title-family wording. It should preserve generic occupations like Consultant and Analyst even when inferred_department remains Other / Unknown.

Allowed inferred_department labels:
- Financial Markets + Corporate Finance
- Sales
- Marketing
- Data Science
- Biz Ops
- Engineering
- Design
- User Ops / Customer Support
- Legal
- Product Management
- Project Management
- Healthcare / Life Sciences
- Education
- Policy
- Security
- People & Recruiting
- IT
- Partnerships
- Comms
- Research
- Other / Unknown

Inferred-department guidelines:
- Financial Markets + Corporate Finance: finance, accounting, audit, treasury, FP&A, investment, portfolio, banking, tax, controller, CFO roles.
- Sales: sales, account executive, account manager, business development, revenue, SDR/BDR, sales engineering when primarily commercial.
- Marketing: brand, growth, demand generation, content marketing, product marketing, SEO, lifecycle, communications only when clearly marketing.
- Data Science: data scientist, data analyst, analytics engineer, machine learning scientist, BI, quantitative analyst when primarily data or ML.
- Biz Ops: operations, strategy, chief of staff, business analyst, general management, consulting, program operations, administrative roles.
- Engineering: software, hardware, infrastructure, platform, systems, DevOps, QA, site reliability, technical engineering roles.
- Design: product design, UX, UI, visual design, creative design, design research.
- User Ops / Customer Support: customer success, support, customer experience, implementation, solutions, technical account management, help desk.
- Legal: legal, counsel, compliance legal, contracts, privacy counsel.
- Product Management: product manager, product owner, product lead, product strategy.
- Project Management: project manager, program manager, delivery manager, scrum master, agile coach, PMO.
- Healthcare / Life Sciences: clinical, medical, physician, nurse, pharmaceutical, biotech, lab, hospital, health science, life science research.
- Education: student, teacher, professor, instructor, faculty, academic, dean, curriculum, school administration.
- Policy: public policy, government affairs, trust and safety policy, regulatory policy, public affairs when policy-focused.
- Security: cybersecurity, information security, security engineering, GRC, threat, risk when security-specific.
- People & Recruiting: HR, people ops, talent, recruiting, compensation, benefits, learning and development.
- IT: IT, systems administration, enterprise applications, workplace technology, service desk, network administration.
- Partnerships: partnerships, alliances, partner management, ecosystem, channel partnerships.
- Comms: communications, PR, media relations, internal comms, corporate communications.
- Research: researcher, scientist, research engineer, research associate, lab research, academic research unless clearly healthcare or education.
- Other / Unknown: seniority-only, malformed, vague, or titles without enough department/function signal.

Allowed role_level labels:
- Student / Trainee / Intern
- Entry / Associate
- Individual Contributor / Professional
- Senior IC / Principal
- Manager / Team Lead
- Director / Head
- VP / Senior Executive
- C-suite / Founder / Owner / Partner
- Contractor / Temporary
- Unknown

Allowed people_manager_signal labels:
- Likely people manager
- Manager title but ambiguous
- Likely individual contributor
- Unknown

Allowed job_title_class labels:
- Consultant / Professional Services
- Analyst
- Engineering / Technical Practitioner
- Data / Analytics Practitioner
- Product Role
- Project / Program Role
- Sales / Account Role
- Customer Success / Support Role
- Finance / Accounting Role
- Legal Role
- Healthcare / Clinical Role
- Education / Academic Role
- Research / Scientist Role
- Design / Creative Role
- Administrative / Executive Support
- People / Recruiting Role
- IT / Systems Role
- Marketing / Communications Role
- Security Role
- Executive / Founder / Partner
- Generic Rank / Seniority Only
- Contractor / Temporary
- Placeholder / Malformed
- Unknown
\end{PromptVerbatim}

The classifier returns a JSON object with the following fields: inferred\_department, confidence, evidence\_source, rationale, role\_level, role\_level\_confidence, people\_manager\_signal, \linebreak people\_manager\_signal\_confidence, role\_rationale, job\_title\_class, job\_title\_class\_confidence, and job\_title\_class\_rationale.

Throughout the paper, before plotting, we collapse several fine-grained job title classifier outputs into coarser display categories and then recompute all totals at the collapsed-category level. For the job title class dimension, Unknown, Generic Rank / Seniority Only, Placeholder / Malformed, Contractor / Temporary, and missing or unclassified job titles are displayed as Other / unknown, while all other job title classes are left unchanged. For the seniority dimension, Student / Trainee / Intern and Entry / Associate are mapped to Early-career / trainee; Individual Contributor / Professional is mapped to ICs / professionals; Senior IC / Principal is mapped to Senior IC / principal; Manager / Team Lead and Director / Head are mapped to Manager / director; VP / Senior Executive and C-suite / Founder / Owner / Partner are mapped to Executive; and Unknown, Contractor / Temporary, and missing or unclassified job titles are displayed as Other / unknown. For the people manager status dimension, Likely individual contributor is mapped to Individual contributors, Likely people manager is mapped to People managers, and Manager title but ambiguous, Unknown, and missing or unclassified job titles are displayed as Other / unknown.

\subsection{Job Title Classifier Validation}

To provide some validation of the values assigned by our job title classifier, we report below the top five most common job titles for each unique inferred job title class, seniority level, and people manager status. To limit disclosure risk, we apply a pre-specified suppression rule to raw job title strings: after normalizing superficial title variants, we display a title only if it is observed in at least two distinct non-OpenAI organizations or in OpenAI’s internal sample. Titles that do not satisfy this rule are redacted. We do not report titles for the labels ``Unknown'' and ``Placeholder / Malformed,'' because their most common values often contained sensitive or identifying information.

\newpage

\section{Task Classifier Details} \label{app:task-classification}

The task classifier assigns each ChatGPT turn to exactly one value in a two-level taxonomy. The first level identifies the broad domain of the user request, and the second level identifies the more specific task type within that domain. The task classifier has been evaluated against an internal benchmark of human- and model-labelled ChatGPT conversations.  Table~\ref{tab:task-classifier-values} lists the possible classifier values.

\end{document}